\documentclass[11pt]{article}
\usepackage[letterpaper,margin=1in]{geometry}
\usepackage[utf8]{inputenc}
\usepackage[T1]{fontenc}
\usepackage{authblk}
\usepackage{dsfont}
\usepackage{lmodern}

\usepackage{amsmath,amssymb,amsthm,mathtools}
\usepackage{microtype}
\usepackage{cite}
\usepackage[dvipsnames]{xcolor}
\definecolor{linkblue}{HTML}{204A76}
\usepackage{hyperref}
\hypersetup{
colorlinks=true,
urlcolor=Cerulean,
linkcolor=NavyBlue,
citecolor=PineGreen,
linktocpage=true,
}

\newtheorem{theorem}{Theorem}
\newtheorem{lemma}{Lemma}

\newtheorem{corollary}[lemma]{Corollary}
\theoremstyle{definition}
\newtheorem{problem}[lemma]{Problem}

\theoremstyle{remark}

\newcommand{\norm}[1]{\left\lVert #1\right\rVert}
\newcommand{\ket}[1]{|#1\rangle}
\newcommand{\bra}[1]{\langle #1|}
\newcommand{\braket}[2]{\langle #1|#2\rangle}
\newcommand{\Adv}{\operatorname{Adv}}
\newcommand{\Tr}{\operatorname{Tr}}
\newcommand{\iu}{\mathrm{i}}
\newcommand{\GSEE}{\mathrm{GSEE}}
\newcommand{\GSP}{\mathrm{GSP}}
\newcommand{\e}{\mathrm e}

\title{Optimal Joint Bounds for Sparse Ground-State\\
Energy Estimation and Ground-State Preparation}
\author{Dhrumil Patel\thanks{\href{mailto:djp265@vt.edu}{djp265@vt.edu}}\\
Department of Computer Science, Virginia Tech, Alexandria, VA 22305, USA}
\date{\today}

\begin{document}

\maketitle

\begin{abstract}
Ground-state energy estimation (GSEE) and ground-state preparation (GSP) are fundamental problems in quantum computing. Given a Hamiltonian $H$ with $\|H\|\le1$ and a guiding state $|\psi\rangle$ with overlap at least $\gamma>0$ with the ground space of $H$, the goal in GSEE is to estimate the ground-state energy of $H$ to additive error $\delta$ with failure probability at most $\varepsilon$, while in GSP it is to prepare a state with ground-space weight at least $1-\varepsilon$, given a gap lower bound $\Delta$.

Their costs are measured by quantum query complexity, i.e., the number of oracle calls needed to access $H$. Recent work proved optimal bounds $\Theta(\log(1/\varepsilon)/(\gamma\delta))$ for GSEE and $\Theta(\log(1/\varepsilon)/(\gamma\Delta))$ for GSP in the time-evolution and block-encoding models. However, these models do not directly account for the sparsity of $H$, measured by the maximum number $s$ of nonzero entries in any row or column. In the sparse-access model, oracles instead return the locations and values of these entries. This introduces the additional parameter $s$, and it remained open how query complexity depends jointly on $s$, $\gamma$, $\delta$ or $\Delta$, and $\varepsilon$.

We resolve this by proving sparse-query lower bounds $\Omega(\sqrt{s}\log(1/\varepsilon)/(\gamma\delta))$ for GSEE and $\Omega(\sqrt{s}\log(1/\varepsilon)/(\gamma\Delta))$ for GSP, even for stoquastic Hamiltonians. We also show that these constructions recover the earlier time-evolution and block-encoding lower-bound scalings. We then give matching upper-bound algorithms in the reversible in-place sparse-location model. Thus, we obtain $\Theta(\sqrt{s}\log(1/\varepsilon)/(\gamma\delta))$ for GSEE and $\Theta(\sqrt{s}\log(1/\varepsilon)/(\gamma\Delta))$ for GSP, showing that $\sqrt{s}$ is necessary and sufficient and completing the joint dependence on all problem parameters.
\end{abstract}

\newpage

\tableofcontents

\newpage

\section{Introduction}\label{sec:intro}

Ground-state energy estimation and ground-state preparation are fundamental problems in quantum computing, with applications to quantum chemistry and many-body physics~\cite{Cao2019}. Let $H$ be a Hamiltonian, which we assume throughout the paper to satisfy the standard normalization $\norm H\le1$, where $\norm H$ denotes its operator norm. Let $E_0(H)$ denote its smallest eigenvalue, called the ground-state energy, and let the eigenspace corresponding to $E_0(H)$ be the ground space of $H$, which need not be one-dimensional. In ground-state energy estimation (GSEE), the goal is to estimate $E_0(H)$ to some desired accuracy, whereas in ground-state preparation (GSP), the goal is to prepare a quantum state supported almost entirely on the ground space. These problems are closely related, since a ground state can be used to estimate its energy, while a sufficiently accurate estimate of $E_0(H)$ can help distinguish the ground space from higher-energy eigenspaces when these eigenspaces are separated by a known spectral gap of $H$.

These tasks are computationally difficult for general Hamiltonians. For example, the local Hamiltonian problem asks whether the ground-state energy of a Hamiltonian specified as a sum of local terms lies below a threshold $a$ or above a threshold $b>a$, under the promise that one of these possibilities holds. When $b-a$ is inverse polynomial in the number of qubits, this decision problem is QMA-complete even for two-local Hamiltonians~\cite{Kempe2006}. The difficulty also applies to preparing ground states of arbitrary local Hamiltonians, since an algorithm that efficiently prepared such a state could be followed by an energy measurement accurate enough to distinguish the two promised cases. Thus we do not expect efficient algorithms for arbitrary Hamiltonians without additional information or promises.

One useful form of such additional information is access to a state
that has a guaranteed overlap with the ground space of $H$. For example, an approximate classical calculation of a molecular wavefunction, such as a Hartree--Fock calculation or a calculation using a small collection of Slater determinants, may suggest a state that can be prepared by an efficient quantum circuit. However, a substantial overlap with the ground space of the molecular Hamiltonian is not guaranteed for every instance~\cite{Fomichev2024}. Rather than requiring an algorithm to find a suitable state for every $H$, we can ask how efficiently GSEE and GSP can be solved when access to such a state is given. This is the setting of the well-studied \emph{guided} versions of GSEE and GSP~\cite{Poulin2009,Ge2019,Lin2020}.

More precisely, in these guided versions, the algorithm is given
access to a unitary $A$ that prepares a state $\ket\psi=A\ket0$,
as well as to $A^\dagger$ and the controlled versions of both unitaries. Here, the state $\ket\psi$ is called a guiding state. Let $\Pi_0(H)$ denote the orthogonal projector onto the ground space of $H$. We assume the overlap promise $\norm{\Pi_0(H)\ket\psi}\ge\gamma>0$, where $\norm v$ denotes the Euclidean norm when $v$ is a vector. Thus $\gamma$ is a known lower bound on the ground-space overlap amplitude of the guiding state, and the cost of guided algorithms generally increases as $\gamma$ decreases. Throughout this paper, a guided algorithm means an algorithm given access to $A$, $A^\dagger$, and their controlled versions. Since we focus only on the guided versions in this paper, we will use GSEE and GSP to mean guided GSEE and guided GSP, respectively.

Given this guiding state and the overlap promise, the goal of GSEE is to output a classical estimate $\widetilde E$ satisfying $\Pr[|\widetilde E-E_0(H)|\le\delta]\ge1-\varepsilon$, where $\delta>0$ is the allowed additive energy-estimation error and $0<\varepsilon<1/2$ is the allowed failure probability. For GSP, we additionally assume the gap promise that a known number $\Delta>0$ satisfies $E-E_0(H)\ge\Delta$ for every eigenvalue $E>E_0(H)$. The goal is then to prepare a quantum state $\rho$ satisfying $\Tr(\Pi_0(H)\rho)\ge1-\varepsilon$. The overlap promise guarantees that the guiding state has a component in the ground space of $H$, while the gap promise allows that component to be distinguished from the higher-energy components. Note that unlike GSP, GSEE does not require the gap promise. We will later give the formal definitions of these problems in Problems~\ref{prob:gsee} and~\ref{prob:gsp}.

Having specified the input promises and the desired outputs, one can now ask how an algorithm accesses $H$ and how much of this access is needed to solve these problems. These questions are usually studied in the standard framework of quantum query complexity. In this framework, a query model specifies the oracles through which the algorithm accesses $H$, and one application of an oracle counts as one query. The query complexity is the minimum number of queries required by an algorithm that succeeds on every input satisfying the promises. Different query models provide different forms of access to $H$, so their query complexities need not be the same. Throughout this paper, we count queries to the oracles for $H$, while treating the supplied guiding-state preparation unitaries $A$ and $A^\dagger$ (and their controlled versions) separately.

One well-studied access model is the block-encoding model. In this model, the algorithm can query a unitary $B$ acting on the system and additional registers, where a specified matrix block of $B$ equals $H/\alpha$ for a known normalization factor $\alpha$. A quantum algorithm can use such a unitary to apply polynomial functions of $H/\alpha$, which can suppress higher-energy components of $\ket\psi$ while retaining its ground-space component~\cite{Gilyen2019}. Ge, Tura, and Cirac developed methods for GSP and high-precision GSEE starting from a guiding state~\cite{Ge2019}, and Lin and Tong subsequently obtained near-optimal algorithms based on polynomial filtering in the block-encoding model~\cite{Lin2020}. More recently, Chen et al.~\cite{Chen2026} established an optimal worst-case GSP query bound in the block-encoding model, assuming a unique ground state and a known threshold in the spectral gap. These results show how the overlap promise and gap promise can be used in algorithms when the corresponding access to $H$ is available.

A second access model is the time-evolution query model. Here the algorithm can apply the unitary $U=\e^{\iu H}$, its inverse $U^\dagger=\e^{-\iu H}$, and their controlled versions, and each application counts as one query to $H$. Dong, Lin, and Tong developed GSEE and GSP algorithms using controlled time evolution through quantum eigenvalue transformation of unitary matrices~\cite{Dong2022}. More recently, Patel, Flammia, and Garc\'{i}a-Patr\'{o}n considered a weaker time-evolution access model in which only $U$ and $U^\dagger$ are supplied, without their controlled versions. They obtained a near-optimal GSP algorithm even without controlled time evolution, showing that these controlled operations are not needed to achieve the optimal dependence on the overlap and gap up to logarithmic factors~\cite{Patel2026a}. These results show how guided algorithms can be developed under
different forms of access to $H$. 

For the time-evolution query
model, matching upper and lower bounds have now established
the optimal query complexities for both problems.
For GSEE, the upper bound of Jeffery and Witteveen~\cite{JW2026} and the lower bound of Somma and de Wolf~\cite{SdW2026} give
\[
 \Theta\!\left(\frac{1}{\gamma\delta}\log\frac1\varepsilon\right)
\]
queries to these unitaries. For GSP, the upper bound of Jeffery, Somma, Witteveen, and de Wolf~\cite{JSWdW2026}, together with the corresponding lower bound~\cite{SdW2026}, gives
\[
 \Theta\!\left(\frac{1}{\gamma\Delta}\log\frac1\varepsilon\right).
\]

The time-evolution query model assumes that an application of $U=\e^{\iu H}$ has unit query cost. This is appropriate for studying the information available through time evolution, but it leaves a different question. How many queries are needed when $H$ is supplied through its matrix entries rather than through $U$? We now turn to the sparse-access model to address this question.

In the standard sparse-access model, $H$ is promised to be $s$-sparse, meaning that each row and each column of $H$ contains at most $s$ nonzero entries. The algorithm accesses $H$ through sparse-matrix oracles that return the locations and values of its entries, as defined precisely in Section~\ref{sec:sparse-matrix-access}. Thus one query to a sparse-matrix oracle provides information about entries of $H$, rather than directly applying the time-evolution unitary $U$. This model is useful when $H$ acts on a large Hilbert space but connects each computational-basis state to only a small number of other basis states. For example, suppose an $n$-qubit Hamiltonian is a sum of $m$ terms, each acting on at most $k$ qubits. Each term has at most $2^k$ nonzero entries in any row in the computational basis, so the sum has sparsity at most $s\le m2^k$. Hence, when $m$ is polynomial in $n$ and $k$ is constant, $s$ is polynomial in $n$, even though the matrix has dimension $2^n$. Many local Hamiltonians and lattice models therefore admit sparse descriptions, provided their nonzero entries and locations can be computed efficiently.

Unlike the time-evolution query model, the sparse-access model introduces the sparsity $s$ as an additional problem parameter. We therefore want to understand how the query complexity depends jointly on $s$ and the parameters already present in the time-evolution query bounds shown above, namely the overlap parameter $\gamma$, the energy-estimation error $\delta$ for GSEE or the gap lower bound $\Delta$ for GSP, and the failure parameter $\varepsilon$.

Sparse-matrix queries can be used to approximate the time-evolution unitary $U$. Indeed, Berry, Ahokas, Cleve, and Sanders developed quantum algorithms for simulating sparse Hamiltonians using these queries~\cite{Berry2006}, and later work improved their dependence on the relevant simulation parameters~\cite{Berry2015}. Consequently, one way to solve GSEE or GSP in the sparse-access model is to start with an algorithm in the time-evolution query model and replace each of its queries to $U$, $U^\dagger$, or their controlled versions by a sparse Hamiltonian simulation. This connects the two access models, but it does not make them equivalent. In particular, a sparse-access algorithm can use the matrix-entry information in a different way without first constructing $U$, so a lower bound on time-evolution queries in the time-evolution query model does not automatically imply the same lower bound on sparse-matrix queries in the sparse-access model. Ref.~\cite[Section~5]{SdW2026}, which proves the time-evolution lower bounds discussed above, explicitly leaves the sparse-access lower bound as an open problem.

Although the time-evolution lower bounds do not directly give sparse-access lower bounds, the sparse Hamiltonian simulation algorithms just discussed allow us to obtain upper bounds for GSEE and GSP. We now examine the number of sparse-matrix queries needed to implement the known time-evolution algorithms. For this purpose, we use the reversible in-place sparse-location model of Ref.~\cite{BRSL2026}. This model provides stronger location oracles than the standard sparse-access model, and we explain the distinction in Section~\ref{sec:sparse-matrix-access}. In the reversible in-place model, Ref.~\cite{BRSL2026} gives an algorithm that approximates one query to $U=\e^{\iu H}$ or $U^\dagger$ with constant operator-norm error using $O(\sqrt{s})$ sparse-matrix queries, under the normalization $\norm H\le1$. It is therefore natural to ask whether an algorithm using $q$ time-evolution queries can be implemented using only $O(q\sqrt{s})$ sparse-matrix queries.

The difficulty is that these $q$ time-evolution queries belong to the same computation, so the errors in their sparse implementations must be controlled together. Suppose we replace each query to $U$, $U^\dagger$, or their controlled versions by an approximation with operator-norm error at most $\zeta$. By replacing the $q$ queries one at a time and applying the triangle inequality, the error of the complete computation is at most $q\zeta$. Thus choosing $\zeta=O(1/q)$ is sufficient to keep the final error constant. With this choice, the precision-dependent simulation bound of Ref.~\cite{BRSL2026} gives a total of
\[
 O\!\left(q\sqrt{s}+q\sqrt{s\log(q+1)}+q\log(q+1)\right)
\]
sparse-matrix queries. This expression differs from $O(q\sqrt{s})$ because the cost of approximating each time-evolution query increases when its required error $\zeta$ decreases.

For constant $\varepsilon$, the optimal algorithms in the time-evolution query model use $q=O(1/(\gamma\delta))$ queries for GSEE and $q=O(1/(\gamma\Delta))$ queries for GSP. Substituting these values of $q$ into the preceding simulation bound yields sparse-query upper bounds $\widetilde O(\sqrt{s}/(\gamma\delta))$ and $\widetilde O(\sqrt{s}/(\gamma\Delta))$, respectively, where $\widetilde O$ hides additional logarithmic factors. We therefore obtain sparse-access algorithms by implementing the known time-evolution algorithms, but this argument does not establish that their dependence on $s$ is optimal. A different sparse-access algorithm might use fewer than $\sqrt{s}$ additional queries, and the extra logarithmic factors from the individual simulations might also be avoidable.

A closely related question was recently resolved for quantum linear systems. In that problem, the input is an $s$-sparse matrix with condition number $\kappa$, and the goal is to prepare a state approximating the solution of a linear system to error $\varepsilon$. Ref.~\cite{Patel2026} proves a joint sparse-query lower bound of $\Omega(\kappa\sqrt{s}\log(1/\varepsilon))$, while independent work by Bravo-Prieto, Harrow, and Kothari gives matching upper and lower bounds~\cite{BravoPrieto2026}. Thus the sparsity factor $\sqrt{s}$ is necessary in that problem, and it multiplies the full dependence on $\kappa$ and $\log(1/\varepsilon)$. For GSEE and GSP, the optimal time-evolution query bounds are already known, but the corresponding joint dependence on sparsity has not been established. This leads to the two questions addressed in this paper.

\bigskip
\noindent\emph{(Q1) Do GSEE and GSP require $\Omega\!\left(\frac{\sqrt{s}}{\gamma\delta}\log\frac1\varepsilon\right)$ and $\Omega\!\left(\frac{\sqrt{s}}{\gamma\Delta}\log\frac1\varepsilon\right)$ sparse-matrix queries, respectively, in the worst case?}

\medskip
\noindent\emph{(Q2) Can we attain matching upper bounds in the reversible in-place sparse-location model, without the additional logarithmic factors introduced by approximating each time-evolution query separately?}

\bigskip

Answering both questions affirmatively would establish the optimal joint dependence of the sparse-query complexity on all the problem parameters. In particular, this would give the following:
\begin{equation}\label{eq:intro-both}
  Q_{\rm sparse}^{\GSEE}
  =\Theta\!\left(
    \frac{\sqrt{s}}{\gamma\delta}\log\frac1\varepsilon
  \right),
  \qquad
  Q_{\rm sparse}^{\GSP}
  =\Theta\!\left(
    \frac{\sqrt{s}}{\gamma\Delta}\log\frac1\varepsilon
  \right),
\end{equation}
where $Q_{\rm sparse}^{\GSEE}$ and $Q_{\rm sparse}^{\GSP}$ denote the minimum numbers of sparse-matrix queries sufficient to solve GSEE and GSP, respectively, in the worst case.

\subsection{Contributions}\label{sec:contributions}

Our first contribution answers the lower-bound question (Q1) affirmatively.
For GSEE, Theorem~\ref{thm:gsee} proves
\begin{equation}\label{eq:intro-gsee}
  Q_{\rm sparse}^{\GSEE}
  =\Omega\!\left(
    \frac{\sqrt{s}}{\gamma\delta}\log\frac1\varepsilon
  \right),
\end{equation}
and for GSP, Theorem~\ref{thm:gsp} proves
\begin{equation}\label{eq:intro-gsp}
  Q_{\rm sparse}^{\GSP}
  =\Omega\!\left(
    \frac{\sqrt{s}}{\gamma\Delta}\log\frac1\varepsilon
  \right).
\end{equation}
We show that both lower bounds hold even for stoquastic Hamiltonians, meaning that $H_{ij}$ is real and $H_{ij}\le0$ for every $i\ne j$ in the basis used by the sparse-matrix oracles.
Our proofs use the Boolean query problem \textsc{Position-Parity}, which was also used to prove the quantum linear systems lower bound in Ref.~\cite{Patel2026}. We will introduce this problem later in Section~\ref{sec:boolean}.

These lower bounds hold in the standard sparse-access model, and they remain valid when the stronger reversible in-place sparse-location oracles are available.
To see why the additional access does not invalidate the lower bounds, consider how our proofs encode a Boolean query problem in a Hamiltonian.
The Boolean input determines the values of certain matrix entries, but the nonzero-entry locations are the same for every input.
The additional location oracles and their inverses can therefore be implemented without querying the Boolean input.
Consequently, the lower bounds remain valid even when these location operations are supplied at no query cost, which allows us to compare them with algorithms that use the stronger model.

Our second contribution answers the upper-bound question (Q2) by establishing matching bounds in the reversible in-place sparse-location model.
Recall that the direct simulation approach introduced additional logarithmic factors because each time-evolution query had to be approximated accurately enough to control the accumulated error.
We show how to avoid this overhead using the sparse-matrix representation of Ref.~\cite{BRSL2026} together with the transducer framework of Ref.~\cite{Belovs2024}.
We explain this approach and its application to both GSEE and GSP in Section~\ref{sec:upper}.
Theorem~\ref{thm:upper} then establishes the resulting upper bounds
\begin{equation}\label{eq:intro-upper}
  Q_{\rm sparse}^{\GSEE}
  =O\!\left(
    \frac{\sqrt{s}}{\gamma\delta}\log\frac1\varepsilon
  \right),
  \qquad
  Q_{\rm sparse}^{\GSP}
  =O\!\left(
    \frac{\sqrt{s}}{\gamma\Delta}\log\frac1\varepsilon
  \right).
\end{equation}

Combining the lower bounds in~\eqref{eq:intro-gsee} and~\eqref{eq:intro-gsp} with the upper bounds in~\eqref{eq:intro-upper}, we obtain matching bounds in the reversible in-place sparse-location model, thereby proving~\eqref{eq:intro-both}. Thus, the factor $\sqrt{s}$ is both necessary and sufficient, and the joint dependence of the sparse-query complexity on all the problem parameters is optimal up to constant factors.

\paragraph{Organization.}
The remainder of the paper is organized as follows.
In Section~\ref{sec:prelim}, we give the formal problem definitions
for GSEE and GSP and describe the sparse-access models. In Sections~\ref{sec:gsee}
and~\ref{sec:gsp}, we establish lower bounds on the sparse-query
complexities of GSEE and GSP, respectively, and in
Section~\ref{sec:access-model-connections}, we show how the same
constructions recover the corresponding query complexity lower
bounds in the block-encoding and time-evolution models.
We then prove the matching sparse-query upper bounds in the
reversible in-place sparse-location model in Section~\ref{sec:upper}
and discuss the scope and limitations of our results in
Section~\ref{sec:discussion}.

\section{Preliminaries and Access Model}\label{sec:prelim}

\paragraph{Notation.} All Hilbert spaces considered below are finite-dimensional. For a vector $v$, we write $\norm{v}$ for its Euclidean norm, and for a matrix $A$, we write $\norm{A}$ for its operator norm and $\norm{A}_1$ for its trace norm. 
We write $X\preceq Y$ when $Y-X$ is positive semidefinite. We use $\Tr(A)$ to denote the trace of $A$, and $A\circ B$ for the Hadamard (entrywise) product of two matrices. 
For a positive integer $N$, let $[N]=\{1,\ldots,N\}$ and $\mathbb Z_N=\{0,\ldots,N-1\}$. 
We index the bits of a Boolean string $x\in\{0,1\}^N$ by $q\in\mathbb Z_N$ and use $\oplus$ for bitwise XOR on bit strings and for direct sums on vector spaces, vectors, and matrices. The meaning will be clear from the operands. 
All logarithms are natural unless otherwise specified. We use $O$, $\Omega$, and $\Theta$ with absolute implicit constants unless a dependence is stated explicitly.

\bigskip

For a Hamiltonian $H\in\mathbb C^{D\times D}$, where $D$ denotes the dimension of the Hilbert space, let $E_0(H)$ denote its smallest eigenvalue, called the ground-state energy, and let $\Pi_0(H)$ denote the orthogonal projector onto its ground space, namely the eigenspace corresponding to $E_0(H)$. The spectral gap of $H$ is the difference between its ground-state energy and the next distinct eigenvalue, which we define as
\begin{equation}\label{eq:gap-definition}
  \operatorname{gap}(H)
  =\min\{E-E_0(H):E\in\operatorname{spec}(H),\ E>E_0(H)\}.
\end{equation}
Here $\operatorname{spec}(H)$ denotes the set of eigenvalues of $H$. If $H$ has only one distinct eigenvalue, we take $\operatorname{gap}(H)=+\infty$. In particular, the ground space need not be one-dimensional, and the gap is measured above the entire ground space.

A Hamiltonian $H$ is called \emph{stoquastic} in a specified orthonormal basis if $H_{ij}\in\mathbb R$ and $H_{ij}\le0$ whenever $i\ne j$. Since a Hermitian matrix has real diagonal entries, such a Hamiltonian is real symmetric. Throughout this paper, stoquasticity refers to the same basis in which the sparse oracles return matrix entries. The property is preserved by direct sums, diagonal shifts, and multiplication by a positive scalar.

We now formally define the two problems considered in this paper.

\begin{problem}[Ground-state energy estimation (GSEE)]\label{prob:gsee}
Let $H\in\mathbb C^{D\times D}$ be a Hamiltonian satisfying
$\norm H\le1$, and let $A$ be a unitary that prepares a guiding
state $\ket\psi=A\ket0$ with ground-space overlap amplitude
$\norm{\Pi_0(H)\ket\psi}\ge\gamma$, where $0<\gamma\le1$ is known.
Given sparse access to $H$, access to $A$ and $A^\dagger$
(and their controlled versions), and parameters $0<\delta\le1$
and $0<\varepsilon<1/2$, the goal is to output an estimate
$\widetilde E$ of the ground-state energy $E_0(H)$ such that
\begin{equation}\label{eq:gsee-error}
  \Pr\bigl[|\widetilde E-E_0(H)|\le\delta\bigr]\ge1-\varepsilon.
\end{equation}
\end{problem}

\bigskip

\begin{problem}[Ground-state preparation (GSP)]\label{prob:gsp}
Under the Hamiltonian, oracle-access, and guiding-state assumptions of Problem~\ref{prob:gsee}, suppose additionally that $\operatorname{gap}(H)\ge\Delta$ for some $\Delta>0$. Given $\Delta$ and $0<\varepsilon<1/2$, output a quantum state $\rho$ satisfying
\begin{equation}\label{eq:gsp-error}
  \Tr\bigl(\Pi_0(H)\rho\bigr)\ge1-\varepsilon.
\end{equation}
\end{problem}

Here $\rho$ denotes the final output state averaged over all
internal randomness and measurement outcomes.
Moreover, the condition in~\eqref{eq:gsp-error} is weaker than requiring the output to be close in trace distance to a state supported on the ground space. Indeed, if $\sigma$ is supported on that space and $\frac12\norm{\rho-\sigma}_1\le\varepsilon$, then $\Tr(\Pi_0(H)\rho)\ge1-\varepsilon$. Thus, any lower bound for Problem~\ref{prob:gsp} also applies when this stronger approximation guarantee is required.

\subsection{Sparse-matrix access}
\label{sec:sparse-matrix-access}

A Hamiltonian $H\in\mathbb C^{D\times D}$ is called $s$-sparse if every row and every column contains at most $s$ nonzero entries. We assume $1\le s\le D$. In the sparse-access model, an algorithm obtains information about $H$ through two oracles: a position oracle that identifies the locations of its nonzero entries and a value oracle that returns its matrix entries.

To define the position oracle, for each row $i\in[D]$, let $\pi_H(i,\ell)$, with $\ell\in[s]$, specify a list of $s$ distinct column indices containing every nonzero entry in that row. 
If the row contains fewer than $s$ nonzero entries, the remaining positions are filled with distinct indices corresponding to zero entries. 
The position oracle is the unitary
\begin{equation}\label{eq:pos}
  O_H^{\rm pos}\ket{i,\ell,j}
  =\ket{i,\ell,j\oplus\pi_H(i,\ell)}.
\end{equation}
Thus, by initializing the third register to zero, one query returns the column index $\pi_H(i,\ell)$.

The value oracle provides the corresponding matrix entries. For row and column indices $i,j\in[D]$, it acts as
\begin{equation}\label{eq:val}
  O_H^{\rm val}\ket{i,j,z}
  =\ket{i,j,z\oplus[H_{ij}]},
\end{equation}
where $[H_{ij}]$ is a fixed finite binary encoding of the matrix entry $H_{ij}$, with real and imaginary parts encoded separately when necessary. The encoded entries define $H$ exactly; approximation of a separately specified real-valued matrix is not included in this query model. Operations that do not query the matrix oracles are not charged to the sparse-query count. 
In our lower-bound construction, all matrix entries are real dyadic numbers, meaning that each entry can be written as $a/2^r$ for integers $a$ and $r\ge0$. Consequently, every entry in the constructed families has an exact finite binary representation.

We measure query complexity by the maximum number of calls to $O_H^{\rm pos}$ and $O_H^{\rm val}$ made by an algorithm over all promised inputs. 
In addition to this, calls to their inverses and controlled versions are also allowed, with each call counted as one query. For our lower bounds, we allow calls to $A$ and $A^\dagger$ (and their controlled versions), as well as all operations independent of the unknown Hamiltonian, at no query cost. This makes the lower bounds stronger, since the algorithm is given these additional operations without charge.

For the upper bounds, we use a stronger sparse-access model that also provides reversible in-place row and column location oracles. We define these oracles explicitly. For each row $i\in[D]$, choose a permutation $\pi_i:[D]\to[D]$ such that $\pi_i(1),\ldots,\pi_i(s)$ list the nonzero column locations of that row, together with distinct zero-entry locations if padding is necessary. The corresponding in-place row oracle acts as
\begin{equation}\label{eq:row-inplace}
  O_{H,\rm row}^{\rm ip}\ket{i,\ell}
  =\ket{i,\pi_i(\ell)},
  \qquad \ell\in[D].
\end{equation}
Since $\pi_i$ is a permutation, its inverse $\pi_i^{-1}$ exists, and the model also provides the inverse oracle
\begin{equation}\label{eq:row-inplace-inverse}
  (O_{H,\rm row}^{\rm ip})^\dagger\ket{i,j}
  =\ket{i,\pi_i^{-1}(j)}.
\end{equation}
Similarly, for each column $j\in[D]$, choose a permutation $\kappa_j:[D]\to[D]$ whose first $s$ outputs contain every nonzero row location in that column, with distinct zero-entry locations used for padding. The in-place column oracle and its inverse are defined by
\begin{equation}\label{eq:column-inplace}
  O_{H,\rm col}^{\rm ip}\ket{j,\ell}
  =\ket{j,\kappa_j(\ell)},
  \qquad
  (O_{H,\rm col}^{\rm ip})^\dagger\ket{j,i}
  =\ket{j,\kappa_j^{-1}(i)}.
\end{equation}
The upper-bound model provides these four location operations, together with the value oracle~\eqref{eq:val}, and counts each oracle call as one query.

The distinction between the two access models is important. The out-of-place oracle~\eqref{eq:pos} does not, in general, provide an implementation of the in-place permutation or its inverse at constant query cost. We therefore prove the upper bounds under the stronger in-place access assumptions. Our lower bounds remain valid even with this additional access, because the Hamiltonians in our construction have the same nonzero-entry locations for every hidden Boolean input. Consequently, their row and column permutations can be chosen independently of that input and supplied without revealing any information about it.

\subsection{Boolean queries and the XOR lemma}\label{sec:boolean-prelim}

Our lower-bound proofs use reductions from Boolean query problems, so we first recall the quantum query model for such problems.
Let $x\in\{0,1\}^n$ be a Boolean string whose bits are accessible through an oracle.
The standard Boolean query oracle $O_x$ acts as
\begin{equation}\label{eq:boolean-oracle}
  O_x\ket{q,b}
  =\ket{q,b\oplus x_q},
  \qquad q\in\mathbb Z_n,\quad b\in\{0,1\}.
\end{equation}
Thus, the first register specifies the position $q$ to be queried, and the oracle adds the corresponding bit $x_q$ to the second register.
In particular, setting $b=0$ allows us to read $x_q$.
Since applying the oracle twice adds the same bit twice, we have $O_x^2=I$, and hence $O_x^\dagger=O_x$.
A Boolean query therefore means one application of $O_x$, which is also its own inverse.

We consider Boolean functions that may be defined only on a subset of all possible inputs.
More precisely, let $f:\mathcal D\to\{0,1\}$, where $\mathcal D\subseteq\{0,1\}^n$ is the set of inputs satisfying a specified promise.
For $0\le\eta\le1/2$, let $Q_\eta(f)$ denote the minimum number of Boolean queries needed by a quantum algorithm to output $f(x)$ with probability at least $1-\eta$ for every $x\in\mathcal D$.
We measure the number of queries in the worst case over all promised inputs. At $\eta=1/2$, random guessing gives $Q_{1/2}(f)=0$.
Our goal is to establish lower bounds on $Q_\eta(f)$ for the Boolean functions used in our Hamiltonian constructions.

We first use the negative-weight adversary method of Ref.~\cite{HLS2007}.
This method gives a lower bound on the number of queries needed to distinguish inputs on which $f$ takes different values.
To define it, let $\Gamma$ be a nonzero real symmetric matrix whose rows and columns are indexed by inputs $x,y\in\mathcal D$.
The matrix $\Gamma$ must satisfy $\Gamma_{xy}=0$ whenever $f(x)=f(y)$, so that its nonzero entries connect only inputs with different function values.
The entries of $\Gamma$ may be positive or negative, which is why this is called the negative-weight adversary method.

For each bit position $q\in\mathbb Z_n$, define a matrix $D_q$ by
\begin{equation}
  (D_q)_{xy}=
  \begin{cases}
    1,&x_q\ne y_q,\\
    0,&x_q=y_q.
  \end{cases}
\end{equation}
Thus, $D_q$ identifies pairs of inputs that differ at position $q$, while $\Gamma\circ D_q$ retains only the corresponding entries of $\Gamma$.
The negative-weight adversary bound is defined as
\begin{equation}\label{eq:adversary}
  \Adv^\pm(f)
  =\max_{\Gamma\ne0}
   \frac{\norm\Gamma}
        {\max_{q\in\mathbb Z_n}\norm{\Gamma\circ D_q}},
\end{equation}
where the maximum is over all real symmetric matrices $\Gamma$ satisfying the condition above. If $f$ is constant on its domain, we set $\Adv^\pm(f)=0$.

The importance of this quantity is that it lower-bounds the quantum query complexity of $f$.
In particular, for any fixed constant error probability strictly below $1/2$, such as $\eta=1/3$, the adversary method gives
$Q_{1/3}(f)=\Omega(\Adv^\pm(f))$~\cite{HLS2007}.
Thus, to prove that a Boolean function requires many queries, it is sufficient to construct an admissible matrix $\Gamma$ for which the ratio in~\eqref{eq:adversary} is large.

The adversary bound will allow us to establish the hardness of one Boolean instance.
However, our GSEE and GSP lower bounds must also capture the dependence on the error parameter $\varepsilon$.
For this purpose, we use the quantum XOR lemma of Ref.~\cite{LR2012}, which relates the difficulty of one Boolean instance to that of computing the XOR of several separate instances.

To state this result, let $\ell\ge1$ and consider inputs $x^{(1)},\ldots,x^{(\ell)}\in\mathcal D$.
We define the Boolean function
\begin{equation}\label{eq:xor-function}
  \left(\bigoplus\circ f^{(\ell)}\right)
  (x^{(1)},\ldots,x^{(\ell)})
  =f(x^{(1)})\oplus\cdots\oplus f(x^{(\ell)}).
\end{equation}
The task is therefore to determine the parity of the $\ell$ output bits, rather than to output each bit separately.

The XOR lemma states that computing this parity with even a small advantage over random guessing requires a number of queries that grows with $\ell$.
More precisely, the result of Ref.~\cite[Corollary~5.3]{LR2012} is expressed in terms of the additive adversary quantity $\Adv^*(F)$, where $F$ is the matrix defined by $F_{xy}=1$ if $f(x)=f(y)$ and $F_{xy}=0$ otherwise.
By Ref.~\cite[Remark~3.4]{LR2012}, we have $\Adv^\pm(f)\le\Adv^*(F)$.
Combining these two results gives the following form of the XOR lemma, which we use in our proofs.

\begin{lemma}[Quantum XOR lemma~\cite{LR2012}]\label{lem:xor}
Let $f:\mathcal D\to\{0,1\}$ be a partial Boolean function, let $\ell\ge1$ be an integer, and let $0\le\rho\le1$.
Then
\begin{equation}\label{eq:xor}
  Q_{(1-\rho^{\ell/2})/2}
  \!\left(\bigoplus\circ f^{(\ell)}\right)
  \ge\frac{\ell\rho}{8}\Adv^\pm(f).
\end{equation}
\end{lemma}

To interpret this bound, observe that an error probability of at most $(1-\rho^{\ell/2})/2$ corresponds to a success probability of at least $(1+\rho^{\ell/2})/2$.
Since random guessing succeeds with probability $1/2$, the required advantage is $\rho^{\ell/2}/2$.
For any fixed $0<\rho<1$, this advantage decreases exponentially with $\ell$, but the lemma still gives a query lower bound of $\Omega(\ell\Adv^\pm(f))$.

\section{Lower Bound for GSEE}\label{sec:gsee}

We now establish a lower bound on the number of sparse-matrix queries needed for GSEE (Problem~\ref{prob:gsee}).

\begin{theorem}[Sparse-query GSEE lower bound]\label{thm:gsee}
There exist absolute constants $s_0\ge1$ and $\delta_0>0$
such that, for every integer $s\ge s_0$,
$0<\gamma\le1/4$, $0<\delta\le\delta_0$, and
$0<\varepsilon\le1/64$, the quantum query complexity
of GSEE (Problem~\ref{prob:gsee}) in both the standard
sparse-access model and the reversible in-place
sparse-location model, even when restricted to
stoquastic Hamiltonians, is
\begin{equation}\label{eq:main-lower}
  Q_{\rm sparse}^{\GSEE} = \Omega\!\left(
    \frac{\sqrt{s}}{\gamma\delta}\log\frac1\varepsilon
  \right).
\end{equation}
\end{theorem}

\paragraph{Proof overview.}
Our proof is based on a reduction from a Boolean query problem. The main idea is to construct a family of Hamiltonians whose ground-state energies encode the answer to this problem. If an algorithm can estimate the ground-state energy accurately enough, it can then recover the Boolean answer. To make this reduction work, we need the two possible ground-state energies to differ by more than $2\delta$, so that an estimate with additive error at most $\delta$ can distinguish them. We also need a known guiding state whose ground-space overlap amplitude is at least $\gamma$ in both cases. These conditions allow us to use any algorithm for GSEE to solve the Boolean problem.

We first construct Hamiltonians with constant sparsity that have the required ground-state energies and guiding-state overlaps. We then add a register of dimension $\Theta(s)$ and construct larger Hamiltonians that preserve these spectral properties while increasing the sparsity to $\Theta(s)$. The additional register also allows us to encode a Boolean input whose length increases by a factor of $\Theta(s)$. We show that solving the Boolean problem requires a number of queries proportional to the square root of its input length, so this increase gives the additional factor $\sqrt{s}$ in the lower bound. By choosing the dimension of the original Hamiltonians according to the overlap parameter $\gamma$ and the energy error $\delta$, we obtain the dependence $\sqrt{s}/(\gamma\delta)$.

It remains to establish the dependence on the failure probability $\varepsilon$. For this, we consider $k$ separate instances of the Boolean problem, with the promise that either all instances have answer zero or at least half of them have answer one. We combine their Hamiltonians into a single Hamiltonian whose ground-state energy distinguishes these two cases. We show that distinguishing them with failure probability exponentially small in $k$ requires $\Omega(k)$ times as many Boolean queries as a single instance. Choosing $k=\Theta(\log(1/\varepsilon))$ therefore gives the additional factor $\log(1/\varepsilon)$. Finally, we show that each sparse-matrix query to the constructed Hamiltonian can be simulated using only a constant number of Boolean queries. Thus, any GSEE algorithm using fewer sparse-matrix queries than our claimed lower bound would also violate the Boolean query lower bound, completing the reduction.

\bigskip

We now develop the proof in the following subsections, beginning
with the Boolean query problem, then constructing the corresponding
Hamiltonians, and finally combining these ingredients to establish
the GSEE lower bound.

\subsection{A Boolean query problem}\label{sec:boolean}

We begin with the Boolean problem \textsc{Position-Parity}, which was also used recently for proving the quantum linear systems lower bound~\cite{Patel2026}.
Let $L$ be an even positive integer, and suppose that we are given oracle access to a bit string $x\in\{0,1\}^L$ containing exactly one bit equal to one. Thus, there is a unique index $q\in\mathbb Z_L$ for which $x_q=1$. The task is not to output $q$, but only to decide whether $q$ is even or odd.

Let $e_q$ denote the bit string with a one at position $q$ and zeros everywhere else, and define the set of promised inputs $\mathcal S_L=\{e_q:q\in\mathbb Z_L\}$. We can then write \textsc{Position-Parity} as the partial Boolean function
\begin{equation}\label{eq:position-parity}
  g_L:\mathcal S_L\to\{0,1\},
  \qquad g_L(e_q)=q\bmod2.
\end{equation}
The promise tells us that exactly one bit of $x$ equals one. We call the position of this bit the \emph{marked} position. The algorithm does not know this position or whether it is even or odd. In particular, an algorithm must obtain information about $x$ through the Boolean oracle $O_x$ defined in~\eqref{eq:boolean-oracle}.

The following result gives the adversary bound for this function.
It was originally proved in Ref.~\cite{Patel2026}.
We include a proof of this result in Appendix~\ref{app:boolean} for completeness. 

\begin{lemma}[\textsc{Position-Parity} adversary bound]\label{lem:parity-adversary}
For every even positive integer $L$,
$\Adv^\pm(g_L)\ge\sqrt{L/2}$.
\end{lemma}

\begin{proof}
    The proof is given in Appendix~\ref{app:boolean}.
\end{proof}

We first establish the query lower bound for a single instance of \textsc{Position-Parity}. Setting $\ell=1$ and $\rho=1/9$ in the XOR lemma (Lemma~\ref{lem:xor}), the allowed error probability becomes $(1-\sqrt{\rho})/2=1/3$. Since the XOR of a single instance is just the function itself, we obtain
\begin{equation}
Q_{1/3}(g_L)
\ge \frac{1}{72}\Adv^\pm(g_L)
\ge \frac{1}{72}\sqrt{\frac L2}
=\Omega(\sqrt L),
\end{equation}
where the second inequality follows from Lemma~\ref{lem:parity-adversary}. Thus, computing \textsc{Position-Parity} with success probability at least $2/3$ requires $\Omega(\sqrt L)$ Boolean queries.

This lower bound $\Omega(\sqrt L)$ for a single \textsc{Position-Parity} instance will allow us to establish the dependence on the problem parameters $s$, $\gamma$, and $\delta$ or $\Delta$ in our query lower bounds for GSEE and GSP, respectively.
More precisely, we will construct a family of $s$-sparse Hamiltonians that encode Boolean inputs of length $L=\Theta(s/(\gamma^2\delta^2))$ for GSEE, giving $\sqrt L=\Theta(\sqrt{s}/(\gamma\delta))$.
The same construction will also be used for GSP, with $\delta=\Theta(\Delta)$, giving $\sqrt L=\Theta(\sqrt{s}/(\gamma\Delta))$.
However, observe that these bounds do not yet capture the additional factor $\log(1/\varepsilon)$ in Theorems~\ref{thm:gsee} and~\ref{thm:gsp}.

To obtain that factor, we consider several \textsc{Position-Parity} instances at once. Let $k\ge1$ and write their joint input as $X=(x^{(1)},\ldots,x^{(k)})\in\mathcal S_L^k$. A query to this tuple is one application of
\begin{equation}\label{eq:joint-boolean-oracle}
  O_X\ket{i,q,b}=\ket{i,q,b\oplus x_q^{(i)}},
  \qquad i\in[k],\quad q\in\mathbb Z_L.
\end{equation}
The instance index may be in superposition, so this is the standard oracle~\eqref{eq:boolean-oracle} for the concatenation of all $k$ strings. For each instance, define its parity bit $z_i=g_L(x^{(i)})$, and let $z=(z_1,\ldots,z_k)$. We now impose an additional promise on $z$: either every parity bit is zero, or at least half of them are one. We want to distinguish these two cases.

Formally, define
\begin{equation}\label{eq:outer}
  F_k(z)=
  \begin{cases}
    1,&z=0^k,\\
    0,&|z|\ge\lceil k/2\rceil,
  \end{cases}
\end{equation}
where $|z|=\sum_{i=1}^kz_i$ is the Hamming weight, and $F_k$ is undefined outside the two promised cases. 
The complete Boolean problem is then $F_k\circ g_L^{(k)}$: given $X$, determine $F_k(g_L(x^{(1)}),\ldots,g_L(x^{(k)}))$.

We need a query lower bound for this composed Boolean problem when the allowed error probability is exponentially small in $k$. 
The result below provides exactly that dependence.

\begin{lemma}\label{lem:outer}
There exists an absolute constant $c_*>0$ such that, for every integer $k\ge1$, every even positive integer $L$, and every $0<\eta\le2^{-2k-1}$,
\begin{equation}\label{eq:outer-lower}
  Q_\eta(F_k\circ g_L^{(k)})\ge c_*k\sqrt L.
\end{equation}
\end{lemma}

\begin{proof}
The proof is given in Appendix~\ref{app:boolean}. The main step is to use an algorithm for $F_k\circ g_L^{(k)}$ with error probability at most $2^{-2k-1}$ to predict the XOR of at least $k/2$ parity bits with an advantage of size $2^{-O(k)}$ over random guessing. 
The XOR lemma (Lemma~\ref{lem:xor}) then gives a lower bound proportional to the number of parity bits times $\Adv^\pm(g_L)$. 
Applying Lemma~\ref{lem:parity-adversary} yields $\Omega(k\sqrt L)$ queries.
\end{proof}

We will eventually choose $k=\Theta(\log(1/\varepsilon))$. Lemma~\ref{lem:outer} will then contribute the factor $\log(1/\varepsilon)$, while the size $L$ of each \textsc{Position-Parity} instance in~\eqref{eq:position-parity} will determine the dependence on $s$, $\gamma$, and $\delta$.

\subsection{Constructing Hamiltonians for \texorpdfstring{\textnormal{\textsc{Position-Parity}}}{Position-Parity}}\label{sec:spectrum}

We next encode the answer to one \textsc{Position-Parity} instance in the ground-state energy of a Hamiltonian. 
Our immediate goal is to construct two types of Hamiltonians: one with ground-state energy $-1$, representing an even marked position, and another with ground-state energy $-1-h$, representing an odd marked position, where $h>0$ is a known parameter. 
The ground-state energy in the second case must be the same regardless of where the odd position occurs. Otherwise, the final reduction would not have just two known energies to distinguish.

We must also preserve the guiding-state promise. In both cases, the ground state should have sufficient overlap with a state that is known before the marked position is revealed. We first solve these spectral requirements using a constant-sparse matrix. The dependence on the target sparsity $s$ will be introduced in the next subsection.

Fix absolute constants $d\ge1$ and rational $0<c\le1$ such that, for every sufficiently large even integer $N$, there is a rational symmetric stochastic matrix $P\in\mathbb R^{N\times N}$ satisfying
\begin{equation}\label{eq:expander}
  P\ket u=\ket u,
  \qquad
  \ket u=\frac1{\sqrt N}\sum_{a=1}^N\ket a,
  \qquad
  \norm P\le1,
  \qquad
  \lambda_2(P)\le1-c.
\end{equation}
Here stochastic means that $P$ has nonnegative entries and every row sums to one. Each row has at most $d$ nonzero entries, and $\lambda_2(P)$ denotes the second-largest eigenvalue of $P$. 
Such matrices can be obtained by dividing the adjacency matrix of a fixed-degree regular expander graph by its degree~\cite{Friedman2008}. The entries are then rational, and a uniform bound below one on the second-largest eigenvalue gives a fixed rational choice of $c>0$. We need only existence: for each allowed $N$, choose one such matrix $P$ independently of the Boolean input. Its description and all computations involving only $P$ are free in the Boolean query reduction.

These properties give us a known low-energy state. Since $P$ is symmetric and stochastic, its normalized uniform vector $\ket u$ has eigenvalue one. The bound on $\lambda_2(P)$ shows that this eigenvalue is unique and that every other eigenvalue is at most $1-c$. Therefore, for
$K^0=-P$, the unique ground state is $\ket u$, the ground-state energy is $-1$, and the spectral gap is at least $c$. We will use $K^0$ to represent the even case.

Now recall that $P$ is an $N\times N$ matrix whose rows and columns are indexed by $a\in[N]$, corresponding to the vertices of the underlying expander graph.
For the odd case, we want to lower the ground-state energy of $K^0=-P$ from $-1$ to exactly $-1-h$ by adding a negative diagonal perturbation at some basis state $\ket a$.
The index $a$ will later be determined by the marked position of the Boolean input and is therefore not known to the algorithm.
We will choose the strength of the perturbation so that the resulting ground-state energy is $-1-h$, independently of the value of $a$.

Let $h>0$ be rational and satisfy $h\le\min\{1,c/4\}$. Define
\begin{equation}\label{eq:resolvents}
  B_-=(1+h)I-P.
\end{equation}
The largest eigenvalue of $P$ is one, so every eigenvalue of $B_-$ is at least $h>0$. Thus $B_-$ is positive definite, and its inverse exists. For each vertex $a\in[N]$, set
\begin{equation}\label{eq:calibration}
  \beta_a^-=\frac1{\bra aB_-^{-1}\ket a},
\end{equation}
and define
\begin{equation}\label{eq:logical}
  K^0=-P,
  \qquad K_a^-=-P-\beta_a^-\ket a\bra a.
\end{equation}
We will use the normalized vector
\begin{equation}\label{eq:calibrated-ground-vector}
 \ket{w_a^-}=\frac{B_-^{-1}\ket a}{\sqrt{\bra aB_-^{-2}\ket a}}.
\end{equation}
Its denominator is the norm of the nonzero vector $B_-^{-1}\ket a$. The denominator defining $\beta_a^-$ in~\eqref{eq:calibration} is positive because $B_-^{-1}$ is positive definite. Hence $\beta_a^->0$. It is also rational, since $B_-$ is an invertible rational matrix. The perturbation changes only one diagonal entry of $-P$, so both $K^0$ and $K_a^-$ remain stoquastic.

We now verify why the choice of $\beta_a^-$ gives an eigenvalue exactly equal to $-1-h$. By the definitions above,
$K_a^-+(1+h)I=B_--\beta_a^-\ket a\bra a$. Applying this operator to $B_-^{-1}\ket a$ gives
\begin{align*}
  \bigl(K_a^-+(1+h)I\bigr)B_-^{-1}\ket a
  &=\bigl(B_--\beta_a^-\ket a\bra a\bigr)B_-^{-1}\ket a\\
  &=\bigl(1-\beta_a^-\bra aB_-^{-1}\ket a\bigr)\ket a
  =0,
\end{align*}
where the last equality follows from~\eqref{eq:calibration}. Thus $B_-^{-1}\ket a$ is an eigenvector of $K_a^-$ with eigenvalue $-1-h$. In particular, this eigenvalue does not depend on $a$, even though the perturbation strength $\beta_a^-$ may depend on $a$.

To use this construction for GSEE, we also need $-1-h$ to be the lowest eigenvalue, to have a spectral gap, and to control the overlap with the known state $\ket u$. These facts are collected in the following lemma.

\begin{lemma}\label{lem:spectral}
The Hamiltonian $K^0$ has unique ground state $\ket u$, ground-state energy $-1$, and spectral gap at least $c$. For every $a\in[N]$, the Hamiltonian $K_a^-$ has unique ground state $\ket{w_a^-}$ with ground-state energy $-1-h$ and spectral gap at least $h$. Its perturbation strength satisfies
\begin{equation}\label{eq:beta-bounds}
  0<\beta_a^-\le3,
\end{equation}
and its squared overlap with $\ket u$ satisfies
\begin{equation}\label{eq:overlap-bounds}
  |\braket u{w_a^-}|^2\ge\frac1{1+Nh^2/c^2}.
\end{equation}
\end{lemma}

\begin{proof}
The proof is given in Appendix~\ref{app:spectrum}.
\end{proof}

Observe that the lower bound in~\eqref{eq:overlap-bounds} becomes larger as $Nh^2$ becomes smaller. In particular, choosing
\begin{equation}\label{eq:N-intuition}
  N=\Theta\!\left(\frac1{\gamma^2h^2}\right)
\end{equation}
with a sufficiently small constant factor ensures that the squared overlap is at least $\gamma^2$ in both cases. We give the exact choice in Section~\ref{sec:gsee-assembly}, where we also leave room for the constant loss from combining several blocks. Since we will set $h=\Theta(\delta)$, this choice gives $\sqrt N=\Theta(1/(\gamma\delta))$.

We have therefore constructed two Hamiltonians with the desired ground-state energies and the same known guiding state $\ket u$.
However, we have not yet encoded the answer to \textsc{Position-Parity} into these Hamiltonians or introduced the dependence on the sparsity parameter $s$.
We now address both by enlarging the Hilbert space while preserving the two possible ground-state energies and the required guiding overlap.

\subsection{Extending the Construction to Sparsity \texorpdfstring{$s$}{s}}\label{sec:lift}

Observe that the Hamiltonians in~\eqref{eq:logical} have at most $d+1$ nonzero entries per row, where $d$ is an absolute constant. Their sparsity therefore does not grow with the parameter $s$ as we need in Theorem~\ref{thm:gsee}. For this, we now add a register of dimension $m=\Theta(s)$ and construct a Hamiltonian whose low-energy subspace still reproduces Hamiltonians $K^0$ or $K_a^-$, depending on the parity of the marked Boolean position.

Let $m\ge4$ be even. 
The original register has Hilbert space $\mathcal A=\mathbb C^N$ with basis $\{\ket a:a\in[N]\}$, and the additional register has Hilbert space $\mathcal B=\mathbb C^m$ with basis $\{\ket b:b\in\mathbb Z_m\}$. Thus the combined space is $\mathcal A\otimes\mathcal B$, with basis $\ket{a,b}=\ket a\otimes\ket b$. We call this second register the auxiliary register. Define the cyclic shift $S$ on $\mathcal B$ by $S\ket b=\ket{b+1\bmod m}$ for $b\in\mathbb Z_m$. 
The normalized uniform state on this register and its projector are as follows:
\begin{equation}\label{eq:auxiliary-uniform}
  \ket v=\frac1{\sqrt m}\sum_{b\in\mathbb Z_m}\ket b,
  \qquad V=\ket v\bra v.
\end{equation}
Note that every power of $S$ preserves the uniform state, so $S^j\ket v=S^{-j}\ket v=\ket v$ for all $j\in\mathbb Z_m$. Also, in the computational basis, every entry of $V$ equals $1/m$.

We take the Boolean input length to be $L=Nm$. Instead of using a single index $q\in\mathbb Z_{Nm}$, we write each position uniquely as a pair $(a,j)\in[N]\times\mathbb Z_m$, with
$q=(a-1)m+j$. We denote the corresponding bit by $x_{a,j}$. Because $m$ is even, $(a-1)m$ is even, and hence $q\bmod2=j\bmod2$. Thus, to determine \textsc{Position-Parity}, it is enough to determine whether the second coordinate $j$ of the unique marked position is even or odd.

Now, for each $a\in[N]$, define
\begin{equation}\label{eq:w}
  W_a(x)=-\frac{\beta_a^-}{2}
  \sum_{\substack{j\in\mathbb Z_m\\j\text{ odd}}}
  x_{a,j}(S^j+S^{-j}).
\end{equation}
Observe that only odd values of $j$ appear in this sum. Recall that $x$ has exactly one nonzero bit, and let us denote its location by $(a_*,j_*)$. If $j_*$ is even, every term in~\eqref{eq:w} vanishes, so $W_a(x)=0$ for all $a$. If $j_*$ is odd, only the operator with $a=a_*$ is nonzero, and
\begin{equation}\label{eq:marked-shift-action}
  W_{a_*}(x)=-\frac{\beta_{a_*}^-}{2}(S^{j_*}+S^{-j_*}),
  \qquad
  W_{a_*}(x)\ket v=-\beta_{a_*}^-\ket v.
\end{equation}
The second identity follows because both cyclic shifts leave the uniform state $\ket v$ unchanged, as shown before. 

Now consider the Boolean-dependent term in the Hamiltonian, $\sum_{a=1}^N\ket a\bra a\otimes W_a(x)$. When $j_*$ is odd, we have $W_a(x)=0$ for every $a\ne a_*$, so this term becomes
\begin{equation}
\ket{a_*}\bra{a_*}\otimes W_{a_*}(x).
\end{equation}
By~\eqref{eq:marked-shift-action}, its action on any state $\ket f\otimes\ket v$ is
\begin{equation}
\bigl(\ket{a_*}\bra{a_*}\otimes W_{a_*}(x)\bigr)
(\ket f\otimes\ket v)
=-\beta_{a_*}^-\ket{a_*}\bra{a_*}f\rangle\otimes\ket v.
\end{equation}
Thus, on states whose auxiliary register is $\ket v$, the Boolean-dependent term acts as $-\beta_{a_*}^-\ket{a_*}\bra{a_*}$ on the original register. This is exactly the perturbation used to define $K_{a_*}^-$ in~\eqref{eq:logical}. When $j_*$ is even, every $W_a(x)$ vanishes, so the Boolean-dependent term is zero, as required for $K^0$.

We now define a Hamiltonian on \(\mathcal A\otimes\mathcal B\) whose ground state lies in the subspace \(\mathcal A\otimes\operatorname{span}\{\ket v\}\), where \(\ket v\) is the uniform state of the auxiliary register defined in~\eqref{eq:auxiliary-uniform}.
Fix $M=17/2$ and set
\begin{equation}\label{eq:lift}
  H_x
  =-P\otimes V
   +M I_N\otimes(I_m-V)
   +\sum_{a=1}^N\ket a\bra a\otimes W_a(x).
\end{equation}
The first term reproduces $-P$ when the auxiliary register is in $\ket v$. The second term vanishes on $\ket v$ but contributes the positive energy $M$ on its orthogonal complement. The final term inserts the Boolean-dependent perturbation described above. We choose $M$ to keep the energies outside the uniform auxiliary subspace well above the desired ground-state energies.

We first check that $H_x$ is stoquastic. The entries of $-P\otimes V$ are nonpositive because $P$ and $V$ have nonnegative entries. The off-diagonal entries of $M I_N\otimes(I_m-V)$ are also nonpositive, since within each diagonal $a$-block they equal $-M/m$. Finally, every $S^j$ is a permutation matrix with nonnegative entries, and the coefficients in~\eqref{eq:w} are nonpositive. Hence the Boolean-dependent term cannot introduce positive off-diagonal entries. All off-diagonal entries of $H_x$ are therefore real and nonpositive in the oracle basis.

Next, we determine the spectrum of $H_x$ near its ground-state energy. Recall that the Hamiltonian acts on $\mathcal A\otimes\mathcal B$, where $\mathcal A=\mathbb C^N$ and $\mathcal B=\mathbb C^m$. Let
\begin{equation}
  \mathcal B_\perp
  =\{\ket g\in\mathcal B:\braket v g=0\}
\end{equation}
denote the subspace of the auxiliary register orthogonal to $\ket v$. Since $\mathcal B=\operatorname{span}\{\ket v\}\oplus\mathcal B_\perp$, the full Hilbert space decomposes as
\begin{equation}\label{eq:lift-decomposition}
  \mathcal A\otimes\mathcal B
  =\underbrace{\mathcal A\otimes\operatorname{span}\{\ket v\}}_{\mathcal V}
  \oplus
  \underbrace{\mathcal A\otimes\mathcal B_\perp}_{\mathcal V^\perp}.
\end{equation}
The orthogonal projectors onto these two subspaces are
\begin{equation}\label{eq:lift-projectors}
  P_{\parallel}=I_N\otimes V,\qquad
  P_{\perp}=I_N\otimes(I_m-V),\qquad
  P_{\parallel}+P_{\perp}=I_{Nm}.
\end{equation}

We first show that $H_x$ does not map states between $\mathcal V$ and $\mathcal V^\perp$. Recall that each $W_a(x)$ in~\eqref{eq:w} is a linear combination of cyclic shifts $S^j$ and $S^{-j}$. Since these shifts leave $\ket v$ unchanged, they satisfy $S^jV=VS^j=V$ for every integer $j$. Thus every $W_a(x)$ commutes with $V$. The other two terms of $H_x$ in~\eqref{eq:lift} also commute with $I_N\otimes V$. Consequently, both subspaces in~\eqref{eq:lift-decomposition} are invariant under $H_x$, and
\begin{equation}\label{eq:lift-no-coupling}
  P_{\perp}H_xP_{\parallel}=0,
  \qquad
  P_{\parallel}H_xP_{\perp}=0.
\end{equation}
We can therefore determine the spectra of the restrictions of $H_x$ to $\mathcal V$ and $\mathcal V^\perp$ separately.

We first consider $\mathcal V$. On this subspace, $V$ acts as the identity on the auxiliary register, while $I_m-V$ acts as zero. Hence the first two terms of~\eqref{eq:lift} act as $-P$ on the original register. As established in~\eqref{eq:marked-shift-action}, the Boolean-dependent term vanishes when $j_*$ is even and acts as $-\beta_{a_*}^-\ket{a_*}\bra{a_*}$ on the original register when $j_*$ is odd.

To express this restriction precisely, define the isometry $J_v:\mathcal A\to\mathcal A\otimes\mathcal B$ by
\begin{equation}
  J_v\ket f=\ket f\otimes\ket v.
\end{equation}
Since $\ket v$ is normalized, $J_v^\dagger J_v=I_N$, and the image of $J_v$ is exactly $\mathcal V$. The restriction of $H_x$ to $\mathcal V$, represented on $\mathcal A$, is therefore
\begin{equation}\label{eq:lift-restriction}
  J_v^\dagger H_x J_v
  =\begin{cases}
    K^0,&j_*\text{ even},\\
    K_{a_*}^-,&j_*\text{ odd}.
  \end{cases}
\end{equation}
By Lemma~\ref{lem:spectral}, the corresponding ground-state energies in $\mathcal V$ are exactly $-1$ and $-1-h$, respectively.

We next consider $\mathcal V^\perp$. On this subspace, $V$ acts as zero and $I_m-V$ acts as the identity. Thus the first term of~\eqref{eq:lift} vanishes, while the second becomes $MI$. The only remaining contribution is the Boolean-dependent term
\begin{equation}
  W(x)=\sum_{a=1}^N\ket a\bra a\otimes W_a(x).
\end{equation}
Since $x$ contains exactly one marked bit, at most one $W_a(x)$ is nonzero. Moreover, by~\eqref{eq:marked-shift-action} and~\eqref{eq:beta-bounds}, its norm is at most $\beta_a^-\le3$. Hence $\norm{W(x)}\le3$.

For every $\ket g\in\mathcal V^\perp$, we consequently have
\begin{align}
  \bra gH_x\ket g
  &=M\norm g^2+\bra gW(x)\ket g\notag\\
  &\ge(M-\norm{W(x)})\norm g^2\notag\\
  &\ge\frac{11}{2}\norm g^2,
  \label{eq:lift-complement-bound}
\end{align}
where we used $M=17/2$.
Applying this inequality to an eigenvector in $\mathcal V^\perp$ shows that every eigenvalue of $H_x$ in this subspace is at least $11/2$. On the other hand, by~\eqref{eq:lift-restriction} and Lemma~\ref{lem:spectral}, the restriction to $\mathcal V$ has a unique ground state with energy either $-1$ or $-1-h$ and spectral gap at least $h$. Since $h\le1$, every eigenvalue in $\mathcal V^\perp$ lies more than $h$ above either possible ground-state energy. Thus $H_x$ has the same unique ground state as its restriction to $\mathcal V$, and its spectral gap is at least $h$.

It remains to bound the operator norm. By~\eqref{eq:lift-restriction}, the restriction to $\mathcal V$ is unitarily equivalent to either $K^0$ or $K_{a_*}^-$. Since $\norm P\le1$ and $\beta_{a_*}^-\le3$, its norm is at most $4$. On $\mathcal V^\perp$, we have $H_x=MI+W(x)$, so
\begin{equation}
  \norm{H_x|_{\mathcal V^\perp}}
  \le M+\norm{W(x)}
  \le\frac{23}{2}.
\end{equation}
Since the two subspaces are invariant and orthogonal, the norm of $H_x$ is the maximum of the norms of its two restrictions. Therefore,
\begin{equation}
  \norm{H_x}\le\frac{23}{2}<12.
\end{equation}
The guiding-state overlap follows from Lemma~\ref{lem:spectral}, since every ground state belongs to $\mathcal V$. Together with the stoquasticity established above, this proves the following lemma.

\begin{lemma}\label{lem:lift-spectrum}
For every input $x\in\mathcal S_{Nm}$, the Hamiltonian $H_x$ is stoquastic, has $\norm{H_x}<12$, a unique ground state, and spectral gap at least $h$. Its ground-state energy is
\begin{equation}\label{eq:lift-energy}
  E_0(H_x)=
  \begin{cases}
    -1,&g_{Nm}(x)=0,\\
    -1-h,&g_{Nm}(x)=1.
  \end{cases}
\end{equation}
With guiding state $\ket u\otimes\ket v$ from~\eqref{eq:expander} and~\eqref{eq:auxiliary-uniform}, its squared ground-state overlap is one in the first case and at least $(1+Nh^2/c^2)^{-1}$ in the second case. The guiding state has nonnegative amplitudes and is independent of $x$.
\end{lemma}

We have shown that the ground-state energy of $H_x$ determines the parity of the marked position. To obtain a sparse-query lower bound, we must also show that each sparse-matrix query to $H_x$ can be simulated using only a constant number of Boolean queries to $x$.

The key properties are that the nonzero-entry locations of $H_x$ are independent of $x$, and that each matrix entry depends on at most two bits of $x$. The first property allows us to implement the sparse-location oracles without querying $x$. For the second, we can query the relevant bits, compute the matrix entry, and uncompute the query registers using at most four Boolean queries. We establish these properties, including the coherent oracle implementations, in Appendix~\ref{app:oracles}.

\begin{lemma}[Fixed support and oracle simulation]\label{lem:oracles}
For the fixed constant $M=17/2$, every $H_x$ has the same nonzero-entry locations, with between $m$ and $(d+1)m$ nonzero entries per row. Each sparse-value query to $H_x$, including a controlled query, can be implemented using at most four Boolean queries to $O_x$. Every position query and every reversible in-place row or column location query, including its inverse, can be implemented without querying $x$.
\end{lemma}

\begin{proof}
The proof is given in Appendix~\ref{app:oracles}.
\end{proof}

We now choose the dimension $m$ of the auxiliary register so that $H_x$ has sparsity at most $s$. Set $s_0=4(d+1)$, and for every integer $s\ge s_0$, define
\begin{equation}\label{eq:m-choice}
  m=2\left\lfloor\frac{s}{2(d+1)}\right\rfloor,
  \qquad
  \frac{s}{2(d+1)}\le m\le\frac{s}{d+1}.
\end{equation}
Thus $m$ is the largest even integer at most $s/(d+1)$, and $m\ge4$. The bounds in~\eqref{eq:m-choice} give $m=\Theta(s)$. By Lemma~\ref{lem:oracles}, every row of $H_x$ has at most $(d+1)m\le s$ nonzero entries. Therefore $H_x$ is $s$-sparse.

Recall that the Boolean input has length $L=Nm$. With $m=\Theta(s)$ and the choice $N=\Theta(1/(\gamma^2h^2))$ indicated in~\eqref{eq:N-intuition}, we obtain
\begin{equation}\label{eq:L-scaling}
  \sqrt L=\sqrt{Nm}
  =\Theta\!\left(\frac{\sqrt{s}}{\gamma h}\right).
\end{equation}
In the final construction, we will choose $h=\Theta(\delta)$, giving
\begin{equation}
  \sqrt L=\Theta\!\left(\frac{\sqrt{s}}{\gamma\delta}\right).
\end{equation}
Since one \textsc{Position-Parity} instance requires $\Omega(\sqrt L)$ Boolean queries, this produces the dependence on $s$, $\gamma$, and $\delta$ in the desired GSEE lower bound. The remaining factor $\log(1/\varepsilon)$ will follow from Lemma~\ref{lem:outer} by combining $k=\Theta(\log(1/\varepsilon))$ Boolean instances.

\subsection{Completing the GSEE lower bound proof}
\label{sec:gsee-assembly}

We now combine the Boolean problem from Section~\ref{sec:boolean} with the Hamiltonian construction of Section~\ref{sec:lift}. Recall that, for each Boolean input $x\in\mathcal S_{Nm}$, the Hamiltonian $H_x$ has ground-state energy $-1$ or $-1-h$, depending on the parity of the marked position. We will combine $k$ such Hamiltonians into a single Hamiltonian whose ground-state energy distinguishes the two cases of $F_k$ defined in~\eqref{eq:outer}. We then verify the norm, sparsity, guiding-state overlap, and spectral gap of the resulting Hamiltonian, and use its sparse oracles to establish the GSEE query lower bound, claimed in Theorem~\ref{thm:gsee}.

We first fix the parameters of the construction. Let $c,d$ be the absolute constants from~\eqref{eq:expander}. Choose $\delta_0>0$ sufficiently small so that the following choices are possible for every $0<\delta\le\delta_0$ and $0<\gamma\le1/4$.

Choose a positive dyadic number $h$ satisfying
\begin{equation}\label{eq:h-choice}
  128\delta\le h\le256\delta,
  \qquad h\le\min\{1,c/4\}.
\end{equation}
Such a number exists because the interval $[128\delta,256\delta]$ contains a power of two, and the upper bound on $h$ holds when $\delta_0$ is sufficiently small. We choose $h=\Theta(\delta)$ so that the two possible ground-state energies remain separated by more than $2\delta$ after the Hamiltonian is normalized.

Next, choose the expander dimension $N$. Set
\begin{equation}\label{eq:N-choice}
  A_0=\frac{c^2}{16\gamma^2h^2},
  \qquad
  N=2\lfloor A_0\rfloor.
\end{equation}
This makes $N$ even. By choosing $\delta_0$ sufficiently small, we may assume that $A_0\ge2$ and that $N$ is large enough for the expander family in~\eqref{eq:expander}. Since $A_0\le2\lfloor A_0\rfloor\le2A_0$ for $A_0\ge2$, we have
\begin{equation}\label{eq:N-bounds}
  \frac{c^2}{16\gamma^2h^2}
  \le N\le
  \frac{c^2}{8\gamma^2h^2},
  \qquad
  N=\Theta\!\left(\frac1{\gamma^2\delta^2}\right).
\end{equation}
We take the auxiliary dimension $m$ as in~\eqref{eq:m-choice}, so that $m=\Theta(s)$ and $(d+1)m\le s$.

Finally, we choose the number of Boolean instances according to the allowed failure probability $\varepsilon$. Set
\begin{equation}\label{eq:k-choice}
  k=\left\lfloor\frac14\log_2\frac1\varepsilon\right\rfloor.
\end{equation}
Since $0<\varepsilon\le1/64$, we have $k\ge1$ and $k=\Theta(\log(1/\varepsilon))$. Moreover, the definition of $k$ gives
\begin{equation}
  \varepsilon\le2^{-4k}\le2^{-2k-1}.
\end{equation}
Thus the error probability $\varepsilon$ satisfies the condition required to apply Lemma~\ref{lem:outer}.

We now construct the final Hamiltonian. Consider a promised Boolean input $X=(x^{(1)},\ldots,x^{(k)}) \in\mathcal S_{Nm}^k$, and let $z_i=g_{Nm}(x^{(i)})$ for every $i\in[k]$. Recall that the promise of $F_k$ is either $z=0^k$ or $|z|\ge\lceil k/2\rceil$.

For each input $x^{(i)}$, construct the Hamiltonian $H_{x^{(i)}}$ defined in~\eqref{eq:lift}, and take their direct sum:
\begin{equation}\label{eq:block-directsum}
  H_X^{\oplus}
  =\bigoplus_{i=1}^kH_{x^{(i)}}.
\end{equation}
More explicitly, $H_X^{\oplus}$ acts on
$\mathbb C^k\otimes\mathcal A\otimes\mathcal B$ as
\begin{equation}\label{eq:block-action}
  H_X^{\oplus}
  =\sum_{i=1}^k
    \ket i\bra i\otimes H_{x^{(i)}}.
\end{equation}
Since $H_X^{\oplus}$ is block diagonal, its eigenvalues are the eigenvalues of its individual blocks. In particular,
\begin{equation}
  E_0(H_X^{\oplus})
  =\min_{i\in[k]}E_0(H_{x^{(i)}}).
\end{equation}
By Lemma~\ref{lem:lift-spectrum}, each block has ground-state energy $-1$ when $z_i=0$ and $-1-h$ when $z_i=1$. Therefore,
\begin{equation}
  E_0(H_X^{\oplus})=
  \begin{cases}
    -1,&z=0^k,\\
    -1-h,&|z|\ge\lceil k/2\rceil.
  \end{cases}
\end{equation}

We also need the final Hamiltonian to have a known spectral gap, which will be useful for the GSP lower bound in Section~\ref{sec:gsp}. By Lemma~\ref{lem:lift-spectrum}, every block $H_{x^{(i)}}$ has spectral gap at least $h$, and the possible block ground-state energies differ by exactly $h$. Hence every eigenvalue of $H_X^{\oplus}$ strictly above its ground-state energy is at least $h$ higher.

To make this gap exactly $h$, independently of the Boolean input, we append a second copy of $H_X^{\oplus}$ shifted upward by $h$ and normalize the resulting Hamiltonian:
\begin{equation}\label{eq:directsum}
  H_X^{\rm rat}
  =\frac1{16}
    \left(H_X^{\oplus}\oplus(H_X^{\oplus}+hI)\right).
\end{equation}
The Hilbert space of this Hamiltonian is
\begin{equation}\label{eq:final-family-space}
  \mathcal H_{\rm fam}
  =\mathbb C^2\otimes\mathbb C^k
    \otimes\mathcal A\otimes\mathcal B,
  \qquad
  \dim\mathcal H_{\rm fam}=2kNm,
\end{equation}
where the first register labels the two copies. Equivalently,
\begin{equation}
  16H_X^{\rm rat}
  =\ket0\bra0 \otimes H_X^{\oplus}
   +\ket1\bra1\otimes(H_X^{\oplus}+hI).
\end{equation}

The shifted copy has smallest eigenvalue $E_0(H_X^{\oplus})+h$. Since the unshifted copy has no eigenvalues strictly between $E_0(H_X^{\oplus})$ and $E_0(H_X^{\oplus})+h$, the first distinct eigenvalue above the ground-state energy of the two-copy Hamiltonian is exactly $h$ higher. Thus
\begin{equation}
  \operatorname{gap}(H_X^{\rm rat})=\frac h{16}.
\end{equation}

We next check the norm, sparsity, and stoquasticity. By Lemma~\ref{lem:lift-spectrum}, every block $H_{x^{(i)}}$ has norm less than $12$, so $\norm{H_X^{\oplus}}<12$. Since $h\le1$, we have
\begin{equation}
  \norm{H_X^{\rm rat}}
  \le\frac{\norm{H_X^{\oplus}}+h}{16}
  <\frac{13}{16}<1.
\end{equation}
Taking direct sums does not increase the maximum number of nonzero entries in any row or column. Moreover, the diagonal entries of $H_x$ are strictly positive, as established in Appendix~\ref{app:oracles}, so adding $hI$ does not change their nonzero status. Hence $H_X^{\rm rat}$ remains $s$-sparse, with nonzero-entry locations independent of $X$.
Every $H_{x^{(i)}}$ is stoquastic by Lemma~\ref{lem:lift-spectrum}. Direct sums, diagonal shifts, and multiplication by a positive scalar preserve nonpositive off-diagonal entries. Therefore $H_X^{\rm rat}$ is also stoquastic.

The entries of $H_X^{\rm rat}$ and its two possible ground-state energies are rational. Applying Lemma~\ref{lem:dyadic} to the finite set containing these numbers for every promised input $X$, we obtain a common scaling factor $\sigma\in(1/2,1]$ such that all matrix entries and both ground-state energies become dyadic after multiplication by $\sigma$. Since the same factor is used for every input, it is independent of $X$. Moreover, this positive rescaling preserves the eigenspaces, nonzero-entry locations, and stoquasticity of the Hamiltonians.

Define the final Hamiltonian and its two possible ground-state energies by
\begin{equation}\label{eq:final-H}
  H_X=\sigma H_X^{\rm rat},
  \qquad
  a=\frac{\sigma(-1-h)}{16},
  \qquad
  b=-\frac{\sigma}{16}.
\end{equation}
Since $h>0$, we have $a<b$. From the construction above,
\begin{equation}
  E_0(H_X)=
  \begin{cases}
    b,&z=0^k,\\
    a,&|z|\ge\lceil k/2\rceil.
  \end{cases}
\end{equation}
The separation between these energies is
\begin{equation}\label{eq:separation}
  4\delta
  \le g_0:=b-a
  =\frac{\sigma h}{16}
  \le16\delta,
\end{equation}
where we used $1/2<\sigma\le1$ and the bounds on $h$ in~\eqref{eq:h-choice}. Thus the two possible ground-state energies are separated by more than $2\delta$, as required for the GSEE reduction.

We next construct a guiding state that satisfies the overlap promise for every input $X$. Recall that the four registers of $\mathcal H_{\rm fam}$ hold the copy label, block label, expander vertex, and auxiliary-register basis label, respectively. Define
\begin{equation}\label{eq:guide}
  \ket\psi
  =\ket0\otimes\frac1{\sqrt k}
    \sum_{i=1}^k
      \ket i\otimes\ket u\otimes\ket v.
\end{equation}
This state lies entirely in the unshifted copy and has equal amplitudes over the $k$ blocks. Since $\ket u$ and $\ket v$ are uniform states, $\ket\psi$ has nonnegative amplitudes. It is also independent of $X$, so the same preparation unitary can be used for every Hamiltonian in the family.

We now verify its overlap with the ground space in the two Boolean cases.

First suppose that $z=0^k$. Every unshifted block of $H_X$ then has ground-state energy $b=-\sigma/16$ and ground state $\ket u\otimes\ket v$. The shifted blocks have strictly larger ground-state energies. Thus $\ket\psi$ lies entirely in the ground space of $H_X$, giving
\begin{equation}
  \norm{\Pi_0(H_X)\ket\psi}^2=1.
\end{equation}

Next suppose that $|z|\ge\lceil k/2\rceil$. In this case, at least half of the unshifted blocks of $H_X$ have ground-state energy $a=-\sigma(1+h)/16$, while the remaining unshifted blocks and all shifted blocks have larger ground-state energies. Hence the ground space of $H_X$ is the direct sum of the ground spaces of precisely those unshifted blocks for which $z_i=1$.

By Lemma~\ref{lem:lift-spectrum}, the squared overlap of $\ket u\otimes\ket v$ with the ground state of each such block is at least $1/(1+Nh^2/c^2)$. Using the upper bound on $N$ from~\eqref{eq:N-bounds}, we obtain
\begin{equation}\label{eq:block-guide-overlap}
  \frac1{1+Nh^2/c^2}
  \ge\frac{8\gamma^2}{1+8\gamma^2}
  \ge4\gamma^2,
\end{equation}
where the last inequality follows from $\gamma\le1/4$.

For each $i$ with $z_i=1$, let $a_i$ denote the expander-vertex coordinate of the marked position in $x^{(i)}$. The corresponding normalized ground-state vectors are $\ket0\otimes\ket i \otimes\ket{w_{a_i}^-}\otimes\ket v$.
These vectors are mutually orthogonal because their block labels differ, and their projectors sum to $\Pi_0(H_X)$. Therefore,
\begin{equation}\label{eq:family-projector-weight}
  \norm{\Pi_0(H_X)\ket\psi}^2
  =\frac1k
    \sum_{i:z_i=1}
      |\braket u{w_{a_i}^-}|^2.
\end{equation}
Since at least $\lceil k/2\rceil$ terms appear in this sum, applying~\eqref{eq:block-guide-overlap} gives
\begin{equation}\label{eq:final-guide-overlap}
  \norm{\Pi_0(H_X)\ket\psi}^2
  \ge\frac{\lceil k/2\rceil}{k}\,4\gamma^2
  \ge2\gamma^2.
\end{equation}
Thus $\norm{\Pi_0(H_X)\ket\psi}\ge\gamma$ in both Boolean cases, as required by Problem~\ref{prob:gsee}.

Finally, the spectral gap of $H_X$ follows from the two-copy construction. We showed above that $H_X^{\rm rat}$ has spectral gap exactly $h/16$. Multiplication by $\sigma$ scales every eigenvalue difference by the same factor, so
\begin{equation}\label{eq:final-gap}
  \operatorname{gap}(H_X)
  =\frac{\sigma h}{16}
  =g_0\in[4\delta,16\delta].
\end{equation}

We have now verified the norm, sparsity, stoquasticity, guiding-state overlap, and spectral gap of the final Hamiltonian family. Its ground-state energy takes one of the two known values $a$ and $b$, with the choice determined by $F_k(z)$. We summarize these properties for later use.

\begin{lemma}[Properties of the final GSEE family]
\label{lem:final-promises}
For every promised Boolean input $X$, the Hamiltonian $H_X$ is stoquastic and has $\norm{H_X}<1$, sparsity at most $s$, exact dyadic entries, and spectral gap $g_0\in[4\delta,16\delta]$. Its ground-state energy is $b$ when $z=0^k$ and $a$ when $|z|\ge\lceil k/2\rceil$, where the dyadic energies $a<b$ satisfy~\eqref{eq:separation}. The guiding state~\eqref{eq:guide} has nonnegative amplitudes, is independent of $X$, and has squared ground-space overlap one in the first case and at least $2\gamma^2$ in the second case. The nonzero-entry locations are independent of $X$, and each sparse-value query to $H_X$ can be simulated using at most four Boolean queries.
\end{lemma}

\begin{proof}[Proof of Theorem~\ref{thm:gsee}]
We now turn the Boolean query lower bound into a GSEE lower bound. Suppose that an algorithm solves GSEE for every Hamiltonian $H_X$ in the family above using at most $Q$ sparse-matrix queries, with energy estimation error at most $\delta$ and failure probability at most $\varepsilon$.

We use this algorithm to construct a Boolean algorithm for $F_k\circ g_{Nm}^{(k)}$. Given a promised Boolean input $X$, simulate the GSEE algorithm on $H_X$ using the fixed guiding state~\eqref{eq:guide}, and let $\widetilde E$ be its output.

Recall that the two possible ground-state energies $a<b$ are known independently of $X$. We can therefore distinguish them by comparing $\widetilde E$ with their midpoint
\begin{equation}
  m_E=\frac{a+b}{2}.
\end{equation}
Output $1$ if $\widetilde E>m_E$, and output $0$ otherwise.

We show that this rule returns $F_k(z)$ whenever the energy estimate is accurate to within $\delta$. By~\eqref{eq:separation}, the separation between the two energies satisfies $b-a\ge4\delta$. Hence each energy is at least $2\delta$ away from their midpoint. In particular,
\begin{equation}\label{eq:gsee-decoding-margins}
  a+\delta\le m_E-\delta<m_E,
  \qquad
  b-\delta\ge m_E+\delta>m_E.
\end{equation}
Thus, if $E_0(H_X)=a$, every estimate satisfying $|\widetilde E-a|\le\delta$ lies below $m_E$, and the Boolean algorithm outputs $0$. Similarly, if $E_0(H_X)=b$, every estimate satisfying $|\widetilde E-b|\le\delta$ lies above $m_E$, and the algorithm outputs $1$.

Since $E_0(H_X)=b$ exactly when $F_k(z)=1$, the Boolean algorithm computes $F_k(z)$ with error probability at most $\varepsilon$.

We next count the Boolean queries needed to simulate the GSEE algorithm. By Lemma~\ref{lem:oracles}, each sparse-value query to an individual Hamiltonian $H_x$ can be simulated using at most four Boolean queries. A value query to $H_X$ first determines the copy and block labels from its row and column indices. The corresponding matrix entry is either zero or a known rescaling and possible diagonal shift of an entry of $H_{x^{(i)}}$. Since $h$ and $\sigma$ are independent of $X$, this requires at most four queries to the joint Boolean oracle $O_X$.

Moreover, the nonzero-entry locations of $H_X$ are independent of $X$. Therefore, its position oracles and the stronger reversible in-place row and column location oracles, including their inverses, require no Boolean queries. The guiding-state preparation unitary is also independent of $X$ and can be implemented without Boolean queries.

Consequently, the GSEE algorithm using at most $Q$ sparse-matrix queries gives a Boolean algorithm for $F_k\circ g_{Nm}^{(k)}$ using at most $4Q$ Boolean queries.

We can now apply Lemma~\ref{lem:outer} with $L=Nm$. Our choice of $k$ in~\eqref{eq:k-choice} ensures that $\varepsilon\le2^{-2k-1}$, so the lemma gives
\begin{equation}
  4Q
  \ge Q_\varepsilon(F_k\circ g_{Nm}^{(k)})
  =\Omega(k\sqrt{Nm}).
\end{equation}

Finally, by~\eqref{eq:k-choice}, \eqref{eq:N-bounds}, and~\eqref{eq:m-choice}, we have
\begin{equation}
  k=\Theta\!\left(\log\frac1\varepsilon\right),
  \qquad
  N=\Theta\!\left(\frac1{\gamma^2\delta^2}\right),
  \qquad
  m=\Theta(s).
\end{equation}
Consequently,
\begin{equation}
  k\sqrt{Nm}
  =\Theta\!\left(
    \frac{\sqrt{s}}{\gamma\delta}
    \log\frac1\varepsilon
  \right).
\end{equation}
Substituting this into the Boolean query lower bound gives
\begin{equation}\label{eq:gsee-final-scaling}
  Q=\Omega\!\left(
    \frac{\sqrt{s}}{\gamma\delta}
    \log\frac1\varepsilon
  \right),
\end{equation}
which proves Theorem~\ref{thm:gsee}.

The dimension of the constructed Hamiltonian is
\begin{equation}
  D=2kNm
  =O\!\left(
    \frac{s}{\gamma^2\delta^2}
    \log\frac1\varepsilon
  \right).
\end{equation}
All the Hamiltonians in this family are stoquastic, and their nonzero-entry locations are independent of $X$. Hence the lower bound also holds when the stronger reversible in-place location oracles are available, completing the proof.
\end{proof}

\section{Lower Bound for GSP}\label{sec:gsp}

We now establish the corresponding sparse-query lower bound for GSP (Problem~\ref{prob:gsp}). We use the Hamiltonian family $H_X$ constructed for the GSEE lower bound in Section~\ref{sec:gsee-assembly}, but modify it so that the Boolean answer can be recovered directly from a prepared ground state.

\begin{theorem}[Sparse-query GSP lower bound]\label{thm:gsp}
There exist absolute constants $s_0\ge1$ and $\Delta_0>0$
such that, for every integer $s\ge s_0$,
$0<\gamma\le1/4$, $0<\Delta\le\Delta_0$, and
$0<\varepsilon\le1/64$, the quantum query complexity
of GSP (Problem~\ref{prob:gsp}) in both the standard
sparse-access model and the reversible in-place
sparse-location model, even when restricted to
stoquastic Hamiltonians, is
\begin{equation}\label{eq:gsp-lower}
  Q_{\rm sparse}^{\GSP} = \Omega\!\left(
    \frac{\sqrt{s}}{\gamma\Delta}
    \log\frac1\varepsilon
  \right).
\end{equation}
\end{theorem}

\paragraph{Proof overview.}
Recall the Hamiltonian family $H_X$ as defined in~\eqref{eq:final-H} for the GSEE lower bound. Also, recall that its ground-state energy is one of two known values $a<b$, depending on the answer to the Boolean problem $F_k\circ g_{Nm}^{(k)}$, where the Boolean functions $F_k$ and $g_{Nm}$ are defined in~\eqref{eq:outer} and~\eqref{eq:position-parity}, respectively. For GSEE, we distinguished these two cases by estimating the ground-state energy. For GSP, however, the algorithm outputs a quantum state rather than an energy estimate. We therefore need to recover the Boolean answer directly from the final prepared state.
We achieve this by adding a one-dimensional subspace containing a known state $\ket{\rm ref}$ with energy $c_{\rm ref}=(a+b)/2$. When the ground-state energy of $H_X$ is $b$, the reference state has lower energy and becomes the unique ground state of the enlarged Hamiltonian. On the other hand, when the ground-state energy of $H_X$ is $a$, the ground space remains in the original Hamiltonian subspace and is orthogonal to $\ket{\rm ref}$. Thus, measuring whether the prepared state lies in the span of $\ket{\rm ref}$ distinguishes the two Boolean cases without additional sparse-matrix queries. We verify that this construction satisfies the GSP promises and preserves the oracle-simulation properties established for $H_X$. Finally, applying the quantum query lower bound for the composed Boolean function $F_k\circ g_{Nm}^{(k)}$ from Lemma~\ref{lem:outer} then gives the GSP lower bound stated in~\eqref{eq:gsp-lower}.

\bigskip

We now provide a complete proof of the theorem. We begin by recalling the properties of the GSEE family that we need. For a promised Boolean input $X=(x^{(1)},\ldots,x^{(k)})$, let $z_i=g_{Nm}(x^{(i)})$ and $z=(z_1,\ldots,z_k)$. By~\eqref{eq:outer}, the Boolean function $F_k$ distinguishes $z=0^k$, where $F_k(z)=1$, from $|z|\ge\lceil k/2\rceil$, where $F_k(z)=0$.

The Hamiltonian $H_X$ defined in~\eqref{eq:final-H} has ground-state energy
\begin{equation}\label{eq:gsp-inherited-energy}
  E_0(H_X)=
  \begin{cases}
    b,&z=0^k,\\
    a,&|z|\ge\lceil k/2\rceil,
  \end{cases}
\end{equation}
where $a<b$ are known independently of $X$. By~\eqref{eq:separation} and~\eqref{eq:final-gap}, the spectral gap of $H_X$ equals the separation between these two energies:
\begin{equation}\label{eq:gsp-energy-bounds}
  \operatorname{gap}(H_X)=b-a=:g_0,
  \qquad
  4\delta\le g_0\le16\delta.
\end{equation}
Furthermore, the guiding state $\ket\psi$ defined in~\eqref{eq:guide} is independent of $X$. By Lemma~\ref{lem:final-promises}, its squared ground-space overlap satisfies
\begin{equation}\label{eq:gsp-inherited-overlap}
  \begin{aligned}
    \norm{\Pi_0(H_X)\ket\psi}^2&=1,
       &&z=0^k,\\
    \norm{\Pi_0(H_X)\ket\psi}^2&\ge2\gamma^2,
       &&|z|\ge\lceil k/2\rceil.
  \end{aligned}
\end{equation}

We now modify $H_X$ so that its ground space identifies the Boolean answer. Define the reference energy $c_{\rm ref}=(a+b)/2$.
Since $a<b$, we have
\begin{equation}\label{eq:gsp-reference-offset}
  a<c_{\rm ref}<b,
  \qquad
  c_{\rm ref}-a=b-c_{\rm ref}=\frac{g_0}{2}.
\end{equation}

Let $\mathcal H_{\rm ref}=\operatorname{span}\{\ket{\rm ref}\}$ be a one-dimensional Hilbert space with normalized basis state $\ket{\rm ref}$. We enlarge the Hilbert space $\mathcal H_{\rm fam}$ defined in~\eqref{eq:final-family-space} to $\mathcal H_{\rm ext} =\mathcal H_{\rm fam}\oplus\mathcal H_{\rm ref}$.
On this enlarged space, define the Hamiltonian $K_X$ and the guiding state $\ket\chi$ by
\begin{equation}\label{eq:reference}
K_X=H_X\oplus[c_{\rm ref}],
\qquad
\ket\chi=\frac1{\sqrt2}
\left(\ket\psi\oplus\ket{\rm ref}\right).
\end{equation}
Here $[c_{\rm ref}]$ denotes the scalar operator on $\mathcal H_{\rm ref}$. The state $\ket\chi$ is normalized because its two components belong to orthogonal subspaces and are individually normalized. When referring to $\ket{\rm ref}$ as a state in $\mathcal H_{\rm ext}$, we identify it with $0\oplus\ket{\rm ref}$.

We first determine the ground space of $K_X$ and its overlap with $\ket\chi$.

Suppose that $z=0^k$, so $F_k(z)=1$. By~\eqref{eq:gsp-inherited-energy}, the ground-state energy of $H_X$ is $b$. On the other hand, the state $\ket{\rm ref}$ is an eigenstate of $K_X$ with eigenvalue $c_{\rm ref}<b$, by~\eqref{eq:gsp-reference-offset} and~\eqref{eq:reference}. Therefore, $\ket{\rm ref}$ is the unique ground state of $K_X$, and $\Pi_0(K_X)=\ket{\rm ref}\bra{\rm ref}$.
Additionally, using the definition of $\ket\chi$ in~\eqref{eq:reference}, its squared ground-state overlap is
\begin{equation}\label{eq:gsp-reference-overlap-one}
  \norm{\Pi_0(K_X)\ket\chi}^2
  =|\braket{\rm ref}{\chi}|^2
  =\frac12
  \ge\gamma^2,
\end{equation}
where we used $\gamma\le1/4$.

Next suppose that $|z|\ge\lceil k/2\rceil$, so $F_k(z)=0$. By~\eqref{eq:gsp-inherited-energy}, the ground-state energy of $H_X$ is now $a<c_{\rm ref}$. Therefore, every ground state of $K_X$ belongs to the original subspace $\mathcal H_{\rm fam}$, giving $\Pi_0(K_X)=\Pi_0(H_X)\oplus0$.
The reference component of $\ket\chi$ is orthogonal to this ground space. Hence, by~\eqref{eq:reference} and~\eqref{eq:gsp-inherited-overlap}, we have
\begin{equation}
  \norm{\Pi_0(K_X)\ket\chi}^2
  =\frac12\norm{\Pi_0(H_X)\ket\psi}^2 \ge\frac12(2\gamma^2)
   =\gamma^2.
  \label{eq:gsp-reference-overlap-zero}
\end{equation}
Thus~\eqref{eq:gsp-reference-overlap-one} and~\eqref{eq:gsp-reference-overlap-zero} establish the guiding-state overlap promise in both Boolean cases.

We now summarize the two possible ground-space projectors as
\begin{equation}\label{eq:reference-projector-cases}
  \Pi_0(K_X)=
  \begin{cases}
    0\oplus I_{\mathcal H_{\rm ref}},
      &z=0^k,\\
    \Pi_0(H_X)\oplus0,
      &|z|\ge\lceil k/2\rceil.
  \end{cases}
\end{equation}
The ground space may be degenerate in the second case. However, by~\eqref{eq:reference-projector-cases}, the entire ground space is orthogonal to $\ket{\rm ref}$ whenever $F_k(z)=0$.

We next determine the spectral gap of $K_X$. Since $K_X$ is the direct sum in~\eqref{eq:reference}, its spectrum consists of the eigenvalues of $H_X$ together with $c_{\rm ref}$.
If $z=0^k$, the ground-state energy of $K_X$ is $c_{\rm ref}$, and the next distinct eigenvalue is $b$, by~\eqref{eq:gsp-inherited-energy}. Thus
\begin{equation}
  \operatorname{gap}(K_X)
  =b-c_{\rm ref}
  =\frac{g_0}{2},
\end{equation}
where the second equality follows from~\eqref{eq:gsp-reference-offset}.
If $|z|\ge\lceil k/2\rceil$, the ground-state energy of $K_X$ is $a$. Since $\operatorname{gap}(H_X)=g_0$ by~\eqref{eq:gsp-energy-bounds}, every eigenvalue of $H_X$ strictly above $a$ is at least $a+g_0$. On the other hand,~\eqref{eq:gsp-reference-offset} gives
\begin{equation}
  c_{\rm ref}=a+\frac{g_0}{2}<a+g_0.
\end{equation}
Therefore, $c_{\rm ref}$ is the next distinct eigenvalue of $K_X$, and the gap is again $g_0/2$.
Combining the two cases and using~\eqref{eq:gsp-energy-bounds}, we obtain
\begin{equation}\label{eq:gsp-reference-gap}
  \operatorname{gap}(K_X)
  =\frac{b-a}{2}
  =\frac{g_0}{2}
  \in[2\delta,8\delta].
\end{equation}
In the proof of Theorem~\ref{thm:gsp}, we will choose $\delta=\Delta/2$, so~\eqref{eq:gsp-reference-gap} guarantees $\operatorname{gap}(K_X)\ge\Delta$.

We also verify that $K_X$ preserves the remaining properties required by the sparse-access model. By Lemma~\ref{lem:final-promises}, $\norm{H_X}<1$. Since $a,b\in(-1,1)$ by~\eqref{eq:final-H}, their midpoint $c_{\rm ref}$ also satisfies $|c_{\rm ref}|<1$. Therefore,
\begin{equation}
  \norm{K_X}
  =\max\{\norm{H_X},|c_{\rm ref}|\}<1.
\end{equation}
The additional one-dimensional block does not increase the sparsity beyond $s$ and preserves stoquasticity because it has no off-diagonal entries. Since $a$ and $b$ are dyadic by Lemma~\ref{lem:final-promises}, so is $c_{\rm ref}=(a+b)/2$. Moreover, the reference block and the nonzero-entry locations of $H_X$ are independent of $X$, so the support of $K_X$ is also independent of $X$. Finally, the guiding state $\ket\chi$ in~\eqref{eq:reference} is independent of $X$ and has nonnegative amplitudes.

We now show how to recover the Boolean answer from the state prepared by a GSP algorithm. By~\eqref{eq:reference-projector-cases}, the reference state belongs to the ground space when $F_k(z)=1$ and is orthogonal to the ground space when $F_k(z)=0$. We can distinguish these cases by measuring the projector
\begin{equation}\label{eq:gsp-decoding}
  P_{\mathrm{ref}} = \ket{\mathrm{ref}}\bra{\mathrm{ref}},
  \qquad
  \{P_{\mathrm{ref}}, I-P_{\mathrm{ref}}\}.
\end{equation}
We output $1$ when the outcome corresponding to $P_{\rm ref}$ occurs and $0$ otherwise. Since $\ket{\rm ref}$ is a known basis state, this measurement requires no sparse-matrix queries.

Suppose that a GSP algorithm outputs a density operator $\rho$ satisfying
\begin{equation}\label{eq:gsp-decoding-promise}
  \Tr(\Pi_0(K_X)\rho)\ge1-\eta.
\end{equation}
We show that the measurement in~\eqref{eq:gsp-decoding} returns $F_k(z)$ with error probability at most $\eta$.
If $z=0^k$, then $F_k(z)=1$ and $\Pi_0(K_X)=P_{\rm ref}$ by~\eqref{eq:reference-projector-cases}. Therefore, the probability of obtaining the correct Boolean answer is
\begin{equation}
  \Pr[\text{output }1]
  =\Tr(P_{\rm ref}\rho)
  =\Tr(\Pi_0(K_X)\rho)
  \ge1-\eta,
\end{equation}
where the inequality follows from~\eqref{eq:gsp-decoding-promise}.
If $|z|\ge\lceil k/2\rceil$, then $F_k(z)=0$. In this case,~\eqref{eq:reference-projector-cases} gives $P_{\rm ref}\Pi_0(K_X)=0$.
Thus $P_{\rm ref}$ and $\Pi_0(K_X)$ project onto orthogonal subspaces, implying $P_{\rm ref}+\Pi_0(K_X)\preceq I$.
Equivalently, $P_{\rm ref}\preceq I-\Pi_0(K_X)$. Taking the trace against $\rho$ and using~\eqref{eq:gsp-decoding-promise}, we obtain
\begin{align}
  \Pr[\text{output }1]
  &=\Tr(P_{\rm ref}\rho)\notag\\
  &\le1-\Tr(\Pi_0(K_X)\rho)\notag\\
  &\le\eta.
  \label{eq:gsp-decoding-error}
\end{align}
Since the correct answer is $0$, this bounds the error probability by $\eta$.
Therefore, in both Boolean cases, the fixed measurement~\eqref{eq:gsp-decoding} returns $F_k(z)$ with error probability at most $\eta$. This conclusion holds for any output density operator $\rho$ satisfying~\eqref{eq:gsp-decoding-promise}, even when the ground space is degenerate.

We summarize the properties established above.

\begin{lemma}\label{lem:reference}
Let $H_X$, $a$, $b$, and $\ket\psi$ be the GSEE family defined in~\eqref{eq:final-H} and~\eqref{eq:guide}, and define $K_X$ and $\ket\chi$ by~\eqref{eq:reference}. For every promised Boolean input $X$, the Hamiltonian $K_X$ is stoquastic, has $\norm{K_X}<1$, sparsity at most $s$, exact dyadic entries, and spectral gap $\operatorname{gap}(K_X)
  =(b-a)/2\in[2\delta,8\delta]$.
Its guiding state $\ket\chi$ has nonnegative amplitudes, is independent of $X$, and satisfies $\norm{\Pi_0(K_X)\ket\chi}\ge\gamma$.
The nonzero-entry locations of $K_X$ are independent of $X$, and any output state $\rho$ satisfying~\eqref{eq:gsp-decoding-promise} yields the Boolean answer $F_k(z)$ with error probability at most $\eta$ through the measurement~\eqref{eq:gsp-decoding}.
\end{lemma}

\begin{proof}[Proof of Theorem~\ref{thm:gsp}]
We now use Lemma~\ref{lem:reference} to turn the Boolean query lower bound into a GSP lower bound.

Recall that the GSEE construction in Section~\ref{sec:gsee-assembly} applies for $0<\delta\le\delta_0$, where $\delta_0>0$ is an absolute constant. Set $\Delta_0=2\delta_0$ and $\delta=\frac{\Delta}{2}$.
For every $0<\Delta\le\Delta_0$, we have $0<\delta\le\delta_0$, so the family $K_X$ in~\eqref{eq:reference} is well defined. By~\eqref{eq:gsp-reference-gap}, we have $\operatorname{gap}(K_X) \in[\Delta,4\Delta]$.
Thus every $K_X$ satisfies the spectral-gap promise of Problem~\ref{prob:gsp}. Lemma~\ref{lem:reference} then establishes the remaining norm, sparsity, and guiding-state overlap promises.

Suppose that an algorithm solves GSP for every Hamiltonian $K_X$ in this family using at most $Q$ sparse-matrix queries. Let $\rho_X$ denote its output density operator, averaged over all internal measurement outcomes. Since the algorithm solves Problem~\ref{prob:gsp}, its output satisfies
\begin{equation}\label{eq:gsp-reduction-output}
  \Tr(\Pi_0(K_X)\rho_X)\ge1-\varepsilon
\end{equation}
for every promised input $X$.

We use this algorithm to solve the Boolean problem $F_k\circ g_{Nm}^{(k)}$ defined in~\eqref{eq:outer}. Given oracle access to a promised Boolean input $X$, simulate the GSP algorithm on $K_X$ using the fixed guiding state $\ket\chi$ from~\eqref{eq:reference}. After it produces $\rho_X$, perform the measurement~\eqref{eq:gsp-decoding} and output the resulting Boolean bit.

Applying Lemma~\ref{lem:reference} to~\eqref{eq:gsp-reduction-output}, with $\eta=\varepsilon$, shows that this bit equals $F_k(z)$ with probability at least $1-\varepsilon$. Hence the resulting Boolean algorithm has error probability at most $\varepsilon$.

We next count the Boolean queries required for this simulation. By Lemmas~\ref{lem:oracles} and~\ref{lem:final-promises}, each sparse-value query to $H_X$ can be simulated using at most four Boolean queries to $O_X$. Since $K_X=H_X\oplus[c_{\rm ref}]$ by~\eqref{eq:reference}, every matrix entry of $K_X$ is either an entry of $H_X$, the known reference energy $c_{\rm ref}$, or zero. Therefore, each sparse-value query to $K_X$ also requires at most four Boolean queries.

Furthermore, the nonzero-entry locations of $K_X$ are independent of $X$, as established in Lemma~\ref{lem:reference}. Thus all sparse-position queries, including the stronger reversible in-place row and column location queries and their inverses, can be implemented without Boolean queries. The guiding-state preparation unitary for $\ket\chi$ and the measurement~\eqref{eq:gsp-decoding} are also independent of $X$.

Consequently, the GSP algorithm using at most $Q$ sparse-matrix queries gives a Boolean algorithm for $F_k\circ g_{Nm}^{(k)}$ using at most $4Q$ Boolean queries.

By the choice of $k$ in~\eqref{eq:k-choice}, we have $\varepsilon\le2^{-2k-1}$. We can therefore apply Lemma~\ref{lem:outer} with Boolean input length $L=Nm$, obtaining
\begin{equation}\label{eq:gsp-query-reduction}
  4Q
  \ge Q_\varepsilon(F_k\circ g_{Nm}^{(k)})
  =\Omega(k\sqrt{Nm}).
\end{equation}

Finally,~\eqref{eq:k-choice}, \eqref{eq:N-bounds}, and~\eqref{eq:m-choice}, together with $\delta=\Delta/2$, give
\begin{equation}
  k=\Theta\!\left(\log\frac1\varepsilon\right),
  \qquad
  N=\Theta\!\left(\frac1{\gamma^2\Delta^2}\right),
  \qquad
  m=\Theta(s).
\end{equation}
Substituting these expressions into~\eqref{eq:gsp-query-reduction} yields
\begin{equation}\label{eq:gsp-final-scaling}
  Q=\Omega\!\left(
    \frac{\sqrt{s}}{\gamma\Delta}
    \log\frac1\varepsilon
  \right),
\end{equation}
which proves Theorem~\ref{thm:gsp}.

The dimension of $H_X$ is $2kNm$ by~\eqref{eq:final-family-space}. Since~\eqref{eq:reference} adds one basis state, the dimension of $K_X$ is
\begin{equation}
  \dim(K_X)=2kNm+1
  =O\!\left(
    \frac{s}{\gamma^2\Delta^2}
    \log\frac1\varepsilon
  \right).
\end{equation}
As in the GSEE construction, adding a known identity block permits padding to any larger dimension without changing the ground space, spectral gap, or oracle-simulation properties. Since every $K_X$ is stoquastic and has input-independent nonzero-entry locations by Lemma~\ref{lem:reference}, the lower bound also holds under these restrictions and in the stronger reversible in-place sparse-location model.
\end{proof}

Theorem~\ref{thm:gsp} establishes the lower bound for the ground-space weight guarantee in~\eqref{eq:gsp-error}. We now show that the same asymptotic lower bound applies to two stronger ways of requiring the prepared state to approximate the ground space.

\begin{corollary}[Stronger state-approximation guarantees]
\label{cor:distance}
Under the parameter assumptions of Theorem~\ref{thm:gsp}, the same lower bound holds if the output density operator $\rho$ must satisfy
\begin{equation}\label{eq:trace-distance-rho-sigma}
  \frac12\norm{\rho-\sigma}_1\le\varepsilon
\end{equation}
for some density operator $\sigma$ supported on the ground space. For pure-state outputs, requiring Euclidean distance at most $\varepsilon$ from some normalized ground state also gives a lower bound
\begin{equation}
  \Omega\!\left(
    \frac{\sqrt{s}}{\gamma\Delta}
    \log\frac1\varepsilon
  \right).
\end{equation}
\end{corollary}

\begin{proof}
We first consider the trace-distance guarantee. Suppose that the output quantum state $\rho$ satisfies~\eqref{eq:trace-distance-rho-sigma} for some quantum state $\sigma$ supported on the ground space of $H$. By the definition of $\Pi_0(H)$ in Section~\ref{sec:prelim}, we have $\Tr(\Pi_0(H)\sigma)=1$. The variational characterization of trace distance then implies that, for every orthogonal projector $P$, we have
\begin{equation}
  |\!\Tr(P\rho)-\Tr(P\sigma)|
  \le\frac12\norm{\rho-\sigma}_1.
\end{equation}
Taking $P=\Pi_0(H)$ therefore gives
\begin{equation}
  \Tr(\Pi_0(H)\rho)
  \ge\Tr(\Pi_0(H)\sigma)
    -\frac12\norm{\rho-\sigma}_1
    \ge1-\varepsilon.
\end{equation}
Thus any algorithm satisfying the trace-distance guarantee also satisfies the GSP guarantee in~\eqref{eq:gsp-error}. The lower bound of Theorem~\ref{thm:gsp} therefore applies.

Next, suppose that the algorithm outputs a normalized pure state $\ket{\widetilde w}$ satisfying $\norm{\ket{\widetilde w}-\ket w}\le\varepsilon$ for some normalized ground state $\ket w$. Write $\Pi=\Pi_0(H)$. Since $\ket w$ belongs to the ground space, $(I-\Pi)\ket w=0$, and hence
\begin{align}
  \norm{(I-\Pi)\ket{\widetilde w}}^2
  &=\norm{(I-\Pi)
    (\ket{\widetilde w}-\ket w)}^2
    \notag\\
  &\le\norm{\ket{\widetilde w}-\ket w}^2
    \notag\\
  &\le\varepsilon^2.
\end{align}
The first inequality follows because an orthogonal projector cannot increase Euclidean norm. Since $\ket{\widetilde w}$ is normalized and $\Pi$ and $I-\Pi$ are complementary orthogonal projectors, we have
\begin{equation}
  \norm{\Pi\ket{\widetilde w}}^2
  =1-\norm{(I-\Pi)\ket{\widetilde w}}^2
  \ge1-\varepsilon^2.
\end{equation}

Thus the pure-state distance guarantee implies the GSP
ground-space weight guarantee in~\eqref{eq:gsp-error}
with error parameter $\varepsilon^2$. Applying
Theorem~\ref{thm:gsp} with this parameter shows that
any algorithm satisfying the pure-state distance
guarantee requires
\[
  \Omega\!\left(
    \frac{\sqrt{s}}{\gamma\Delta}
    \log\frac1{\varepsilon^2}
  \right)
\]
sparse-matrix queries. Since
$\log(1/\varepsilon^2)=2\log(1/\varepsilon)$,
this is the claimed lower bound.

\end{proof}

\section{Connections to Other Access Models}
\label{sec:access-model-connections}

Theorems~\ref{thm:gsee} and~\ref{thm:gsp} establish lower bounds for GSEE and GSP when the Hamiltonian is accessed through sparse-matrix oracles. We now explain how these results are related to the lower bounds previously established in the block-encoding and time-evolution query models. In particular, we show that our constructions also recover the optimal dependence on the problem parameters in both of these models. We write
$Q_{\rm BE}^{\GSEE}$ and $Q_{\rm BE}^{\GSP}$ for the
query complexities in the block-encoding model, and
$Q_{\rm TE}^{\GSEE}$ and $Q_{\rm TE}^{\GSP}$ for those
in the time-evolution query model.

Recall that the sparse-access model provides the locations and values of the matrix entries of $H$, whereas the block-encoding and time-evolution models provide unitaries that encode spectral information about $H$. To obtain a lower bound in one of these unitary-access models from our sparse-access construction, we must show that its oracle can be implemented using a constant number of sparse-matrix queries. If this is possible, an algorithm making $q$ queries to the unitary oracle would give an algorithm making $O(q)$ sparse-matrix queries. The sparse-query lower bounds would then apply to the unitary-access algorithm.

We will use this observation with Hamiltonians of constant sparsity. The important point is that Theorems~\ref{thm:gsee} and~\ref{thm:gsp} hold for every sufficiently large $s$, so we may fix $s$ to be an absolute constant. The factor $\sqrt{s}$ then becomes a constant, while the joint dependence on $\gamma$, $\delta$ or $\Delta$, and $\varepsilon$ remains unchanged.

We first construct an exact block encoding from constant-sparse access. We then use this block encoding to construct a time-evolution unitary whose eigenphases preserve the ordering and separations of the eigenvalues of the original Hamiltonian. The construction will also preserve the guiding-state overlap up to an absolute constant. These properties will allow us to apply our Boolean reductions to both access models.

\subsection{An exact block encoding from constant-sparse access}
\label{sec:constant-sparse-block}

Fix an integer $s_\ast\ge s_0$, where $s_0$ is the absolute constant in Theorems~\ref{thm:gsee} and~\ref{thm:gsp}. Throughout this section, $s_\ast$ is fixed independently of the problem parameters. Let $M\in\mathbb R^{D\times D}$ be any real symmetric, $s_\ast$-sparse Hamiltonian satisfying $\norm M\le1$. The Hamiltonians $H_X$ and $K_X$ constructed in Sections~\ref{sec:gsee-assembly} and~\ref{sec:gsp} satisfy these assumptions when we set $s=s_\ast$.

We first show that sparse access to $M$ permits an exact block encoding with a constant normalization factor. More precisely, we construct a Hermitian unitary $B_M$ and a known isometry $J$ satisfying
\begin{equation}\label{eq:connection-block}
J^\dagger B_MJ=\frac{M}{\alpha},
\qquad
B_M^\dagger=B_M,
\qquad
B_M^2=I,
\end{equation}
where $\alpha=s_\ast$. Since $s_\ast$ is fixed, $\alpha$ is an absolute constant. We will also show that $B_M$, its inverse, and their controlled versions can be implemented using $O(1)$ sparse-matrix queries.

We begin with two families of states that encode the entries of $M$. Let $\pi_i(1),\ldots,\pi_i(s_\ast)$ be the row-support list of row $i$, padded by distinct zero-entry locations when necessary, as in Section~\ref{sec:sparse-matrix-access}. Similarly, let $\kappa_j(1),\ldots,\kappa_j(s_\ast)$ be the corresponding column-support list for column $j$.

Since $\norm M\le1$, every matrix entry satisfies $|M_{ij}|\le1$. We can therefore define the normalized vectors
\begin{align}
\ket{r_i}
&=\frac1{\sqrt{\alpha}}
  \sum_{\ell=1}^{\alpha}
  \ket{i,\pi_i(\ell)}
  \left(
    \sqrt{|M_{i,\pi_i(\ell)}|}\ket0
    +\sqrt{1-|M_{i,\pi_i(\ell)}|}\ket1
  \right),
  \label{eq:connection-row-state}\\
\ket{c_j}
&=\frac1{\sqrt{\alpha}}
  \sum_{\ell=1}^{\alpha}
  \ket{\kappa_j(\ell),j}
  \left(
    \operatorname{sgn}(M_{\kappa_j(\ell),j})
    \sqrt{|M_{\kappa_j(\ell),j}|}\ket0
    +\sqrt{1-|M_{\kappa_j(\ell),j}|}\ket2
  \right),
  \label{eq:connection-column-state}
\end{align}
where we set $\operatorname{sgn}(0)=1$. The final register has three orthogonal basis states $\ket0,\ket1,\ket2$. Each vector is normalized because its $\alpha$ summands have distinct position labels and each parenthesized vector has norm one. Moreover, the vectors $\ket{r_i}$ are orthonormal as $i$ varies, since their first registers differ, and the vectors $\ket{c_j}$ are orthonormal as $j$ varies, since their second registers differ.

The reason for using different flag states $\ket1$ and $\ket2$ is that they are orthogonal. Consequently, only the components carrying flag $\ket0$ contribute to the inner product between a row state and a column state. For fixed $i,j$ with $M_{ij}\ne0$, the label $(i,j)$ occurs in both sums and contributes $M_{ij}/\alpha$. If $M_{ij}=0$, its contribution is zero even when both lists contain $(i,j)$ as padding. Hence
\begin{equation}\label{eq:connection-overlap}
\braket{r_i}{c_j}=\frac{M_{ij}}{\alpha}.
\end{equation}

We now turn these state families into a unitary block encoding. Since both families are orthonormal, there exist unitaries $U_R$ and $U_C$ that prepare them from the same known input subspace. Let $J_0$ be the isometry identifying the original $D$-dimensional system with this input subspace, so that
\[
U_RJ_0\ket i=\ket{r_i},
\qquad
U_CJ_0\ket j=\ket{c_j}.
\]
Define $U_M=U_R^\dagger U_C$. Taking matrix elements and using~\eqref{eq:connection-overlap}, we obtain
\begin{equation}\label{eq:connection-unitary-block}
\bra iJ_0^\dagger U_MJ_0\ket j
=\braket{r_i}{c_j}
=\frac{M_{ij}}{\alpha}.
\end{equation}
Thus $J_0^\dagger U_MJ_0=M/\alpha$.

We next verify the query cost. To prepare $\ket{r_i}$, we first prepare the uniform superposition over the $\alpha$ row positions. The reversible in-place row-location oracle then maps each position to the corresponding column index. A value query returns the matrix entry, which determines the amplitudes in~\eqref{eq:connection-row-state}. The value register can be uncomputed using the inverse query. The column states are prepared in the same way using the column-location oracle. Since $\alpha=s_\ast$ is an absolute constant, each state-preparation circuit uses $O(1)$ sparse-matrix queries. The same holds for the inverse and controlled circuits. Consequently, $U_M$ and $U_M^\dagger$ can each be implemented using $O(1)$ sparse queries.

The unitary $U_M$ need not be Hermitian. To obtain a Hermitian unitary, introduce one additional qubit and define
\begin{equation}\label{eq:connection-hermitian-block}
B_M=
\begin{pmatrix}
0&U_M\\
U_M^\dagger&0
\end{pmatrix},
\qquad
J\ket\xi=\ket+\otimes J_0\ket\xi,
\end{equation}
where $\ket+=(\ket0+\ket1)/\sqrt2$. Since $U_M$ is unitary, direct multiplication gives $B_M^\dagger=B_M$ and $B_M^2=I$. Moreover,
\[
J^\dagger B_MJ
=\frac12J_0^\dagger(U_M+U_M^\dagger)J_0
=\frac{M}{\alpha},
\]
where the last equality follows from~\eqref{eq:connection-unitary-block} and the Hermiticity of $M$. This establishes~\eqref{eq:connection-block}.

Notice that the preparation isometry $J$ is known independently of the matrix entries of $M$. The dependence on $M$ appears only in $B_M$, whose queries have constant sparse-query cost. In particular, if $\ket\psi$ is a guiding state for $M$, then $J\ket\psi$ can be prepared using the same guiding-state unitary together with operations independent of the unknown matrix entries.

For the GSP family $K_X=H_X\oplus[c_{\rm ref}]$ in~\eqref{eq:reference}, we choose the block encoding to preserve this known direct-sum decomposition. This can be done by constructing the block encoding of $H_X$ as above and a separate, input-independent block encoding of the scalar $c_{\rm ref}$, using the same normalization $\alpha$. Their direct sum gives a block encoding of $K_X/\alpha$ that commutes with the projector onto the reference block. This property will be useful when we recover the Boolean answer from a prepared ground state.

\subsection{Recovering the block-encoding lower bounds}
\label{sec:connection-be-lower}

We now apply the construction above to the hard Hamiltonian families of Sections~\ref{sec:gsee} and~\ref{sec:gsp}. Fixing $s=s_\ast$ in Theorems~\ref{thm:gsee} and~\ref{thm:gsp} gives the sparse-query lower bounds
\begin{equation}\label{eq:connection-constant-sparse}
\Omega\!\left(
\frac1{\gamma\delta}\log\frac1\varepsilon
\right)
\quad\text{and}\quad
\Omega\!\left(
\frac1{\gamma\Delta}\log\frac1\varepsilon
\right),
\end{equation}
respectively, since $\sqrt{s_\ast}$ is an absolute constant.

For GSEE, take the Hamiltonian $H_X$ defined in~\eqref{eq:final-H} and construct its Hermitian block encoding $B_{H_X}$ using~\eqref{eq:connection-hermitian-block}. Consider the normalized Hamiltonian $G_X=H_X/\alpha$. By~\eqref{eq:connection-block}, the unitary $B_{H_X}$ is an exact unit-normalized block encoding of $G_X$. Its ground space is the same as that of $H_X$, and its ground-state energy is $E_0(H_X)/\alpha$.

Suppose an algorithm estimates the ground-state energy of $G_X$ to additive error $\delta_{\rm BE}$ with failure probability at most $\varepsilon$, using $q$ queries to $B_{H_X}$ and its inverse. Each such query can be implemented using $O(1)$ sparse queries to $H_X$. Moreover, multiplying the reported estimate by $\alpha$ gives an estimate of $E_0(H_X)$ to error $\alpha\delta_{\rm BE}$. Thus, by applying Theorem~\ref{thm:gsee} with $\delta=\alpha\delta_{\rm BE}$, we obtain
\[
Q_{\rm BE}^{\GSEE} =\Omega\!\left(
\frac1{\gamma\delta_{\rm BE}}
\log\frac1\varepsilon
\right),
\]
where the constant factor $\alpha$ is absorbed into the $\Omega$ notation.

For GSP, use the family $K_X$ defined in~\eqref{eq:reference}, and again set $G_X=K_X/\alpha$. Its ground space is unchanged, while its spectral gap is divided by $\alpha$. The guiding state $\ket\chi$ is also unchanged. If an algorithm prepares a state with ground-space weight at least $1-\varepsilon$ using $q$ queries to the block encoding of $G_X$, we simulate those queries using the sparse oracles of $K_X$. By choosing the original gap parameter to be $\Delta=\alpha\Delta_{\rm BE}$, we obtain from Theorem~\ref{thm:gsp}
\[
Q_{\rm BE}^{\GSP} =\Omega\!\left(
\frac1{\gamma\Delta_{\rm BE}}
\log\frac1\varepsilon
\right).
\]
The reference-block measurement in~\eqref{eq:gsp-decoding} continues to recover the Boolean answer, since the rescaling does not change the ground space.

We have therefore recovered the joint block-encoding lower bounds for both problems. The reduction uses exact block encodings, so it introduces no approximation error and no additional dependence on $\varepsilon$.

\subsection{Recovering the time-evolution lower bounds}
\label{sec:connection-te-lower}

We next turn to the time-evolution query model. Here the algorithm is given a unitary of the form $\e^{\iu L}$, together with its inverse and controlled versions, for some Hamiltonian $L$ satisfying $\norm L\le1$. Our goal is to construct such a unitary from the exact block encoding $B_M$ above, while preserving the spectral information needed for GSEE and GSP.

A direct approach would be to simulate $\e^{\iu M}$ using sparse-matrix queries. However, an approximate simulation of each time-evolution query would introduce an error that must be controlled across the complete algorithm. We instead construct a different Hamiltonian whose exact time-evolution unitary can be implemented with $O(1)$ sparse queries.

Let $B_M$ and $J$ satisfy~\eqref{eq:connection-block}, and define the projector $P=JJ^\dagger$ onto the image of $J$. Fix the absolute constants $t=1/2, r=1/16$, and $c_0=1/8$, and define
\begin{equation}\label{eq:connection-evolution}
V_M=\e^{\iu c_0}
     \e^{\iu t(I-P)}
     \e^{\iu rB_M}.
\end{equation}

We first show that $V_M$, its inverse, and their controlled versions can be implemented exactly using $O(1)$ sparse queries. The projector $P$ is known independently of $M$, so $\e^{\iu t(I-P)}$ requires no matrix queries. Also, since $B_M^2=I$, its eigenvalues are $\pm1$. A controlled application of $B_M$ can therefore be used to write its eigenvalue into an ancilla qubit. Applying the corresponding phase $\e^{\pm\iu r}$ and then uncomputing the ancilla implements $\e^{\iu rB_M}$ exactly using two controlled applications of $B_M$. Since each application of $B_M$ requires $O(1)$ sparse queries, so does the complete unitary~\eqref{eq:connection-evolution}. The known phase $\e^{\iu c_0}$ requires no queries.

We now determine the eigenphases of $V_M$. Let $\ket{u_E}$ be a normalized eigenvector of $M$ with eigenvalue $E$, and set $\ket{\phi_E}=J\ket{u_E}$. Since $J^\dagger B_MJ=M/\alpha$, we have
\[
PB_M\ket{\phi_E}
=\frac{E}{\alpha}\ket{\phi_E}.
\]
Write $x=E/\alpha$. Because $|E|\le1$ and $\alpha=s_\ast>1$, we have $|x|<1$. We can therefore define the normalized vector
\begin{equation}\label{eq:connection-second-vector}
\ket{\phi_E^\perp}
=\frac{B_M\ket{\phi_E}-x\ket{\phi_E}}
       {\sqrt{1-x^2}}.
\end{equation}
The relation above shows that $P\ket{\phi_E^\perp}=0$, so this vector is orthogonal to $\ket{\phi_E}$. Using $B_M^2=I$, we also obtain
\[
B_M\ket{\phi_E}
=x\ket{\phi_E}
 +\sqrt{1-x^2}\ket{\phi_E^\perp},
\]
and
\[
B_M\ket{\phi_E^\perp}
=\sqrt{1-x^2}\ket{\phi_E}
 -x\ket{\phi_E^\perp}.
\]
Hence the two-dimensional subspace
$\mathcal C_E=\operatorname{span}
\{\ket{\phi_E},\ket{\phi_E^\perp}\}$
is invariant under both $B_M$ and $P$.

In the ordered orthonormal basis
$(\ket{\phi_E},\ket{\phi_E^\perp})$,
their restrictions are
\[
B_M\big|_{\mathcal C_E}
=
\begin{pmatrix}
x&\sqrt{1-x^2}\\
\sqrt{1-x^2}&-x
\end{pmatrix},
\qquad
(I-P)\big|_{\mathcal C_E}
=
\begin{pmatrix}
0&0\\
0&1
\end{pmatrix}.
\]
Substituting into~\eqref{eq:connection-evolution}, we find that the restriction of $V_M$ is
\begin{equation}\label{eq:connection-two-dimensional}
V_M\big|_{\mathcal C_E}
=\e^{\iu c_0}
\begin{pmatrix}1&0\\0&\e^{\iu t}\end{pmatrix}
\left[
\cos r\,I+
\iu\sin r
\begin{pmatrix}
x&\sqrt{1-x^2}\\
\sqrt{1-x^2}&-x
\end{pmatrix}
\right].
\end{equation}

We next calculate its eigenphases. The determinant of the matrix in~\eqref{eq:connection-two-dimensional} is $\e^{\iu(2c_0+t)}$, since the matrix multiplying $\iu\sin r$ has eigenvalues $\pm1$. Its trace is
\[
2\e^{\iu(c_0+t/2)}
\left(
\cos\frac t2\cos r
+x\sin\frac t2\sin r
\right).
\]
Therefore, its two eigenphases are
\begin{equation}\label{eq:connection-two-phases}
\vartheta_\pm(E)
=c_0+\frac t2\pm\omega(E),
\end{equation}
where
\begin{equation}\label{eq:connection-omega}
\omega(E)=\arccos\!\left(
\cos\frac t2\cos r
+\frac E\alpha\sin\frac t2\sin r
\right).
\end{equation}

We now determine which of these eigenphases corresponds to the ground-state energy. Since $-1\le E/\alpha\le1$, the expression inside the arccosine lies between $\cos(t/2+r)$ and $\cos(t/2-r)$. As $0<r<t/2<\pi/2$, and arccosine is decreasing on $[-1,1]$, we obtain
\[
\frac t2-r\le\omega(E)\le\frac t2+r.
\]
Substituting this into~\eqref{eq:connection-two-phases} gives
\begin{equation}\label{eq:connection-phase-bands}
\vartheta_-(E)\in
\left[\frac1{16},\frac3{16}\right],
\qquad
\vartheta_+(E)\in
\left[\frac9{16},\frac{11}{16}\right].
\end{equation}

To account for the rest of the enlarged Hilbert space, choose an orthonormal eigenbasis of $M$ and construct the corresponding two-dimensional spaces $\mathcal C_E$. Their direct sum contains the image of $J$ and is invariant under $B_M$. Since $B_M$ is Hermitian, its orthogonal complement is also invariant. On that complement, $P=0$ and $B_M$ has eigenvalues $\pm1$. Thus the remaining eigenphases of $V_M$ are $c_0+t+r$ and $c_0+t-r$, both of which belong to the second interval in~\eqref{eq:connection-phase-bands}.

Consequently, every eigenphase of $V_M$ lies in $[1/16,11/16]$, and there is no ambiguity from phase wraparound. We may therefore define the Hermitian operator $L_M$ by taking the eigenphases of $V_M$ as its eigenvalues. Equivalently,
\begin{equation}\label{eq:connection-effective-hamiltonian}
V_M=\e^{\iu L_M},
\qquad
\frac1{16}I\preceq L_M
\preceq\frac{11}{16}I.
\end{equation}
In particular, $\norm{L_M}\le11/16<1$, as required by the time-evolution query model. Notice that we need not construct the matrix entries of $L_M$: the oracle supplied to the time-evolution algorithm is precisely the unitary $V_M$, which we have already shown how to implement exactly.

We next show that the ground-state energy and spectral gap of $M$ are preserved up to absolute constants. By~\eqref{eq:connection-phase-bands}, every eigenphase in the minus branch lies below every eigenphase in the plus branch or the remaining subspace. Thus the ground-state energy of $L_M$ is the minimum value of $\vartheta_-(E)$. Define
$f(E)=c_0+t/2-\omega(E)$.
Differentiating~\eqref{eq:connection-omega} gives
\begin{equation}\label{eq:connection-derivative}
f'(E)=
\frac{\sin(t/2)\sin r}
{\alpha\sin\omega(E)}.
\end{equation}
Since $t/2-r\le\omega(E)\le t/2+r<\pi/2$, the denominator is positive and bounded above and below by absolute constants. Hence there are absolute constants $c_{\rm map},C_{\rm map}>0$ such that
\begin{equation}\label{eq:connection-energy-comparison}
c_{\rm map}|E-E'|
\le |f(E)-f(E')|
\le C_{\rm map}|E-E'|
\end{equation}
for all $E,E'\in[-1,1]$. For example, we may take
\[
c_{\rm map}=
\frac{\sin(t/2)\sin r}{\alpha},
\qquad
C_{\rm map}=
\frac{\sin(t/2)\sin r}
{\alpha\sin(t/2-r)}.
\]
Because $\alpha=s_\ast$ is fixed, these are absolute constants independent of the problem parameters.

In particular, $f$ is strictly increasing. Thus the smallest eigenvalue of $M$ gives the smallest eigenphase in the minus branch, and
\begin{equation}\label{eq:connection-ground-energy}
E_0(L_M)=f(E_0(M)).
\end{equation}
Furthermore, every excited eigenvalue of $M$ lies at least
$\operatorname{gap}(M)$ above $E_0(M)$. By~\eqref{eq:connection-energy-comparison},
the corresponding minus-branch eigenphase is at least
$c_{\rm map}\operatorname{gap}(M)$ above $E_0(L_M)$.
All eigenphases outside the minus branch are separated from it
by at least $t-2r=3/8$. Consequently,
$\operatorname{gap}(L_M)\ge
\min\{c_{\rm map}\operatorname{gap}(M),3/8\}$.
For the Hamiltonian families $H_X$ and $K_X$ used below,
the spectral gap is finite and at most $2$, since their norms
are at most one. As $2c_{\rm map}<3/8$, we obtain
\begin{equation}\label{eq:connection-gap}
\operatorname{gap}(L_M)
\ge c_{\rm map}\operatorname{gap}(M)
\end{equation}
for both families.
Thus the transformation preserves both energy differences and the spectral gap up to absolute constant factors.

It remains to verify the guiding-state overlap. This is important because the eigenvectors of $L_M$ belong to the enlarged Hilbert space, and are not simply the embedded eigenvectors $J\ket{u_E}$.

Let $\Pi_-^{(E)}$ denote the projector onto the eigenvector of~\eqref{eq:connection-two-dimensional} with eigenphase $\vartheta_-(E)$, and write
\[
w_E=\norm{\Pi_-^{(E)}\ket{\phi_E}}^2.
\]
We show that $w_E\ge1/2$ for every $E$. To see this, multiply the matrix in~\eqref{eq:connection-two-dimensional} by the known phase $\e^{-\iu(c_0+t/2)}$. Its eigenvalues are then $\e^{\pm\iu\omega(E)}$. Therefore, taking its expectation in $\ket{\phi_E}$ and using the spectral decomposition gives
\[
\operatorname{Im}\!\left(
\e^{-\iu(c_0+t/2)}
\bra{\phi_E}V_M\ket{\phi_E}
\right)
=(1-2w_E)\sin\omega(E).
\]
On the other hand, evaluating the upper-left matrix element of~\eqref{eq:connection-two-dimensional} gives
\[
\operatorname{Im}\!\left(
\e^{-\iu(c_0+t/2)}
\bra{\phi_E}V_M\ket{\phi_E}
\right)
=
\frac E\alpha\cos\frac t2\sin r
-\sin\frac t2\cos r.
\]
Since $E/\alpha\le1$ and $r<t/2$, the last expression is at most
\[
\cos\frac t2\sin r
-\sin\frac t2\cos r
=-\sin\left(\frac t2-r\right)<0.
\]
As $\sin\omega(E)>0$, the preceding identities imply $1-2w_E<0$, and hence $w_E\ge1/2$.

Now let $\ket\psi$ be any guiding state satisfying $\norm{\Pi_0(M)\ket\psi}\ge\gamma$. Its embedded version is $J\ket\psi$, which can be prepared using the supplied guiding-state preparation followed by the fixed embedding $J$. For the hard instances considered below, the guiding-state preparation is independent of the Boolean input. Applying the preceding weight bound to an orthonormal basis of the ground space of $M$, we obtain
\begin{equation}\label{eq:connection-guiding-overlap}
\norm{\Pi_0(L_M)J\ket\psi}^2
\ge\frac12\norm{\Pi_0(M)\ket\psi}^2
\ge\frac{\gamma^2}{2}.
\end{equation}
Thus the new guiding-state overlap amplitude is at least $\gamma/\sqrt2$. This argument does not require the ground space of $M$ to be one-dimensional.

We have now established the three properties needed for the reduction: one query to $\e^{\iu L_M}$ costs $O(1)$ sparse queries to $M$, the energy and gap parameters change by only absolute constant factors, and the guiding-state overlap decreases by at most a factor of $\sqrt2$.

We first apply these properties to GSEE. Take the family $M=H_X$ from~\eqref{eq:final-H}, whose ground-state energy is one of the two known values $a<b$. By~\eqref{eq:connection-ground-energy}, the corresponding time-evolution Hamiltonian $L_{H_X}$ has ground-state energy $f(a)$ or $f(b)$. Since $f$ is strictly increasing, these are two distinct known values. Moreover, their separation satisfies
\[
f(b)-f(a)
\ge c_{\rm map}(b-a)
\ge4c_{\rm map}\delta,
\]
where the last inequality follows from~\eqref{eq:separation}.

Suppose a time-evolution algorithm estimates the ground-state energy of $L_{H_X}$ with additive error $\delta_{\rm TE}$ and failure probability at most $\varepsilon$. Choose the original sparse-family parameter to be $\delta=\delta_{\rm TE}/c_{\rm map}$. Then the two possible ground-state energies are separated by at least $4\delta_{\rm TE}$. As in the proof of Theorem~\ref{thm:gsee}, comparing the estimate with the midpoint of $f(a)$ and $f(b)$ therefore determines the Boolean answer whenever the estimate is accurate.

By~\eqref{eq:connection-guiding-overlap}, a desired time-evolution overlap promise $\gamma_{\rm TE}$ is obtained by choosing the original sparse-family overlap parameter to be $\gamma=\sqrt2\gamma_{\rm TE}$. Every time-evolution oracle query is simulated using $O(1)$ sparse queries, and all other operations are independent of the Boolean input. Applying Theorem~\ref{thm:gsee} gives
\begin{equation}\label{eq:connection-te-gsee}
Q_{\rm TE}^{\GSEE}
=\Omega\!\left(
\frac1{\gamma_{\rm TE}\delta_{\rm TE}}
\log\frac1\varepsilon
\right).
\end{equation}
Here the absolute constants $c_{\rm map}$ and $\sqrt2$ have been absorbed into the $\Omega$ notation.

For GSP, take the reference-level Hamiltonian family $M=K_X$ from~\eqref{eq:reference}. As established in Lemma~\ref{lem:reference}, its ground space belongs entirely to the reference block when $F_k(z)=1$, and entirely to the original Hamiltonian block when $F_k(z)=0$. Recall that we chose the block encoding of $K_X$ to preserve this known direct-sum decomposition. Hence $B_{K_X}$, the projector $P$, and the unitary~\eqref{eq:connection-evolution} all preserve the reference-block label. The same is therefore true of the Hamiltonian $L_{K_X}$ defined through its eigenphases.

Since~\eqref{eq:connection-ground-energy} preserves the ordering of the ground-state energies, every ground state of $L_{K_X}$ belongs to the reference block when $F_k(z)=1$, and to its orthogonal complement when $F_k(z)=0$. Therefore, measuring this known block label on the output of a GSP algorithm recovers the Boolean answer without any further time-evolution queries.

By~\eqref{eq:connection-gap}, the new Hamiltonian has spectral gap at least $c_{\rm map}\operatorname{gap}(K_X)$. For a target gap lower bound $\Delta_{\rm TE}$, choose the original sparse-family gap parameter $\Delta=\Delta_{\rm TE}/c_{\rm map}$. Equation~\eqref{eq:connection-guiding-overlap} again gives the required overlap promise after choosing $\gamma=\sqrt2\gamma_{\rm TE}$. Thus a time-evolution GSP algorithm whose unconditional output has ground-space weight at least $1-\varepsilon$ yields, through the reference-block measurement, a Boolean algorithm with error probability at most $\varepsilon$.

Finally, every query to $\e^{\iu L_{K_X}}$ or its inverse is implemented using $O(1)$ sparse-matrix queries to $K_X$. Every sparse query to $K_X$ costs $O(1)$ Boolean queries.
Thus a $q$-query time-evolution algorithm gives an $O(q)$-query
algorithm for $F_k\circ g_{Nm}^{(k)}$ with error at most $\varepsilon$.
Lemma~\ref{lem:outer} gives $q=\Omega(k\sqrt{Nm})$.
Using the choices in the proof of Theorem~\ref{thm:gsp} with
$s=s_\ast$, $\gamma=\sqrt2\gamma_{\rm TE}$, and
$\Delta=\Delta_{\rm TE}/c_{\rm map}$ gives
\begin{equation}\label{eq:connection-te-gsp}
Q_{\rm TE}^{\GSP}
=\Omega\!\left(
\frac1{\gamma_{\rm TE}\Delta_{\rm TE}}
\log\frac1\varepsilon
\right).
\end{equation}
This proves the corresponding time-evolution lower bound. As in our sparse-access construction, the ground space is allowed to be degenerate, and the guarantee concerns the unconditional output state.

\subsection{Consequences and scope}
\label{sec:connection-scope}

We have shown that the hard Hamiltonian families used to prove Theorems~\ref{thm:gsee} and~\ref{thm:gsp} also yield lower bounds in the block-encoding and time-evolution query models. The key step was to fix the sparsity at an absolute constant and construct the corresponding unitary oracles using only a constant number of sparse-matrix queries. For block encodings, this follows from~\eqref{eq:connection-block}. For time evolution, the exact unitary~\eqref{eq:connection-evolution} gives a normalized Hamiltonian whose spectral parameters are related to those of the original Hamiltonian by~\eqref{eq:connection-energy-comparison}, \eqref{eq:connection-gap}, and~\eqref{eq:connection-guiding-overlap}.

Our constructions give lower bounds of $\Omega(\log(1/\varepsilon)/(\gamma\delta))$ for GSEE and $\Omega(\log(1/\varepsilon)/(\gamma\Delta))$ for GSP in both unitary-access models. These agree with the optimal dependences established in earlier work~\cite{SdW2026}. In this sense, the sparse-access lower bounds extend the earlier parameter picture: when the Hamiltonian is accessed through its matrix entries, the same dependence on overlap, spectral resolution, and error probability necessarily incurs the additional multiplicative factor $\sqrt{s}$.

There are two qualifications. First, the full logarithmic dependence in our lower-bound construction allows a degenerate ground space. The reductions in this section preserve that possible degeneracy, so they do not by themselves establish the corresponding unique-ground-state lower bounds of Ref.~\cite{SdW2026}. Second, the time-evolution reduction concerns queries to the fixed unitary $\e^{\iu L_M}$, its inverse, and their controlled versions. It does not establish a lower bound for arbitrary fractional-time evolutions with cost measured by the total evolution time. Those stronger statements require separate arguments.

Thus, the results of this section recover the earlier joint lower-bound scalings under the stated oracle conventions, while Theorems~\ref{thm:gsee} and~\ref{thm:gsp} establish the additional, necessary dependence on sparsity.

\section{Matching Upper Bounds}\label{sec:upper}

We now establish sparse-query upper bounds for GSEE (Problem~\ref{prob:gsee}) and GSP (Problem~\ref{prob:gsp}) that match the lower bounds of Theorems~\ref{thm:gsee} and~\ref{thm:gsp}. Throughout this section, we use the reversible in-place sparse-location model defined in Section~\ref{sec:sparse-matrix-access}. Recall that the query bounds count only calls to the sparse-matrix oracles, including their controlled and inverse versions. Calls to the guiding-state preparation unitary $A$, its inverse $A^\dagger$, and their controlled versions are not included in the sparse-query count.

\begin{theorem}[Sparse-query GSEE and GSP upper bounds]\label{thm:upper}
Let $H$ be an $s$-sparse Hamiltonian satisfying $\norm H\le1$, and let $\ket\psi=A\ket0$ be a guiding state satisfying $\norm{\Pi_0(H)\ket\psi}\ge\gamma$, where $0<\gamma\le1$.
With access to $A$, $A^\dagger$, and their controlled versions, for $0<\delta,\Delta,\varepsilon\le1/64$, the reversible in-place sparse-location model admits
\begin{align}
  Q_{\rm sparse}^{\GSEE}
  &=O\!\left(
    \frac{\sqrt{s}}{\gamma\delta}
    \log\frac1\varepsilon
  \right),
  \label{eq:gsee-upper-main}\\
  Q_{\rm sparse}^{\GSP}
  &=O\!\left(
    \frac{\sqrt{s}}{\gamma\Delta}
    \log\frac1\varepsilon
  \right).
  \label{eq:gsp-upper-main}
\end{align}
For GSP, we additionally assume $\operatorname{gap}(H)\ge\Delta$, with the gap defined in~\eqref{eq:gap-definition}. The GSP bound achieves the ground-space weight guarantee in~\eqref{eq:gsp-error}. The same asymptotic bound also suffices to prepare a state within trace distance $\varepsilon$ of a state supported on the ground space. Neither result requires a unique ground state.
\end{theorem}

\paragraph{Proof overview.}
As discussed in Section~\ref{sec:intro}, the guided GSEE and GSP algorithms of Refs.~\cite{JW2026,JSWdW2026} use queries to the time-evolution unitary $\e^{\iu H}$ and its inverse, with query complexities $O(\log(1/\varepsilon)/(\gamma\delta))$ and $O(\log(1/\varepsilon)/(\gamma\Delta))$, respectively. A natural approach to obtaining sparse-query upper bounds is therefore to run these algorithms and implement each time-evolution query using the sparse Hamiltonian simulation algorithm of Ref.~\cite{BRSL2026}. However, if the guided algorithm makes $q$ time-evolution queries and each is approximated with operator-norm error at most $\zeta$, the error of the complete computation can be as large as $q\zeta$. To keep this error constant, it is sufficient to choose $\zeta=O(1/q)$, and simulating each query to this accuracy introduces additional factors depending on $q$. We will instead use a different unitary whose eigenphases encode the energies of $H$, and show how to implement the complete guided computation without this additional simulation-precision overhead.

We begin with the Cayley transform $U_H$ of the rescaled Hamiltonian $H/(4\sqrt{s})$, defined in~\eqref{eq:phase-map}. On an eigenvector of $H$ with energy $E$, the eigenphase of $U_H$ is $\theta(E)=-2\arctan(E/(4\sqrt{s}))$. Thus $U_H$ has the same eigenspaces as $H$, while its eigenphases are related to the energies by a known, strictly decreasing function $\theta$. For GSEE, we use the phase-shifted unitary $V_H=\e^{\iu\pi/4}U_H$ from~\eqref{eq:shifted-cayley}. The phase shift places its eigenphases in $(0,\pi/2)$, avoiding phase wraparound. Its maximum-eigenphase eigenspace is exactly the ground space of $H$, so the guided phase-estimation algorithm of Ref.~\cite{JW2026} can estimate its maximum eigenphase and thereby recover $E_0(H)$. For GSP, we instead use $W_H=\e^{\iu\pi/4}U_H^\dagger$, also defined in~\eqref{eq:shifted-cayley}. We can write $W_H=\e^{\iu H_{\rm eff}}$, where
  $H_{\rm eff}=\frac{\pi}{4}I
    +2\arctan\!\left(H/(4\sqrt{s})\right)$.
The function $E\mapsto\pi/4+2\arctan(E/(4\sqrt{s}))$ is strictly increasing, so $H_{\rm eff}$ has the same ground space as $H$. Moreover, for every excited energy $E>E_0(H)$, the eigenphase difference of $W_H$ between $E$ and $E_0(H)$ is at least $8(E-E_0(H))/(17\sqrt{s})$ by~\eqref{eq:phase-energy-comparison}. Consequently, $\operatorname{gap}(H)\ge\Delta$ implies $\operatorname{gap}(H_{\rm eff})\ge8\Delta/(17\sqrt{s})$. We can therefore apply the guided GSP algorithm of Ref.~\cite{JSWdW2026} to $H_{\rm eff}$ using queries to $W_H$ and its inverse, once we have a sufficiently accurate estimate of its minimum eigenphase. The analysis of the algorithm of Ref.~\cite{JSWdW2026} assumes a unique ground state, whereas our GSP problem allows a degenerate ground space. We therefore address this by restricting the algorithm to the subspace (see~\eqref{eq:cyclic}), which contains the guiding state $\ket\psi$ and has a unique ground state with overlap at least $\gamma$ with $\ket\psi$.

We next show how to implement the queries to $V_H$, $V_H^\dagger$, $W_H$, and $W_H^\dagger$ using the sparse-matrix oracles of $H$. Since these unitaries differ from $U_H$ or $U_H^\dagger$ only by known phase factors, it suffices to consider queries to $U_H$ and $U_H^\dagger$. The Cayley-transform construction of Ref.~\cite{BRSL2026} gives an exact \emph{transducer} representation of these queries: an auxiliary component is used to implement the desired unitary on the system and is returned unchanged. Unlike an ordinary approximate implementation, such representations can be composed exactly using the transducer framework of Ref.~\cite{Belovs2024}. In Lemma~\ref{lem:cayley-upper}, we show that each query to $U_H$, $U_H^\dagger$, or their controlled versions admits an exact transducer using $O(1)$ sparse-matrix queries, and that an algorithm making $q$ such queries can be implemented with $O(q)$ sparse-matrix queries at any fixed constant final error. Thus, the Cayley transform and transducer framework are existing tools, and here we use them to implement the guided algorithms without approximating each of their queries separately.

For GSEE, we apply the guided phase-estimation algorithm of Ref.~\cite{JW2026} to $V_H$. By~\eqref{eq:phase-energy-comparison}, we show that an estimate of its maximum eigenphase with additive error at most $8\delta/(17\sqrt{s})$ gives, upon inverting the aforementioned function $\theta$, an estimate of $E_0(H)$ with additive error at most $\delta$. The guided phase-estimation algorithm achieves this precision with constant success probability using $q=O(\sqrt{s}/(\gamma\delta))$ queries to $V_H$ and $V_H^{\dagger}$. By Lemma~\ref{lem:cayley-upper}, we can implement the complete algorithm with $O(q)$ sparse queries and sufficiently small constant error that its success probability remains strictly greater than $1/2$. We therefore obtain an energy estimate accurate to $\delta$ with probability at least $1/2+\beta$, for some absolute constant $\beta>0$, using $O(\sqrt{s}/(\gamma\delta))$ sparse queries. To reduce the failure probability to $\varepsilon$, we repeat this procedure independently $R$ times and take the median of the resulting estimates. The median is accurate whenever more than half of the individual estimates are accurate, and Hoeffding's inequality bounds the probability of the opposite event by $\e^{-2\beta^2R}$. Choosing $R=O(\log(1/\varepsilon))$ therefore gives the total sparse-query complexity $O(\sqrt{s}\log(1/\varepsilon)/(\gamma\delta))$, establishing~\eqref{eq:gsee-upper-main}.

For GSP, the guided GSP algorithm of Ref.~\cite{JSWdW2026} requires a sufficiently accurate estimate of the minimum eigenphase of the time-evolution unitary before its preparation procedure can be applied. Since this estimate is not supplied as part of our GSP problem, we first use the GSEE algorithm established above to obtain an estimate $\widetilde E_0$ satisfying $|\widetilde E_0-E_0(H)|\le\Delta/32$. Recall that we apply the guided GSP algorithm to the effective Hamiltonian $H_{\rm eff}$, whose time-evolution unitary is $W_H=\e^{\iu\pi/4}U_H^\dagger$ by~\eqref{eq:shifted-cayley}. Since the eigenphases of $W_H$ are known functions of the eigenvalues of $H$, the estimate $\widetilde E_0$ also determines an estimate of the minimum eigenphase of $W_H$. By~\eqref{eq:gsp-phase-estimate}, this phase estimate has error less than one-third of the guaranteed spectral gap $8\Delta/(17\sqrt{s})$ of $H_{\rm eff}$, and therefore satisfies the accuracy requirement of Ref.~\cite{JSWdW2026}. Applying that algorithm to $H_{\rm eff}$ then produces a candidate state $\rho_{\rm cand}$ satisfying $\Tr(\Pi_0(H)\rho_{\rm cand})\ge p_0$ for some absolute constant $p_0>0$, using $O(\sqrt{s}/(\gamma\Delta))$ queries to $W_H$ and $W_H^\dagger$. To implement these queries using sparse-matrix oracles, recall that $W_H$ differs from $U_H^\dagger$ only by a known phase factor. We can therefore apply the transducer construction of Lemma~\ref{lem:cayley-upper} to the complete guided GSP algorithm, just as we did previously for GSEE. This gives the same asymptotic sparse-query cost, that is, $O(\sqrt{s}/(\gamma\Delta))$ sparse queries. However, GSP requires ground-space weight at least $1-\varepsilon$, so we must further increase the weight of this candidate in the ground space. For this purpose, we construct a \emph{verifier} that accepts ground-state eigenvectors and rejects excited-state eigenvectors, with the probability of either incorrect decision at most $O(\e^{-2r})$, where $r\ge1$ controls the accuracy of the verifier.

Unlike the candidate-preparation procedure, which produces a state $\rho_{\rm cand}$ with constant ground-space weight using queries to the Cayley unitary through the transducer framework, the verifier uses controlled time evolutions under $H$ directly. More precisely, it uses $R=\Theta(r)$ repeated phase-estimation tests, each involving controlled evolutions for times $O(1/\Delta)$. Instead of simulating these controlled evolutions separately, we use the fact that they are functions of the same Hamiltonian and therefore commute. In particular, for fixed time parameters $t_1,\ldots,t_R$, their product satisfies $\prod_{j=1}^R\e^{\iu t_jH}=\exp(\iu H\sum_{j=1}^R t_j)$. We use this identity to express the combined action of all controlled evolutions as a single evolution $\e^{\iu T\widehat H}$, where $T=O(r/\Delta)$ and the Hamiltonian $\widehat H$ has diagonal blocks that are scalar multiples of $H$ (see~\eqref{eq:clock-H}). Since $\widehat H$ has norm at most one, remains $s$-sparse, and has sparse-matrix oracles implementable with a constant number of queries to those of $H$, we can simulate the combined evolution only once. Applying the simulation bound of Ref.~\cite{BRSL2026} with error $\e^{-r}$ then gives a verifier with sparse-query cost $O(\sqrt{s}r/\Delta)$ without an additional logarithmic overhead from the repetitions.

Finally, we combine the constant-quality candidate obtained through the Cayley-transducer construction with the sparse ground-space verifier. Starting with a fresh candidate, we apply verifiers with increasing accuracy, discarding the candidate and preparing a new one whenever a test rejects. At verification stage $j$, the accuracy parameter is $r_j=r_0 2^j$, so the sparse-query cost of that stage is $O(\sqrt{s}r_j/\Delta)$. We show in~\eqref{eq:late-rejection} that the probability of an attempt first rejecting at stage $j\ge1$ is $O(\e^{-r_j})$, which decreases exponentially even though the cost of that stage increases linearly with $r_j$. This relation allows us to control the total cost of rejected attempts and impose a deterministic cutoff without introducing an additional factor of $\log(1/\varepsilon)$. Accounting also for the possible failure of the initial energy estimate and for all cutoff branches, the resulting procedure prepares a state $\rho$ satisfying $\Tr(\Pi_0(H)\rho)\ge1-\varepsilon$ with sparse-query complexity $O(\sqrt{s}\log(1/\varepsilon)/(\gamma\Delta))$, proving~\eqref{eq:gsp-upper-main}.

\bigskip

We now give the complete proof, beginning with the Cayley transform
and its sparse-query implementation. We then establish the GSEE
upper bound, followed by the GSP upper bound, where we first
construct a constant-quality ground-state candidate and then
develop the verification procedure needed to achieve the desired
ground-space weight.

\subsection{Exact Cayley transduction}\label{sec:cayley}

We begin by constructing the unitary $U_H$ used in the proof overview and determining how accurately its eigenphases must be estimated to recover the energies of $H$. For the $s$-sparse Hamiltonian $H$, with $\norm H\le1$, define $U_H$ as follows:
\begin{equation}\label{eq:phase-map}
 U_H=\left(I-\frac{\iu H}{4\sqrt{s}}\right)
       \left(I+\frac{\iu H}{4\sqrt{s}}\right)^{-1}
     =\exp\!\left(-2\iu\arctan\frac{H}{4\sqrt{s}}\right).
\end{equation}
The inverse exists because the eigenvalues of the Hermitian matrix $H$ are real. To understand the action of $U_H$, let $\ket{\phi_E}$ be any eigenvector of $H$ with eigenvalue $E$. The scalar identity $(1-\iu x)/(1+\iu x)=\e^{-2\iu\arctan x}$ for real $x$ then gives
\begin{equation}\label{eq:cayley-eigenphase}
 U_H\ket{\phi_E}=\e^{\iu\theta(E)}\ket{\phi_E},
 \qquad \theta(E)=-2\arctan\frac{E}{4\sqrt{s}}.
\end{equation}
In particular, $U_H$ is unitary and preserves every eigenspace of $H$. Since $\theta$ is strictly decreasing, smaller energies of $H$ correspond to larger eigenphases of $U_H$, where the eigenphases are understood in the interval specified below.

We next relate the difference between two energies to the difference between their eigenphases. Differentiating the function in~\eqref{eq:cayley-eigenphase}, and using $|E|\le1$ and $s\ge1$, we obtain
\begin{equation}\label{eq:phase-derivative}
 \theta'(E)=-\frac{1}{2\sqrt{s}}\frac1{1+E^2/(16s)},
 \qquad \frac8{17\sqrt{s}}\le|\theta'(E)|\le\frac1{2\sqrt{s}}.
\end{equation}
The mean value theorem therefore implies that, for all $E,E'\in[-1,1]$,
\begin{equation}\label{eq:phase-energy-comparison}
 \frac8{17\sqrt{s}}|E-E'|
 \le|\theta(E)-\theta(E')|
 \le\frac1{2\sqrt{s}}|E-E'|.
\end{equation}
Thus resolving two energies separated by $x$ requires eigenphase precision of order $x/\sqrt{s}$. The same inequality shows that a spectral gap $\Delta$ of $H$ gives an eigenphase separation of at least $8\Delta/(17\sqrt{s})$ between the ground space and every excited eigenspace.

The guided algorithms will be applied to two phase-shifted versions of $U_H$. Define
\begin{equation}\label{eq:shifted-cayley}
 V_H=\e^{\iu\pi/4}U_H,
 \qquad W_H=\e^{\iu\pi/4}U_H^\dagger.
\end{equation}
By~\eqref{eq:cayley-eigenphase}, $|\theta(E)|\le2\arctan(1/4)<\pi/4$, so the eigenphases of both unitaries lie strictly between $0$ and $\pi/2$ and cannot wrap around $2\pi$. The ground space of $H$ is the maximum-eigenphase eigenspace of $V_H$ and the minimum-eigenphase eigenspace of $W_H$. The phase factors in~\eqref{eq:shifted-cayley} are known independently of $H$, and therefore, they require no sparse-matrix queries, even when the unitaries are controlled.

We now show how to implement queries to these unitaries without approximating each one separately. We use the transducer framework of Ref.~\cite{Belovs2024}. Let $\mathcal H_{\rm pub}$ be the Hilbert space on which we want to implement a unitary $U$, and let $\mathcal H_{\rm priv}$ be an additional Hilbert space used by the transducer. The subscripts ``pub'' and ``priv'' stand for public and private, respectively. A transducer for $U$ consists of a unitary $S$ acting on the direct sum $\mathcal H_{\rm pub}\oplus\mathcal H_{\rm priv}$ and a linear map $\Gamma:\mathcal H_{\rm pub}\to\mathcal H_{\rm priv}$ satisfying
\begin{equation}\label{eq:transduction}
 S\bigl(\ket\xi\oplus\Gamma\ket\xi\bigr)
 =U\ket\xi\oplus\Gamma\ket\xi
 \qquad(\ket\xi\in\mathcal H_{\rm pub}).
\end{equation}
Thus, for every public input $\ket\xi$, the unitary $S$ applies $U$ to the public component while leaving the private component $\Gamma\ket\xi$ unchanged. This private component is called the \emph{catalyst}. It is part of the mathematical representation and need not be prepared by the algorithm.

To obtain an ordinary quantum circuit, we use the transducer conversion theorem of Ref.~\cite[Theorem~3.2]{Belovs2024}. If $\norm{\Gamma\ket\xi}^2\le W\norm\xi^2$ for every public input, the theorem gives an implementation of the public action with Euclidean error at most $\tau>0$, using $O(1+W/\tau^2)$ applications of $S$ and its inverse. The advantage of this representation is that several transducers can be composed exactly before applying the conversion theorem. Consequently, the approximation error is introduced only when the complete computation is converted into an ordinary circuit.

The following lemma follows from the sparse Cayley-transform
construction of Ref.~\cite{BRSL2026} and the transducer
composition and conversion results of Ref.~\cite{Belovs2024}.
We state it in the form needed for our algorithms and include
a proof in Appendix~\ref{app:cayley-transducer} for completeness.

\begin{lemma}\label{lem:cayley-upper}
Let $H$ be an $s$-sparse Hermitian matrix with $\norm H\le1$. In the reversible in-place sparse-location model, the unitary $U_H$ in~\eqref{eq:phase-map} admits an exact transducer satisfying
\begin{equation}
 \norm{\Gamma\ket\xi}^2\le\frac{17}{16}\norm\xi^2
 \qquad(\ket\xi\in\mathcal H_{\rm pub}).
\end{equation}
The transducer unitary $S$, its inverse $S^\dagger$, and their controlled versions can each be implemented using $O(1)$ queries to the sparse-matrix oracles of $H$.
Consequently, any circuit making $q\ge1$ queries to $U_H$, $U_H^\dagger$, and their controlled versions, interleaved with operations independent of the sparse representation of $H$, can be implemented with $O(q)$ sparse-matrix queries and any fixed constant Euclidean error in a purification of its output. The claim also allows untouched reference registers, intermediate measurements, and classical control. If $q=0$, no sparse-matrix queries are required.
\end{lemma}

\begin{proof}
    The proof is given in Appendix~\ref{app:cayley-transducer}.
\end{proof}

In the proof of Lemma~\ref{lem:cayley-upper}, we establish the exact transducer construction, its implementation using sparse-matrix oracles, and the transducer-composition bound. The importance of this lemma is that it allows us to implement the complete guided algorithms without approximating each of their unitary queries separately. Recall from~\eqref{eq:shifted-cayley} that the unitaries $V_H$ and $W_H$, used for GSEE and GSP, respectively, differ from $U_H$ or $U_H^\dagger$ only by known phase factors. Thus every query to these unitaries, including their controlled and inverse versions, can be represented using the Cayley transducers of Lemma~\ref{lem:cayley-upper}. If a guided algorithm makes $q$ such queries, we can compose their exact transducers into a single transducer and then convert the complete computation into an ordinary quantum circuit with $O(q)$ sparse-matrix queries and constant final error. In particular, we do not need to approximate each individual unitary query to error $O(1/q)$, which would introduce the additional simulation overhead discussed in Section~\ref{sec:intro}. We will use this result in Section~\ref{sec:gsee-upper} to implement guided phase estimation for GSEE, and in Section~\ref{sec:gsp-candidate} to prepare a constant-quality candidate for GSP. The subsequent ground-space verification procedure will instead use the sparse Hamiltonian simulation method developed in Section~\ref{sec:verifier}.

\subsection{Upper bound for GSEE}\label{sec:gsee-upper}

We now prove the GSEE upper bound using the phase-shifted Cayley unitary $V_H$ defined in~\eqref{eq:shifted-cayley}. Recall that the guiding state $\ket\psi=A\ket0$ satisfies $\norm{\Pi_0(H)\ket\psi}\ge\gamma$, as assumed in Problem~\ref{prob:gsee}. Our first step is to estimate the maximum eigenphase of $V_H$, which determines the ground-state energy of $H$. We then convert the phase estimate into an energy estimate, implement the resulting computation using sparse-matrix queries, and reduce its failure probability to $\varepsilon$.

The algorithm is summarized below. The analysis following it establishes the phase precision, the effect of the sparse-query implementation, and the success probability of the median.

\medskip
\noindent\rule{\linewidth}{0.4pt}\hypertarget{alg:gsee}{}
\par\noindent\textbf{Algorithm 1 (Ground-state energy estimation).}

\smallskip
\textbf{Input:} Reversible in-place sparse-location and value access to $H$, access to $A$, $A^\dagger$
(and their controlled versions), overlap lower bound $\gamma$,
additive energy-estimation error $\delta$, and failure probability~$\varepsilon$.
\begin{enumerate}
\setlength{\itemsep}{2pt}
\item Define $V_H=\e^{\iu\pi/4}U_H$ using~\eqref{eq:phase-map} and~\eqref{eq:shifted-cayley}, and set $h_{\rm ph}=8\delta/(17\sqrt{s})$.
\item Run the guided maximum-eigenphase algorithm of Ref.~\cite{JW2026} with unitary $V_H$, guiding state $\ket\psi$, overlap lower bound $\gamma$, and phase precision $h_{\rm ph}$. Use the Cayley-transducer construction of Lemma~\ref{lem:cayley-upper} to implement the complete algorithm using sparse-matrix queries to $H$, with the final Euclidean error chosen to be at most $1/24$.
\item Clamp the phase estimate $\widehat\vartheta_0$ obtained in the previous step to $[\pi/4+\theta(1),\pi/4+\theta(-1)]$, denote the result by $\vartheta_{\rm cl}$, and record $\widetilde E=-4\sqrt{s}\tan((\vartheta_{\rm cl}-\pi/4)/2)$ as an energy estimate.
\item Perform Steps 2--3 independently a total of $R$ times, where $R$ is the smallest odd integer satisfying $R\ge32\log(1/\varepsilon)$, and output the median of the $R$ energy estimates, denoted by~$\widehat E_{\rm med}$.
\end{enumerate}
\textbf{Guarantee:} $\Pr[|\widehat E_{\rm med}-E_0(H)|\le\delta]\ge1-\varepsilon$, using $O(\sqrt{s}\log(1/\varepsilon)/(\gamma\delta))$ sparse-matrix queries.
\par\noindent\rule{\linewidth}{0.4pt}
\medskip

By~\eqref{eq:cayley-eigenphase} and~\eqref{eq:shifted-cayley}, an eigenvector $\ket{\phi_E}$ of $H$ with energy $E$ is an eigenvector of $V_H$ with eigenphase $\vartheta(E)=\pi/4+\theta(E)$, where $\theta(E)=-2\arctan(E/(4\sqrt{s}))$. Since $\theta$ is strictly decreasing, $E_0(H)$ corresponds to the largest eigenphase, namely $\vartheta_0=\pi/4+\theta(E_0(H))$. Thus the maximum-eigenphase eigenspace of $V_H$ is exactly the ground space of $H$, including when that space is degenerate. In particular, the overlap of the guiding state with this eigenspace is $\norm{\Pi_0(H)\ket\psi}\ge\gamma$. Moreover, all eigenphases of $V_H$ lie in $(0,\pi/2)$, as established after~\eqref{eq:shifted-cayley}. We can therefore apply the guided phase-estimation algorithm of Ref.~\cite[Theorem~1.1]{JW2026}: for phase precision $0<h_{\rm ph}\le1/8$, it returns an estimate $\widehat\vartheta_0$ satisfying $|\widehat\vartheta_0-\vartheta_0|\le h_{\rm ph}$ with probability at least $2/3$, using $O(1/(\gamma h_{\rm ph}))$ queries to $V_H$, $V_H^\dagger$, and their controlled versions.

We next determine the phase precision needed to obtain an energy estimate accurate to $\delta$. By~\eqref{eq:cayley-eigenphase}, the relation between energy and phase can be inverted as $E=-4\sqrt{s}\tan(\theta(E)/2)$. However, the phase estimate $\widehat\vartheta_0$ need not lie in the interval corresponding to energies in $[-1,1]$, so we first restrict it to this known interval. Since $\norm H\le1$ and $\theta$ is decreasing, the true phase $\vartheta_0$ lies in $[\vartheta_{\min},\vartheta_{\max}]$, where $\vartheta_{\min}=\pi/4+\theta(1)$ and $\vartheta_{\max}=\pi/4+\theta(-1)$. Define the clamped phase and its associated energy estimate by
\begin{equation}\label{eq:phase-inversion}
 \vartheta_{\rm cl}
 =\min\{\vartheta_{\max},\max\{\vartheta_{\min},\widehat\vartheta_0\}\},
 \qquad
 \widetilde E=-4\sqrt{s}\tan\frac{\vartheta_{\rm cl}-\pi/4}{2}.
\end{equation}
Because $\vartheta_{\rm cl}$ belongs to the image of $[-1,1]$ under $E\mapsto\pi/4+\theta(E)$, the inverse formula gives $\widetilde E\in[-1,1]$ and $\theta(\widetilde E)=\vartheta_{\rm cl}-\pi/4$. Clamping cannot increase the error relative to $\vartheta_0$: if $\widehat\vartheta_0$ already lies in the interval it is unchanged, while otherwise it is replaced by the nearer endpoint, which lies between $\widehat\vartheta_0$ and $\vartheta_0$. Hence $|\vartheta_{\rm cl}-\vartheta_0|\le|\widehat\vartheta_0-\vartheta_0|$.

We can now apply the lower bound in~\eqref{eq:phase-energy-comparison} to $\widetilde E$ and $E_0(H)$, both of which belong to $[-1,1]$. Using the definitions above, we obtain
\begin{equation}
 \frac8{17\sqrt{s}}|\widetilde E-E_0(H)|
 \le |\theta(\widetilde E)-\theta(E_0(H))|
 =|\vartheta_{\rm cl}-\vartheta_0|
 \le|\widehat\vartheta_0-\vartheta_0|.
\end{equation}
Thus, whenever the phase-estimation error is at most $h_{\rm ph}$, the energy-estimation error is at most $17\sqrt{s}\,h_{\rm ph}/8$. To make this no larger than $\delta$, choose
\begin{equation}\label{eq:gsee-phase-precision}
 h_{\rm ph}=\frac{8\delta}{17\sqrt{s}}.
\end{equation}
This precision satisfies $h_{\rm ph}\le1/8$ because $\delta\le1/64$ and $s\ge1$. Consequently, the ideal guided phase-estimation algorithm produces an estimate $\widetilde E$ with $|\widetilde E-E_0(H)|\le\delta$ with probability at least $2/3$. Its number of queries to $V_H$, $V_H^\dagger$, and their controlled versions is
\begin{equation}
 q=O\!\left(\frac1{\gamma h_{\rm ph}}\right)
  =O\!\left(\frac{\sqrt{s}}{\gamma\delta}\right).
\end{equation}

These are queries to $V_H$, rather than sparse-matrix queries to $H$. To obtain an algorithm in our access model, recall from~\eqref{eq:shifted-cayley} that $V_H=\e^{\iu\pi/4}U_H$. The known phase factor can be applied without querying $H$, including for controlled queries. Therefore, each query to $V_H$ or $V_H^\dagger$ can be expressed using the corresponding query to $U_H$ or $U_H^\dagger$ together with operations independent of $H$. Lemma~\ref{lem:cayley-upper} applies to the entire $q$-query guided algorithm: by composing the exact Cayley transducers before converting the computation into a quantum circuit, it gives an implementation using $O(q)=O(\sqrt{s}/(\gamma\delta))$ sparse-matrix queries, with any chosen fixed constant Euclidean error in a purification of the final state.

We must still account for the effect of this final implementation error on the probability of obtaining a correct energy estimate. Choose the Euclidean error of the complete implementation to be at most $1/24$, and let $\ket{\Psi_{\rm id}}$ and $\ket{\Psi_{\rm sp}}$ be normalized purifications of the ideal and implemented final states, respectively. Lemma~\ref{lem:cayley-upper} guarantees $\norm{\ket{\Psi_{\rm id}}-\ket{\Psi_{\rm sp}}}\le1/24$. For any normalized pure states $\ket u,\ket v$, the trace distance satisfies
\begin{equation}\label{eq:pure-trace-distance}
 \frac12\norm{\ket u\bra u-\ket v\bra v}_1
 =\sqrt{1-|\braket uv|^2}
 \le\norm{\ket u-\ket v}.
\end{equation}
It follows that the ideal and implemented purifications have trace distance at most $1/24$. Discarding work registers and measuring the energy-estimate register cannot increase this distance, so the probability of any event defined by the reported estimate changes by at most $1/24$. In particular, the implemented circuit satisfies
\begin{equation}
 \Pr\bigl[|\widetilde E-E_0(H)|\le\delta\bigr]
 \ge\frac23-\frac1{24}
 =\frac58=\frac12+\beta,
 \qquad \beta=\frac18.
\end{equation}
Thus one implemented run returns a $\delta$-accurate estimate with probability strictly greater than one half, using $O(\sqrt{s}/(\gamma\delta))$ sparse-matrix queries.

Finally, we reduce the failure probability from a constant to $\varepsilon$. Run the implemented procedure independently an odd number $R$ of times, obtaining estimates $\widetilde E_1,\ldots,\widetilde E_R$, and let $\widehat E_{\rm med}$ be their median. Define $X_j=1$ if $|\widetilde E_j-E_0(H)|\le\delta$ and $X_j=0$ otherwise. The variables $X_1,\ldots,X_R$ are independent and satisfy $\Pr(X_j=1)\ge1/2+\beta$. Whenever more than half of them equal one, more than half of the estimates lie in the interval $[E_0(H)-\delta,E_0(H)+\delta]$, so their median lies in that interval as well. Consequently, Hoeffding's inequality gives
\begin{equation}\label{eq:gsee-median-bound}
 \Pr\bigl[|\widehat E_{\rm med}-E_0(H)|>\delta\bigr]
 \le\Pr\!\left[\sum_{j=1}^R X_j\le\frac R2\right]
 \le\e^{-2\beta^2R}.
\end{equation}
Choosing $R$ to be the smallest odd integer satisfying $R\ge\log(1/\varepsilon)/(2\beta^2)$ makes the last expression at most $\varepsilon$. Since each run uses $O(\sqrt{s}/(\gamma\delta))$ sparse-matrix queries and $R=O(\log(1/\varepsilon))$, the complete procedure uses
\begin{equation}\label{eq:gsee-upper}
 Q_{\rm sparse}^{\GSEE}=O\!\left(
    \frac{\sqrt{s}}{\gamma\delta}\log\frac1\varepsilon
 \right)
\end{equation}
sparse-matrix queries, proving~\eqref{eq:gsee-upper-main}.

\subsection{Upper bound for GSP}\label{sec:gsp-upper}

We now describe the GSP algorithm before proving its guarantees. A candidate is a newly prepared state whose ground-space weight is at least $p_0=1/200$ when the initial energy estimate is accurate. A verifier performs repeated phase-estimation tests and accepts when a majority of the reported energies lie below the threshold $E_{\rm th}$. Otherwise, it rejects. An attempt starts with a fresh candidate and applies a sequence of verifiers. It succeeds only if all verifiers accept. If one rejects, the candidate is discarded and a new attempt begins. To ensure a worst-case query bound, we also stop when the prescribed number of attempts or verification budget is reached.

\medskip
\noindent\rule{\linewidth}{0.4pt}\hypertarget{alg:gsp}{}
\par\noindent\textbf{Algorithm 2 (Ground-state preparation).}

\smallskip
\textbf{Input:} Reversible in-place sparse-location and value access to $H$, access to $A$, $A^\dagger$
and their controlled versions, overlap lower bound $\gamma$,
spectral-gap lower bound $\Delta$, and error parameter $\varepsilon$.

\begin{enumerate}
\setlength{\itemsep}{4pt}

\item Run the GSEE algorithm (\hyperlink{alg:gsee}{Algorithm~1}) with
$(\delta,\varepsilon')=(\Delta/32,\varepsilon/4)$
to obtain an estimate $\widetilde E_0$ of the ground-state
energy. Set
\[
\overline E_0=\min\{1,\max\{-1,\widetilde E_0\}\},
\qquad
\Delta_{\rm ph}=\frac{8\Delta}{17\sqrt{s}},
\]
\[
\widetilde\varphi_0=
\frac{\pi}{4}
+2\arctan\!\left(\frac{\overline E_0}{4\sqrt{s}}\right),
\qquad
E_{\rm th}=\widetilde E_0+\frac{\Delta}{2}.
\]

\item Set $p_0=1/200$ and $p=p_0/4$. Choose the
verification parameters
\[
r_0=\left\lceil\log\frac8{\sqrt{p_0}}\right\rceil,
\qquad
r_j=2^jr_0,
\]
\[
M=\left\lceil
\frac{\log(4/\varepsilon)}{-\log(1-p)}
\right\rceil,
\qquad
J=\min\{j\ge0:4M\e^{-2r_j}\le\varepsilon/4\}.
\]

\item Define
\[
K_*=1+\e^{r_0/2}
+16\sum_{j\ge1}\e^{-r_j/2},
\qquad
B_*=4\left(M\log K_*+\log\frac4\varepsilon\right).
\]
Set the verification budget $B=B_*+2r_J$ and initialize
a counter $b=0$, which records the sum of the verification
parameters used across all attempts.

\item Repeat the following for at most $M$ attempts:

\begin{enumerate}
\setlength{\itemsep}{3pt}

\item Prepare a fresh candidate state. If $\gamma\ge1/3$,
prepare $A\ket0$. Otherwise, run the guided GSP algorithm
of Ref.~\cite[Section~2.1]{JSWdW2026} with unitary
$W_H=\e^{\iu\pi/4}U_H^\dagger$, guiding state $\ket\psi$,
overlap lower bound $\gamma$, minimum-eigenphase estimate
$\widetilde\varphi_0$, and gap lower bound $\Delta_{\rm ph}$.
Use the Cayley-transducer construction of
Lemma~\ref{lem:cayley-upper} to implement the complete
guided algorithm using sparse-matrix queries to $H$,
with the final Euclidean error chosen to be at most $1/200$.

\item Starting with $j=0$, apply the following verification
procedure for $j=0,1,\ldots,J$:

\begin{enumerate}
\setlength{\itemsep}{3pt}

\item Check whether $b+r_j>B$. If so, stop the algorithm
and return a fixed normalized fallback state.

\item Apply the sparse ground-space verifier $V_{r_j}$
of Lemma~\ref{lem:sparse-verifier}. 
This verifier coherently performs $\lceil7r_j\rceil$ phase-estimation tests using $\e^{\iu(H+2I)}$, compares their energy estimates with $E_{\rm th}$, and sets an acceptance flag if a strict majority of the estimates are below $E_{\rm th}$.
The complete verifier is implemented using sparse-matrix queries to $H$, as described in the lemma.

\item Set $b\leftarrow b+r_j$ and measure the acceptance flag.

\item If the verifier rejects, discard the candidate and
begin the next attempt with a freshly prepared state,
keeping the current value of $b$.

If the verifier accepts, continue to stage $j+1$.
If all $J+1$ verifiers accept, return the system register
of the candidate as the output state.

\end{enumerate}

\end{enumerate}

\item If none of the $M$ attempts is accepted, return
the fixed normalized fallback state.

\end{enumerate}

\textbf{Guarantee:} The output state $\rho$
satisfies $\Tr(\Pi_0(H)\rho)\ge1-\varepsilon$, using
$O(\sqrt{s}\log(1/\varepsilon)/(\gamma\Delta))$
sparse-matrix queries in the worst case.

\par\noindent\rule{\linewidth}{0.4pt}
\medskip

The rest of this subsection proves the guarantees used in \hyperlink{alg:gsp}{Algorithm~2}. We first construct a constant-quality candidate, then establish the error and query bounds for the verifier, and finally analyze the restart procedure and its cutoff.

\subsubsection{A constant-quality candidate for GSP}\label{sec:gsp-candidate}

Recall that the guiding state $\ket\psi=A\ket0$ satisfies
$\norm{\Pi_0(H)\ket\psi}\ge\gamma$, so its ground-space
weight is at least $\gamma^2$. Our first goal is to prepare a state whose ground-space weight is bounded below by an absolute positive constant, independently of $\gamma$. We will then use the verifier constructed in Section~\ref{sec:verifier} to reduce the remaining weight outside the ground space to at most $\varepsilon$.

The guided GSP algorithm of Ref.~\cite[Section~2.1]{JSWdW2026} requires an estimate of the smallest eigenphase of the unitary on which it acts. Such an estimate is not part of the input to Problem~\ref{prob:gsp}, so we first obtain an estimate of $E_0(H)$ using the GSEE algorithm (\hyperlink{alg:gsee}{Algorithm~1}). For the present subsection, assume that we have obtained a number $\widetilde E_0$ satisfying
\begin{equation}\label{eq:gsp-energy-estimate}
 |\widetilde E_0-E_0(H)|\le\frac\Delta{32}.
\end{equation}
The GSEE bound in~\eqref{eq:gsee-upper} allows us to obtain this estimate with any desired failure probability.

We next identify the unitary to which the guided GSP algorithm will be applied. Recall from~\eqref{eq:shifted-cayley} that $W_H=\e^{\iu\pi/4}U_H^\dagger$. By~\eqref{eq:cayley-eigenphase}, if $H\ket{\phi_E}=E\ket{\phi_E}$, then $W_H\ket{\phi_E}=\e^{\iu\varphi(E)}\ket{\phi_E}$, where $\varphi(E)=\pi/4-\theta(E)=\pi/4+2\arctan(E/(4\sqrt{s}))$. Equivalently, we can write
\begin{equation}
 W_H=\e^{\iu H_{\rm eff}},\qquad
 H_{\rm eff}=\frac\pi4I+2\arctan\!\left(\frac{H}{4\sqrt{s}}\right).
\end{equation}
Since $\varphi$ is strictly increasing, the smallest eigenvalue of $H_{\rm eff}$ is $\varphi(E_0(H))$, and its ground-space projector is $\Pi_0(H)$. Moreover, if $E>E_0(H)$ is any excited energy, the spectral-gap promise and the lower inequality in~\eqref{eq:phase-energy-comparison} imply
\begin{equation}
 \varphi(E)-\varphi(E_0(H))
 \ge\frac8{17\sqrt{s}}(E-E_0(H))
 \ge\frac{8\Delta}{17\sqrt{s}}.
\end{equation}
Thus $H_{\rm eff}$ has the same ground space as $H$, and its spectral gap is at least $\Delta_{\rm ph}:=8\Delta/(17\sqrt{s})$. As noted after~\eqref{eq:shifted-cayley}, all eigenphases of $W_H$ lie in $(0,\pi/2)$, so they can be treated as ordinary real numbers without ambiguity from phase wraparound.

We must now show that~\eqref{eq:gsp-energy-estimate} gives a sufficiently accurate estimate of the smallest eigenphase $\varphi(E_0(H))$. Since $E_0(H)\in[-1,1]$, first clamp $\widetilde E_0$ to this interval, setting $\overline E_0=\min\{1,\max\{-1,\widetilde E_0\}\}$, and define $\widetilde\varphi_0=\varphi(\overline E_0)$. The clamping cannot increase the distance to $E_0(H)$, so $|\overline E_0-E_0(H)|\le\Delta/32$. Both energies now belong to $[-1,1]$, and hence the upper inequality in~\eqref{eq:phase-energy-comparison} gives
\begin{equation}\label{eq:gsp-phase-estimate}
 \begin{aligned}
 |\widetilde\varphi_0-\varphi(E_0(H))|
 &\le\frac1{2\sqrt{s}}|\overline E_0-E_0(H)|\\
 &\le\frac\Delta{64\sqrt{s}}
 =\frac{17}{512}\Delta_{\rm ph}
 <\frac{\Delta_{\rm ph}}3.
 \end{aligned}
\end{equation}
The estimate $\widetilde\varphi_0$ is therefore within one-third of the known lower bound $\Delta_{\rm ph}$ on the spectral gap of $H_{\rm eff}$. This is precisely the energy-estimate condition needed for the guided GSP procedure of Ref.~\cite[Section~2.1]{JSWdW2026} when applied to $W_H=\e^{\iu H_{\rm eff}}$.

We first explain what that procedure gives when the ground state is unique, as assumed in Ref.~\cite{JSWdW2026}; we will remove this assumption below. The procedure uses the guiding state, the estimate $\widetilde\varphi_0$, and the gap lower bound $\Delta_{\rm ph}$ to distinguish the ground-state eigenphase from the excited eigenphases, and then increases the ground-state overlap by amplitude amplification with error reduction. In particular, Section~2.1 of Ref.~\cite{JSWdW2026} gives, for $\gamma<1/3$, a coherent preparation procedure whose final pure state has ground-state overlap amplitude at least $1/10$. Thus, after its work registers are discarded, the output has ground-space weight at least $1/100$. The procedure makes
\begin{equation}\label{eq:gsp-candidate-queries}
 q=O\!\left(\frac1{\gamma\Delta_{\rm ph}}\right)
  =O\!\left(\frac{\sqrt{s}}{\gamma\Delta}\right)
\end{equation}
queries to $W_H$, $W_H^\dagger$, and their controlled versions. Here the second equality follows by substituting $\Delta_{\rm ph}=8\Delta/(17\sqrt{s})$. When $\gamma\ge1/3$, the original guiding state already has ground-space weight at least $\gamma^2\ge1/9$, so no amplification is necessary. We may therefore assume $\gamma<1/3$ in the remaining discussion.

The guided GSP algorithm of Ref.~\cite[Section~2.1]{JSWdW2026} is analyzed under the assumption that the ground state is unique, whereas our GSP formulation in Problem~\ref{prob:gsp} allows the ground space of $H$ to be degenerate. We now show that the same algorithm can be used in our setting. The key observation is that, although the ground space may contain several orthogonal eigenvectors, the guiding state $\ket\psi$ determines a single normalized vector in that space. We will construct an invariant subspace containing $\ket\psi$ in which this vector is the unique ground state, and then show that the preparation algorithm acts entirely within this subspace.

Let $H=\sum_E E\Pi_E$ be the spectral decomposition of $H$, where the sum runs over its distinct eigenvalues. For each $E$, define $p_E=\norm{\Pi_E\ket\psi}^2$. Whenever $p_E>0$, let $\ket{\psi_E}=\Pi_E\ket\psi/\sqrt{p_E}$. These vectors are normalized and mutually orthogonal, since $\Pi_E\Pi_{E'}=0$ for $E\ne E'$. Moreover, because $\sum_E\Pi_E=I$, the guiding state can be written as $\ket\psi=\sum_{E:p_E>0}\sqrt{p_E}\ket{\psi_E}$. We therefore define
\begin{equation}\label{eq:cyclic}
 \mathcal C_\psi
 =\operatorname{span}\{\ket{\psi_E}:p_E>0\}.
\end{equation}
By construction, $\ket\psi\in\mathcal C_\psi$, and $H\ket{\psi_E}=E\ket{\psi_E}$ for every basis vector defining this subspace. Consequently, $\mathcal C_\psi$ is invariant under $H$ and every function of $H$, including $W_H$ and its inverse.

We next determine the ground state of $H$ restricted to $\mathcal C_\psi$. The guiding-state promise in Problem~\ref{prob:gsp} gives $p_{E_0(H)}=\norm{\Pi_0(H)\ket\psi}^2\ge\gamma^2>0$.
Hence $\mathcal C_\psi$ contains the normalized vector
\begin{equation}
 \ket{\psi_{E_0(H)}}
 =\frac{\Pi_0(H)\ket\psi}
        {\norm{\Pi_0(H)\ket\psi}}.
\end{equation}
Although the full ground space of $H$ may have dimension greater than one, $\ket{\psi_{E_0(H)}}$ is the only vector in the orthonormal basis defining~\eqref{eq:cyclic} with eigenvalue $E_0(H)$. Thus the intersection of $\mathcal C_\psi$ with the ground space of $H$ is exactly $\operatorname{span}\{\ket{\psi_{E_0(H)}}\}$. In particular, the restriction of $H$ to $\mathcal C_\psi$ has a unique ground state, whose overlap with the guiding state satisfies
\begin{equation}
 |\braket{\psi_{E_0(H)}}{\psi}|
 =\norm{\Pi_0(H)\ket\psi}\ge\gamma.
\end{equation}
Every other eigenvalue appearing in this restriction is an excited eigenvalue of $H$, and is therefore at least $\Delta$ above $E_0(H)$. Similarly, the restriction of $H_{\rm eff}$ to $\mathcal C_\psi$ has a unique ground state with the same overlap and spectral gap at least $\Delta_{\rm ph}$.

It remains to show that we can apply the guided GSP algorithm to this restricted Hamiltonian without changing the operations used by the algorithm. Recall that the phase-estimation and eigenphase-marking operations of Ref.~\cite[Section~2.1]{JSWdW2026} are constructed using controlled powers of $W_H$ and $W_H^\dagger$, together with operations on additional work registers. Let $\mathcal H_{\rm anc}$ denote the Hilbert space of these work registers, including the phase-estimation registers and flag qubits. Since $W_H$ and $W_H^\dagger$ preserve $\mathcal C_\psi$, these operations preserve $\mathcal C_\psi\otimes\mathcal H_{\rm anc}$. The recursive procedure also uses reflections about states prepared at previous stages. If a normalized state $\ket\Phi$ belongs to this subspace, then for every $\ket v\in\mathcal C_\psi\otimes\mathcal H_{\rm anc}$,
\begin{equation}
 (2\ket\Phi\bra\Phi-I)\ket v
 =2\braket\Phi v\ket\Phi-\ket v
 \in\mathcal C_\psi\otimes\mathcal H_{\rm anc}.
\end{equation}
Thus these reflections also preserve the subspace. This statement concerns the complete reflection, and the individual gates used to implement it, including calls to the guiding-state preparation unitary $A$ and its inverse, need not preserve $\mathcal C_\psi$ separately.

Since the initial state belongs to $\mathcal C_\psi$, induction over the recursive preparation stages shows that all states prepared by the ideal algorithm have their system registers supported on $\mathcal C_\psi$. The final selection among candidate registers also preserves this property, since it involves only flag-controlled swaps between registers supported on $\mathcal C_\psi$. The algorithm therefore has the same action on these states as the unique-ground-state procedure applied to the restriction of $H_{\rm eff}$ to $\mathcal C_\psi$. We can consequently apply the constant-overlap guarantee of Ref.~\cite[Section~2.1]{JSWdW2026} to this restricted computation. Let $\rho_{\rm id}$ denote its reduced output state after discarding the work registers. The guarantee gives
\begin{equation}
 \bra{\psi_{E_0(H)}}\rho_{\rm id}
 \ket{\psi_{E_0(H)}}\ge\frac1{100}.
\end{equation}
Finally, since $\ket{\psi_{E_0(H)}}$ belongs to the ground space of $H$, we have
$\ket{\psi_{E_0(H)}}\bra{\psi_{E_0(H)}}
\preceq\Pi_0(H)$.
Taking the trace against $\rho_{\rm id}$ therefore yields
\begin{equation}
 \Tr\bigl(\Pi_0(H)\rho_{\rm id}\bigr)
 \ge\bra{\psi_{E_0(H)}}\rho_{\rm id}
       \ket{\psi_{E_0(H)}}
 \ge\frac1{100}.
\end{equation}
This proves that the ideal guided GSP procedure produces a state with constant ground-space weight even when the ground space of $H$ is degenerate.

Finally, the query count in~\eqref{eq:gsp-candidate-queries} is expressed in queries to $W_H$, rather than sparse-matrix queries to $H$. To obtain an algorithm in our access model, recall again from~\eqref{eq:shifted-cayley} that $W_H=\e^{\iu\pi/4}U_H^\dagger$. Its queries, including controlled queries, therefore reduce to queries to $U_H$ or $U_H^\dagger$ and known phase operations. Lemma~\ref{lem:cayley-upper} applies to the complete guided GSP computation, allowing its $q=O(\sqrt{s}/(\gamma\Delta))$ Cayley-unitary queries to be implemented using $O(q)$ sparse-matrix queries at any fixed constant Euclidean error in a purification of the final state. Choose that error to be at most $1/200$. By~\eqref{eq:pure-trace-distance} and the fact that discarding work registers cannot increase trace distance, the ground-space weight of the implemented state $\rho_{\rm cand}$ differs from that of $\rho_{\rm id}$ by at most $1/200$. Thus its weight is at least $1/100-1/200=1/200$ when $\gamma<1/3$, while for $\gamma\ge1/3$ we may use $\ket\psi$ directly. In both cases there are absolute constants $p_0>0$ and $C>0$ such that
\begin{equation}\label{eq:candidate}
 \Tr\bigl(\Pi_0(H)\rho_{\rm cand}\bigr)\ge p_0,
 \qquad C_{\rm cand}\le C\frac{\sqrt{s}}{\gamma\Delta}.
\end{equation}
Here $C_{\rm cand}$ counts the sparse-matrix queries needed to prepare one candidate, and the weight guarantee applies to the unconditional output state, averaged over all internal measurement outcomes. In particular, we may choose $p_0=1/200$ independently of $s$, $\gamma$, and $\Delta$. This is the candidate-preparation guarantee used in Section~\ref{sec:restart}.

\subsubsection{Sparse ground-space verification}\label{sec:verifier}

The candidate state constructed in~\eqref{eq:candidate} has ground-space weight at least $p_0$, where $p_0>0$ is an absolute constant. However, recall that the GSP problem formulation in Problem~\ref{prob:gsp} requires a state whose ground-space weight is at least $1-\varepsilon$. We therefore need a procedure that distinguishes the ground space of $H$ from its orthogonal complement, so that we can test whether a candidate should be retained. We will construct such a verifier by estimating energies and comparing them with a known threshold.

As in Section~\ref{sec:gsp-candidate}, we assume for now that the estimate $\widetilde E_0$ satisfies~\eqref{eq:gsp-energy-estimate}, namely $|\widetilde E_0-E_0(H)|\le\Delta/32$. Since every excited energy is at least $E_0(H)+\Delta$, a threshold near $\widetilde E_0+\Delta/2$ should separate the ground-state energy from all excited energies. Define
\begin{equation}\label{eq:verification-threshold}
 E_{\rm th}=\widetilde E_0+\frac\Delta2.
\end{equation}
We now verify this separation without assuming that $\widetilde E_0$ equals $E_0(H)$. From~\eqref{eq:gsp-energy-estimate}, we have $\widetilde E_0\ge E_0(H)-\Delta/32$, and therefore $E_{\rm th}-E_0(H)\ge\Delta/2-\Delta/32=15\Delta/32$. On the other hand, if $E>E_0(H)$ is any excited eigenvalue, then $E\ge E_0(H)+\Delta$ by the gap promise, while $\widetilde E_0\le E_0(H)+\Delta/32$. Consequently, $E-E_{\rm th}\ge\Delta-\Delta/32-\Delta/2=15\Delta/32$. Together, these inequalities give
\begin{equation}\label{eq:gsp-threshold-separation}
 E_{\rm th}-E_0(H)\ge\frac{15\Delta}{32},
 \qquad E-E_{\rm th}\ge\frac{15\Delta}{32}
 \quad(E>E_0(H)).
\end{equation}
In particular, suppose an energy-estimation test is performed on an eigenvector of energy $E$ and returns $\widehat E$ with $|\widehat E-E|\le\Delta/8$. If $E=E_0(H)$, then $\widehat E\le E_0(H)+\Delta/8<E_{\rm th}$, and if $E>E_0(H)$, then $\widehat E\ge E-\Delta/8>E_{\rm th}$. Thus an estimate accurate to $\Delta/8$ determines which side of the ground-space decomposition contains the input eigenvector. Our verifier will repeat this test and accept when a majority of the estimates lie below $E_{\rm th}$.

We state the guarantee in operator norm because the candidate to be verified need not be an energy eigenvector. It may be an arbitrary superposition of ground and excited eigenvectors, and it may also be entangled with registers prepared by an earlier computation. The operator-norm formulation ensures that the verification errors are bounded for all such inputs.

\begin{lemma}[Sparse ground-space verification]\label{lem:sparse-verifier}
Suppose that $H$ is an $s$-sparse Hamiltonian satisfying $\norm H\le1$ and $\operatorname{gap}(H)\ge\Delta$, where $0<\Delta\le1$, and that the known estimate $\widetilde E_0$ satisfies~\eqref{eq:gsp-energy-estimate}. For every $r\ge1$, there is an implemented isometry $V_r:\mathbb C^D\to\mathbb C^D\otimes\mathcal W_r\otimes\mathbb C_F^2$, where $\mathcal W_r$ contains the work registers and the qubit $F$ records acceptance or rejection, with sparse-matrix query cost
\begin{equation}\label{eq:verifier-cost}
 C_r=O\!\left(\frac{\sqrt{s}}\Delta r\right).
\end{equation}
There is an ideal isometry $V_r^{\rm id}$ satisfying $\norm{V_r-V_r^{\rm id}}\le\e^{-r}$ and preserving each eigenspace of $H$: for every spectral projector $\Pi_E$,
\begin{equation}\label{eq:verifier-intertwining}
 (\Pi_E\otimes I_{\mathcal W_r\otimes\mathbb C_F^2})V_r^{\rm id}
 =V_r^{\rm id}\Pi_E.
\end{equation}
Let $P_{\rm acc}=I_D\otimes I_{\mathcal W_r}\otimes\ket1\bra1_F$ and $P_{\rm rej}=I_D\otimes I_{\mathcal W_r}\otimes\ket0\bra0_F$. Then
\begin{equation}\label{eq:ideal-verifier}
 \norm{P_{\rm rej}V_r^{\rm id}\Pi_0(H)}\le\e^{-r},
 \qquad
 \norm{P_{\rm acc}V_r^{\rm id}(I-\Pi_0(H))}\le\e^{-r}.
\end{equation}
The same operator-norm bounds hold if the input is entangled with a reference register.
\end{lemma}

\begin{proof}
We first construct an ideal verifier using exact time-evolution unitaries, prove that it makes the correct decision with high probability on every energy eigenvector, and then extend that guarantee to arbitrary quantum states. Finally, we replace the exact evolutions by a single sparse Hamiltonian simulation and establish the query bound~\eqref{eq:verifier-cost}.

We begin with one phase-estimation test. Since $\norm H\le1$, every eigenvalue $E$ of $H$ lies in $[-1,1]$. Hence $E+2\in[1,3]\subset(0,2\pi)$, so the eigenphase of $U_{\rm pe}=\e^{\iu(H+2I)}$ can be represented by the real number $E+2$ without crossing the phase boundary at $0$ or $2\pi$. We use this known shift so that an estimate of the eigenphase can be converted directly into an estimate of $E$ by subtracting two.

Let $N_{\rm pe}=2^b$ be the smallest power of two satisfying $N_{\rm pe}\ge256\pi/\Delta$. Because $N_{\rm pe}$ is the smallest such power, $256\pi/\Delta\le N_{\rm pe}<512\pi/\Delta$, and therefore $N_{\rm pe}=\Theta(1/\Delta)$. Prepare a $b$-qubit phase-estimation register in the uniform state $N_{\rm pe}^{-1/2}\sum_{t=0}^{N_{\rm pe}-1}\ket t$. For any normalized eigenvector $\ket{u_E}$ satisfying $H\ket{u_E}=E\ket{u_E}$, the controlled power of $U_{\rm pe}$ acts as
\begin{equation*}
 \frac1{\sqrt{N_{\rm pe}}}\sum_{t=0}^{N_{\rm pe}-1}\ket t\ket{u_E}
 \longmapsto
 \frac1{\sqrt{N_{\rm pe}}}\sum_{t=0}^{N_{\rm pe}-1}
 \e^{\iu t(E+2)}\ket t\ket{u_E}.
\end{equation*}
Here we used $U_{\rm pe}^t\ket{u_E}=\e^{\iu t(E+2)}\ket{u_E}$. Thus the controlled evolution changes only the phase-estimation register and leaves the system eigenvector unchanged.

Apply the inverse quantum Fourier transform to the phase-estimation register. Its amplitude at $y\in\mathbb Z_{N_{\rm pe}}$ is the average of the $N_{\rm pe}$ phases in the preceding expression, with the Fourier phase $\e^{-2\pi\iu ty/N_{\rm pe}}$ included. The probability of obtaining $y$ is therefore
\begin{equation}\label{eq:phase-estimation-distribution}
 p_y=\left|\frac1{N_{\rm pe}}\sum_{t=0}^{N_{\rm pe}-1}
 \e^{\iu t(E+2-2\pi y/N_{\rm pe})}\right|^2.
\end{equation}
We next bound the probability that this outcome is far from the phase $E+2$. Because phases are defined modulo $2\pi$, let $d_y$ denote the circular distance, measured in Fourier-grid units, between the integer $y$ and the real number $N_{\rm pe}(E+2)/(2\pi)$. Explicitly,
\begin{equation*}
 d_y=\min_{k\in\mathbb Z}\left|
 y-\frac{N_{\rm pe}(E+2)}{2\pi}+kN_{\rm pe}\right|,
 \qquad 0\le d_y\le\frac{N_{\rm pe}}2.
\end{equation*}
Choose $a_y\in[-\pi,\pi]$ congruent to $E+2-2\pi y/N_{\rm pe}$ modulo $2\pi$. Then $|a_y|=2\pi d_y/N_{\rm pe}$, and the sum in~\eqref{eq:phase-estimation-distribution} is unchanged if its phase difference is replaced by $a_y$. For $d_y>0$, the geometric-series identity gives
\begin{align*}
 p_y
 &=\frac1{N_{\rm pe}^2}\left|\frac{1-\e^{\iu N_{\rm pe}a_y}}
 {1-\e^{\iu a_y}}\right|^2\\
 &\le\frac1{N_{\rm pe}^2\sin^2(a_y/2)}
 =\frac1{N_{\rm pe}^2\sin^2(\pi d_y/N_{\rm pe})}.
\end{align*}
The inequality uses $|1-\e^{\iu N_{\rm pe}a_y}|\le2$ and $|1-\e^{\iu a_y}|=2|\sin(a_y/2)|$. Since $0\le\pi d_y/N_{\rm pe}\le\pi/2$, the elementary inequality $\sin x\ge2x/\pi$ on $[0,\pi/2]$ now yields
\begin{equation}\label{eq:phase-estimation-point-bound}
 p_y\le\frac1{N_{\rm pe}^2\sin^2(\pi d_y/N_{\rm pe})}
 \le\frac1{4d_y^2}\qquad(d_y>0).
\end{equation}
This bounds the probability of any individual outcome in terms of its distance from the correct phase. For each integer $\ell\ge1$, at most two outcomes have $d_y\in[\ell,\ell+1)$, one on each side of the correct phase around the Fourier circle. Therefore, summing~\eqref{eq:phase-estimation-point-bound} over all outcomes with $d_y\ge16$ gives
\begin{equation}\label{eq:phase-estimation-tail}
 \Pr(d_y\ge16)
 \le\sum_{\ell=16}^{\infty}\frac2{4\ell^2}
 =\frac12\sum_{\ell=16}^{\infty}\frac1{\ell^2}
 \le\frac12\int_{15}^{\infty}\frac{dx}{x^2}
 =\frac1{30}<\frac1{10}.
\end{equation}
Thus the phase-estimation outcome satisfies $d_y<16$ with probability greater than $9/10$.

On this event, the circular distance between $2\pi y/N_{\rm pe}$ and $E+2$ is less than $32\pi/N_{\rm pe}\le\Delta/8$. Since $E+2\in[1,3]$ and $\Delta/8\le1/8$, such a small distance cannot cross either endpoint of $[0,2\pi)$: both endpoints are at least one unit away from the possible true phases. The circular error is therefore the ordinary absolute difference $|2\pi y/N_{\rm pe}-(E+2)|$. Defining $\widehat E=2\pi y/N_{\rm pe}-2$, we conclude that
\begin{equation*}
 \Pr\!\left[|\widehat E-E|\le\frac\Delta8\right]\ge\frac9{10}.
\end{equation*}
By~\eqref{eq:gsp-threshold-separation}, comparing $\widehat E$ with $E_{\rm th}$ consequently identifies a ground-state eigenvector as a ground state, and an excited-state eigenvector as excited, with probability at least $9/10$ in either case.

One test has a constant error probability, so we now reduce it by repetition. Let $R=\lceil7r\rceil$ and perform $R$ copies of the phase-estimation test with independent work registers, accepting if a strict majority of the estimates satisfy $\widehat E_j<E_{\rm th}$. On a fixed energy eigenvector $\ket{u_E}$, each test acts on a fresh phase-estimation register while leaving $\ket{u_E}$ unchanged. Consequently, the $R$ outcomes would be independent if the registers were measured. Each outcome is classified correctly with probability at least $9/10$, so an incorrect majority requires at least half the tests to give an incorrect decision. Hoeffding's inequality therefore gives
\begin{equation}\label{eq:verifier-hoeffding}
 \Pr[\text{incorrect majority}]
 \le\exp\!\left[-2R\left(\frac12-\frac1{10}\right)^2\right]
 =\e^{-8R/25}\le\e^{-2r},
\end{equation}
where the last inequality follows from $R\ge7r$ and $56/25>2$. The bound holds for every energy eigenvalue $E$, with no dependence on the dimension of its eigenspace.

For use inside a quantum algorithm, we perform the repeated tests coherently rather than measuring their registers. After the $R$ inverse Fourier transforms, we reversibly compare each stored estimate with $E_{\rm th}$ and compute the majority decision into a fresh qubit $F$. We use $F=1$ for acceptance and $F=0$ for rejection, leaving the work registers unmeasured. Let $\mathcal W_r$ be the Hilbert space of these phase-estimation and comparison registers, and let $V_r^{\rm id}$ be the isometry obtained by starting the work registers in their prescribed initial state and carrying out this ideal circuit.

For any normalized $\ket{u_E}$ in the eigenspace of $H$ with eigenvalue $E$, the preceding controlled evolutions act on the system only through the scalar phases $\e^{\iu tE}$. All other operations act on the work and flag registers. Thus the system vector is unchanged, and the final work-and-flag state depends on $E$ but not on which vector in that eigenspace was supplied. We can write
\begin{equation}\label{eq:verifier-eigenaction}
 V_r^{\rm id}\ket{u_E}=\ket{u_E}\otimes\ket{a_E},
\end{equation}
where $\ket{a_E}\in\mathcal W_r\otimes\mathbb C_F^2$ is normalized. Writing the spectral decomposition of $H$ as $H=\sum_E E\Pi_E$, linearity gives, for every input $\ket u$,
\begin{equation}\label{eq:verifier-spectral-map}
 V_r^{\rm id}\ket u
 =\sum_E(\Pi_E\ket u)\otimes\ket{a_E},
 \qquad\text{or equivalently}\qquad
 V_r^{\rm id}=\sum_E\Pi_E\otimes\ket{a_E}.
\end{equation}
The last expression denotes a map from the system space into the system tensored with the work-and-flag space. In particular, the system component associated with eigenvalue $E$ is still supported on the range of $\Pi_E$. Applying $\Pi_E$ to the output of~\eqref{eq:verifier-spectral-map} and using $\Pi_E\Pi_{E'}=0$ for $E\ne E'$ proves the eigenspace-preservation identity~\eqref{eq:verifier-intertwining}. This reasoning also applies to degenerate eigenspaces, since~\eqref{eq:verifier-eigenaction} holds for every vector within them.

We next convert the incorrect-decision probability in~\eqref{eq:verifier-hoeffding} into the operator-norm estimates of the lemma. If $E=E_0(H)$, a wrong decision means that $F=0$, whereas if $E>E_0(H)$, a wrong decision means that $F=1$. Hence, by~\eqref{eq:verifier-eigenaction} and~\eqref{eq:verifier-hoeffding},
\begin{equation*}
 \norm{P_{\rm rej}V_r^{\rm id}\ket{u_{E_0}}}^2\le\e^{-2r},
 \qquad
 \norm{P_{\rm acc}V_r^{\rm id}\ket{u_E}}^2\le\e^{-2r}
 \quad(E>E_0(H))
\end{equation*}
for every normalized vector in the indicated eigenspace. The quantities on the left are squared norms because the probability of a flag outcome equals the squared norm of the component selected by its projector.

Now let $\ket u$ be any normalized input, not necessarily an eigenvector, and write $\ket u=\sum_E\Pi_E\ket u$. By~\eqref{eq:verifier-spectral-map}, the output terms arising from distinct energies remain orthogonal: although the work states $\ket{a_E}$ need not be orthogonal, their system components lie in orthogonal eigenspaces. Their squared norms therefore add. For the ground-space part and the excited-space part, respectively, we obtain
\begin{align*}
 \norm{P_{\rm rej}V_r^{\rm id}\Pi_0(H)\ket u}^2
 &\le\e^{-2r}\norm{\Pi_0(H)\ket u}^2,\\
 \norm{P_{\rm acc}V_r^{\rm id}(I-\Pi_0(H))\ket u}^2
 &\le\e^{-2r}\sum_{E>E_0(H)}\norm{\Pi_E\ket u}^2
 \le\e^{-2r}\norm{\ket u}^2.
\end{align*}
Taking square roots and maximizing over normalized inputs proves~\eqref{eq:ideal-verifier}. Notice that $\e^{-2r}$ bounds an incorrect-decision probability, while $\e^{-r}$ bounds the norm of the corresponding amplitude component; these are different statements. The same estimates hold for a system entangled with a reference register because the relevant operators are tensored with the identity on that register, which does not change their operator norms.

It remains to implement $V_r^{\rm id}$ using sparse-matrix queries. The matrix-dependent operation in one phase-estimation test is the controlled evolution $\e^{\iu t(H+2I)}$, for $t\in\{0,\ldots,N_{\rm pe}-1\}$. Since $\e^{\iu t(H+2I)}=\e^{2\iu t}\e^{\iu tH}$, the shift $2I$ contributes only the known phase $\e^{2\iu t}$ on the control register, and requires no sparse-matrix queries. We therefore need to implement the controlled evolutions $\e^{\iu tH}$ appearing in the $R$ tests.

Let $\mathcal T=(\mathbb C^{N_{\rm pe}})^{\otimes R}$ be the joint space of their control registers, and write $\ket{\mathbf t}=\ket{t_1,\ldots,t_R}$ for its standard basis. The $R$ controlled evolutions act on the same system register, but on distinct control registers. Since each is a function of $H$, they commute. On the basis state $\ket{\mathbf t}$, their product acts on the system as
\begin{equation*}
 \e^{\iu t_1H}\cdots\e^{\iu t_RH}
 =\exp\!\left(\iu H\sum_{j=1}^R t_j\right).
\end{equation*}
Here we used the fact that $t_1H,\ldots,t_RH$ commute, so the product of their exponentials equals the exponential of their sum. Since the equality holds for every control basis state, the complete controlled operation is
\begin{equation}\label{eq:batched-evolution}
 \sum_{\mathbf t}\ket{\mathbf t}\bra{\mathbf t}\otimes
 \exp\!\left(\iu H\sum_{j=1}^R t_j\right).
\end{equation}
Thus we do not need to simulate the $R$ evolutions separately. Instead, we will express their joint action as time evolution under one larger sparse Hamiltonian.

Choose $T$ to be the smallest power of two with $T\ge R(N_{\rm pe}-1)$. Since $0\le t_j\le N_{\rm pe}-1$, we have $0\le\sum_jt_j\le T$. Furthermore, $R=\Theta(r)$ and $N_{\rm pe}=\Theta(1/\Delta)$, so $T=\Theta(r/\Delta)$. Define the clock-controlled Hamiltonian on $\mathcal T\otimes\mathbb C^D$ by
\begin{equation}\label{eq:clock-H}
 \widehat H=\sum_{\mathbf t}\ket{\mathbf t}\bra{\mathbf t}\otimes c_{\mathbf t}H,
 \qquad c_{\mathbf t}=\frac{\sum_{j=1}^Rt_j}{T}.
\end{equation}
For each fixed $\mathbf t$, the Hamiltonian $\widehat H$ acts as $c_{\mathbf t}H$ on the system. The coefficients satisfy $0\le c_{\mathbf t}\le1$, so the block-diagonal form gives $\norm{\widehat H}=\max_{\mathbf t}c_{\mathbf t}\norm H\le1$. Since distinct clock labels are not coupled and each block is a scalar multiple of $H$, every row and column of $\widehat H$ has at most $s$ nonzero entries. Moreover, its row and column Euclidean norms are at most one: for each standard basis vector $\ket z$, the corresponding column norm is $\norm{\widehat H\ket z}\le\norm{\widehat H}\le1$, and the row bound follows because $\widehat H$ is Hermitian.

We can now see directly why this Hamiltonian generates the required controlled evolution. On $\ket{\mathbf t}$, evolution for time $T$ is $\e^{\iu T c_{\mathbf t}H}$, and $Tc_{\mathbf t}=\sum_jt_j$ by definition. Hence
\begin{equation*}
 \e^{\iu T\widehat H}
 =\sum_{\mathbf t}\ket{\mathbf t}\bra{\mathbf t}\otimes
 \e^{\iu T c_{\mathbf t}H}
 =\sum_{\mathbf t}\ket{\mathbf t}\bra{\mathbf t}\otimes
 \exp\!\left(\iu H\sum_{j=1}^Rt_j\right),
\end{equation*}
which is exactly~\eqref{eq:batched-evolution}. Therefore, implementing the $R$ controlled evolutions reduces to simulating $\widehat H$ once for time $T=\Theta(r/\Delta)$.

Before applying the sparse Hamiltonian simulation theorem, we must show that the sparse-matrix oracles of $\widehat H$ can be implemented using a constant number of queries to those of $H$. For clock indices $\mathbf t,\mathbf t'$ and system indices $i,j$,~\eqref{eq:clock-H} gives
\begin{equation*}
 \bra{\mathbf t,i}\widehat H\ket{\mathbf t',j}
 =\begin{cases}
 c_{\mathbf t}H_{ij},&\mathbf t=\mathbf t',\\
 0,&\mathbf t\ne\mathbf t'.
 \end{cases}
\end{equation*}
The coefficient $c_{\mathbf t}$ is known once the clock index is given, and can be computed reversibly without querying $H$. In fact, it has an exact finite binary representation because $T$ is a power of two and the $t_j$ are integers. A value query to $\widehat H$ can therefore read $H_{ij}$ using the value oracle~\eqref{eq:val}, multiply its encoded value by $c_{\mathbf t}$, write the result into the output register, and uncompute the temporary value using another query to $H$. If the clock labels differ, the entry is zero and no matrix entry is needed. All these operations can be performed coherently on superpositions of indices, using $O(1)$ sparse-matrix queries.

For the in-place location oracles, we use the row and column permutations defined in Section~\ref{sec:sparse-matrix-access}. Let $C=N_{\rm pe}^R$ be the number of clock configurations and enumerate them by $a\in\mathbb Z_C$. Fix a row $(a,i)$ of $\widehat H$. Its nonzero entries, if any, have the same clock label $a$, and their system-column labels belong to the row-support list of $H$ at row $i$. Let $\pi_i:[D]\to[D]$ be the in-place row permutation in~\eqref{eq:row-inplace}. Index the enlarged column labels by $(b,\ell)\in\mathbb Z_C\times[D]$, with the first $s$ input slots represented by $(0,1),\ldots,(0,s)$, and define the permutation
\begin{equation*}
 (b,\ell)\longmapsto
 (b+a\bmod C,\pi_i(\ell)).
\end{equation*}
It is a permutation because addition by $a$ modulo $C$ and $\pi_i$ are both permutations. Its first $s$ outputs are $(a,\pi_i(1)),\ldots,(a,\pi_i(s))$, which contain every nonzero column of the row $(a,i)$. Its inverse subtracts $a$ from the clock label and applies $\pi_i^{-1}$. Hence the in-place row-location oracle of $\widehat H$ and its inverse each need only one corresponding in-place row-location query to $H$, together with reversible arithmetic independent of $H$. The column-location oracle and its inverse are obtained in exactly the same way using the column permutation $\kappa_j$ from~\eqref{eq:column-inplace}. This construction remains valid when $c_{\mathbf t}=0$, because the sparse-access model permits zero-entry locations as padding. We have therefore shown that every sparse-matrix query to $\widehat H$ requires only $O(1)$ queries to $H$.

We are now ready to bound the cost of implementing the verifier. Apply the sparse Hamiltonian simulation theorem of Ref.~\cite[Theorem~2.1]{BRSL2026} to $\widehat H$, with evolution time $-T$ to obtain $\e^{\iu T\widehat H}$ and operator-norm simulation error $\zeta=\e^{-r}$. The theorem applies because $\widehat H$ is Hermitian, is $s$-sparse, and has row and column Euclidean norms at most one. With these parameters, its query bound becomes
\begin{align}\label{eq:verifier-simulation-cost}
 C_r
 &=O\!\left(\sqrt{s}T+\sqrt{sT\log(1/\zeta)}
          +\log(1/\zeta)\right)\notag\\
 &=O\!\left(\frac{\sqrt{s}}\Delta r
          +r\sqrt{\frac{s}{\Delta}}+r\right)
  =O\!\left(\frac{\sqrt{s}}\Delta r\right),
\end{align}
where we used $T=\Theta(r/\Delta)$, $\log(1/\zeta)=r$, $s\ge1$, and $0<\Delta\le1$. The last equality follows because $\sqrt{s/\Delta}\le\sqrt{s}/\Delta$ and $1\le\sqrt{s}/\Delta$. By the oracle constructions above, the same bound counts queries to the original Hamiltonian $H$, not merely to $\widehat H$.

All other operations in the ideal verifier, including the known phase shifts from $2I$, inverse Fourier transforms, comparisons with $E_{\rm th}$, and reversible majority computation, are independent of the sparse-matrix oracles. Let $S$ denote the simulation circuit and let $J_{\rm sim}$ append
its workspace in the zero state. The simulation guarantee is
\[
 \norm{S J_{\rm sim}-J_{\rm sim}\e^{\iu T\widehat H}}\le\zeta .
\]
Include this workspace in $\mathcal W_r$, and append it in the zero
state to the ideal output. Composing the two maps with the same
verifier input isometry and the same subsequent gates gives
$\norm{V_r-V_r^{\rm id}}\le\zeta=\e^{-r}$. Together with~\eqref{eq:ideal-verifier} and~\eqref{eq:verifier-simulation-cost}, this proves the lemma.
\end{proof}

The lemma gives bounds for the ideal verifier and an independent bound on the error of implementing it. Combining them also controls the verifier that we actually use. For example, with $\Pi=\Pi_0(H)$, the triangle inequality gives $\norm{P_{\rm rej}V_r\Pi}\le2\e^{-r}$ and $\norm{P_{\rm acc}V_r(I-\Pi)}\le2\e^{-r}$. Thus an implemented verifier rejects a state supported on the ground space, or accepts a state supported on the excited subspace, with probability at most $4\e^{-2r}$. These conclusions hold for arbitrary states within the respective subspaces, including states entangled with untouched registers. We will use these estimates when analyzing the restart procedure in Section~\ref{sec:restart}.

\subsubsection{A bounded restart procedure}\label{sec:restart}

We now combine the candidate state from~\eqref{eq:candidate} with the verifier of Lemma~\ref{lem:sparse-verifier}, adapting the bounded-restart strategy of Refs.~\cite[Section~2.2]{JSWdW2026} and~\cite[Section~7.2]{Chen2026} to the sparse-access setting. Recall that a fresh candidate has ground-space weight at least $p_0>0$, where $p_0$ is an absolute constant. We will apply a sequence of verifiers to this candidate, with each verifier more accurate than the preceding one. If every test accepts, we return the system register. If a test rejects, we discard the candidate and prepare a new one. We must prove that the returned state has ground-space weight at least $1-\varepsilon$ and that the number of sparse-matrix queries is bounded for every possible sequence of outcomes. The second requirement is important because restarting until acceptance, without a cutoff, gives no deterministic bound on the number of attempts.

Throughout this subsection, assume that the energy estimate satisfies~\eqref{eq:gsp-energy-estimate}. We will account for the probability that this estimate is inaccurate in the next subsection. Write $\Pi=\Pi_0(H)$ and $\Pi_\perp=I-\Pi$ for the ground-space and excited-space projectors. For $r\ge1$, let $V_r$ be the implemented verifier of Lemma~\ref{lem:sparse-verifier}, and define its accepting and rejecting maps by
\begin{equation}\label{eq:verification-branches}
 A_r=P_{\rm acc}V_r,\qquad B_r=P_{\rm rej}V_r.
\end{equation}
For a normalized input $\ket u$, the probabilities of acceptance and rejection are $\norm{A_r\ket u}^2$ and $\norm{B_r\ket u}^2$. Since $V_r$ is an isometry and $P_{\rm acc}+P_{\rm rej}=I$, we have $A_r^\dagger A_r+B_r^\dagger B_r=I$, which implies $\norm{A_r},\norm{B_r}\le1$. We will keep the output vector of each measurement branch unnormalized. Its squared norm then includes the probability of all measurement outcomes leading to that branch. Each verifier uses fresh work registers, and $\Pi$ and $\Pi_\perp$ act on the system register as the identity on all other registers.

We first derive three bounds for the maps in~\eqref{eq:verification-branches}. Set $\eta=\e^{-r}$ and let $V_r^{\rm id}$ be the ideal verifier. Lemma~\ref{lem:sparse-verifier} gives $\norm{V_r-V_r^{\rm id}}\le\eta$. In particular, the definition of the operator norm implies $\norm{(V_r-V_r^{\rm id})\ket u}\le\eta\norm u$ for every input vector $\ket u$, whether or not it is normalized. The lemma also gives
\[
 \norm{P_{\rm rej}V_r^{\rm id}\Pi}\le\eta,
 \qquad
 \norm{P_{\rm acc}V_r^{\rm id}\Pi_\perp}\le\eta.
\]
Finally,~\eqref{eq:verifier-intertwining} states that the ideal verifier preserves the eigenspaces of $H$. Hence it cannot map a ground-space input into the excited space, or an excited-space input into the ground space. We now use these facts to study the two measurement outcomes.

First consider the excited-state component remaining when a verifier accepts. Decompose an arbitrary input as $\ket u=\Pi\ket u+\Pi_\perp\ket u$. Since the ideal verifier preserves the ground space, $\Pi_\perp P_{\rm acc}V_r^{\rm id}\Pi=0$. We may therefore separate the accepting output into the ideal action on the excited component and the implementation error. Taking norms and applying the triangle inequality gives
\begin{align}\label{eq:accepted-leakage}
 \norm{\Pi_\perp A_r\ket u}
 &\le \norm{P_{\rm acc}V_r^{\rm id}\Pi_\perp\ket u}
       +\norm{(V_r-V_r^{\rm id})\ket u}\notag\\
 &\le\eta\norm{\Pi_\perp\ket u}+\eta\norm u
 \le2\eta\norm u.
\end{align}
The first term is bounded by the incorrect-acceptance guarantee of the ideal verifier, the second comes from the implementation error, and the last inequality uses $\norm{\Pi_\perp\ket u}\le\norm u$. Thus an accepting test with parameter $r$ leaves excited-state amplitude at most $2\e^{-r}\norm u$, even if the input is mostly excited.

Next suppose that the input lies in the ground space. For such an input, the ideal verifier rejects with amplitude at most $\eta$. Using the same decomposition of $V_r$ into its ideal part and its implementation error, we obtain
\begin{align}\label{eq:ground-rejection}
 \norm{B_r\Pi\ket u}
 &\le\norm{P_{\rm rej}V_r^{\rm id}\Pi\ket u}
      +\norm{(V_r-V_r^{\rm id})\Pi\ket u}\notag\\
 &\le2\eta\norm{\Pi\ket u}.
\end{align}
In particular, the probability of rejecting a normalized ground-space state is at most $4\e^{-2r}$.

We also need to know how much ground-space amplitude survives when a test accepts. For the ideal verifier, the accepting and rejecting branches are orthogonal, so its isometry property gives
\[
 \norm{P_{\rm acc}V_r^{\rm id}\Pi\ket u}^2
 =\norm{\Pi\ket u}^2
   -\norm{P_{\rm rej}V_r^{\rm id}\Pi\ket u}^2
 \ge(1-\eta^2)\norm{\Pi\ket u}^2.
\]
The ideal accepting output remains in the ground space, and the implemented accepting output differs from the ideal one by a vector of norm at most $\eta\norm u$. Moreover, the ideal verifier cannot transfer the excited part of the input into the ground space. Applying the reverse triangle inequality therefore yields
\begin{align}\label{eq:ground-survival}
 \norm{\Pi A_r\ket u}
 &\ge\sqrt{1-\eta^2}\,\norm{\Pi\ket u}-\eta\norm u\notag\\
 &\ge\norm{\Pi\ket u}-(\eta+\eta^2)\norm u.
\end{align}
Here we used $\sqrt{1-\eta^2}\ge1-\eta^2$ and $\norm{\Pi\ket u}\le\norm u$. Equation~\eqref{eq:ground-survival} shows that each accepting test decreases the ground-space amplitude by at most $(\eta+\eta^2)\norm u$. Together with~\eqref{eq:accepted-leakage} and~\eqref{eq:ground-rejection}, this gives the three estimates needed for the restart analysis. They also hold for inputs entangled with reference registers, by the operator-norm guarantees of Lemma~\ref{lem:sparse-verifier}.

We next choose the verifier accuracies. Let
\begin{equation}\label{eq:verification-schedule}
 r_0=\left\lceil\log\frac8{\sqrt{p_0}}\right\rceil,
 \qquad r_j=2^jr_0,
 \qquad \eta_j=\e^{-r_j}\quad(j\ge0).
\end{equation}
The quantity $r_0$ is an absolute constant because $p_0$ is an absolute constant. The parameter $r_j$ doubles at each stage, while the corresponding error amplitude $\eta_j$ becomes much smaller. To see why this choice is useful, set $a_0=\e^{-r_0}\le\sqrt{p_0}/8\le1/8$. Then $\eta_j=a_0^{2^j}\le a_0^{j+1}$ because $2^j\ge j+1$ and $0<a_0<1$. Since $\eta_j^2\le\eta_j$, summing a geometric series gives
\begin{equation}\label{eq:r0}
 \sum_{j\ge0}(\eta_j+\eta_j^2)
 \le2\sum_{j\ge0}a_0^{j+1}
 =\frac{2a_0}{1-a_0}
 \le\frac{2\sqrt{p_0}}7
 <\frac{\sqrt{p_0}}2.
\end{equation}
Thus, even if a candidate passes arbitrarily many tests, the total possible loss of its ground-space amplitude is less than half of the amplitude $\sqrt{p_0}$ that it had initially.

We now analyze one attempt with verification stages $0,1,\ldots,J$. Let $\ket{u_{-1}}$ be a normalized purification of the candidate state. The extra purification register is not acted on by the verifiers. By~\eqref{eq:candidate}, its ground-space component satisfies $\norm{\Pi\ket{u_{-1}}}^2=\Tr(\Pi\rho_{\rm cand})\ge p_0$. For each stage define
\[
 \ket{u_j}=A_{r_j}\ket{u_{j-1}}\qquad(0\le j\le J).
\]
The vector $\ket{u_j}$ describes the branch on which stages $0$ through $j$ have all accepted. It is not normalized, and its squared norm is the probability of this sequence of acceptances. Since each $A_{r_j}$ is a contraction, $\norm{u_j}\le1$. Applying~\eqref{eq:ground-survival} successively and then using~\eqref{eq:r0}, we obtain
\[
 \begin{aligned}
 \norm{\Pi\ket{u_J}}
 &\ge\norm{\Pi\ket{u_{-1}}}
      -\sum_{j=0}^J(\eta_j+\eta_j^2)\norm{u_{j-1}}\\
 &\ge\sqrt{p_0}-\sum_{j\ge0}(\eta_j+\eta_j^2)
 >\frac{\sqrt{p_0}}2.
 \end{aligned}
\]
The probability that every verifier accepts is $\norm{u_J}^2$. Since $\Pi$ is an orthogonal projector, $\norm{u_J}^2\ge\norm{\Pi\ket{u_J}}^2$, so
\begin{equation}\label{eq:attempt-success}
 \Pr[\text{one attempt accepts}]
 =\norm{u_J}^2
 \ge\norm{\Pi\ket{u_J}}^2
 \ge p,
 \qquad p:=\frac{p_0}{4}>0.
\end{equation}
This lower bound does not depend on $J$. We can therefore make the final verifier very accurate without making acceptance arbitrarily unlikely.

We also bound the excited-state component of an attempt that accepts every stage. Applying~\eqref{eq:accepted-leakage} to the last verifier gives
\begin{equation}\label{eq:attempt-error}
 \norm{\Pi_\perp\ket{u_J}}^2
 =\norm{\Pi_\perp A_{r_J}\ket{u_{J-1}}}^2
 \le4\e^{-2r_J}\norm{u_{J-1}}^2
 \le4\e^{-2r_J}.
\end{equation}
Because $\ket{u_J}$ is unnormalized, the left-hand side is the joint probability that this attempt accepts and its output is found outside the ground space. It already includes the probability of accepting every earlier stage. This is why we will be able to add these contributions over several attempts without dividing by the probability of acceptance.

We first choose how many attempts to allow. Each attempt begins with a fresh candidate and has acceptance probability at least $p$ by~\eqref{eq:attempt-success}. For
\begin{equation}\label{eq:attempt-number}
 M=\left\lceil\frac{\log(4/\varepsilon)}{-\log(1-p)}\right\rceil
 =O\!\left(\log\frac1\varepsilon\right),
\end{equation}
the probability that all $M$ attempts reject is at most $(1-p)^M\le\varepsilon/4$. Next, choose $J$ to be the smallest nonnegative integer for which
\begin{equation}\label{eq:last-stage}
 4M\e^{-2r_J}\le\frac\varepsilon4.
\end{equation}
This is equivalent to $r_J\ge\tfrac12\log(16M/\varepsilon)$. Since $r_j$ doubles, the first $r_J$ meeting this condition satisfies $r_J=O(\log(M/\varepsilon))=O(\log(1/\varepsilon))$. By~\eqref{eq:attempt-error}, the unnormalized excited-state weight of an accepting attempt is at most $4\e^{-2r_J}$. The algorithm returns the first accepting attempt, if there is one. For each possible attempt, the probability that all earlier attempts rejected is at most one. Thus, summing the excited-state contributions over at most $M$ possible accepting attempts gives
\[
 \Pr[\text{output from an accepting attempt lies outside the ground space}]
 \le4M\e^{-2r_J}\le\frac\varepsilon4.
\]
We have now controlled the two kinds of error associated with complete attempts. All attempts reject with probability at most $\varepsilon/4$, and the excited-state weight of accepted outputs is at most $\varepsilon/4$. What remains is to ensure that the number of queries is bounded in the worst case.

A single attempt that reaches stage $J$ uses $O(\sqrt{s}r_J/\Delta)$ verifier queries, because Lemma~\ref{lem:sparse-verifier} gives cost $O(\sqrt{s}r_j/\Delta)$ for stage $j$ and $\sum_{j=0}^Jr_j<2r_J$. If every one of the $M$ attempts paid this full cost, we would obtain a bound proportional to $Mr_J$, which could be of order $\log^2(1/\varepsilon)$. We avoid this extra factor by showing that attempts rejected at expensive stages are very unlikely. This allows us to limit the total verification cost of rejected attempts, while reserving enough additional queries for one attempt that reaches the final stage.

Let $p_j$ be the probability that a fresh attempt accepts stages $0,\ldots,j-1$ and first rejects at stage $j$, for $j\ge1$. The unnormalized state just before this rejection is $\ket{u_{j-1}}$, so $p_j=\norm{B_{r_j}\ket{u_{j-1}}}^2$. Decomposing this state into its ground-space and excited-space components gives
\[
 B_{r_j}\ket{u_{j-1}}
 =B_{r_j}\Pi\ket{u_{j-1}}
  +B_{r_j}\Pi_\perp\ket{u_{j-1}}.
\]
The first term has norm at most $2\e^{-r_j}$ by~\eqref{eq:ground-rejection} and $\norm{u_{j-1}}\le1$. The second term has norm at most $\norm{\Pi_\perp\ket{u_{j-1}}}$ because $\norm{B_{r_j}}\le1$. Since stage $j-1$ accepted,~\eqref{eq:accepted-leakage} bounds this remaining excited-space amplitude by $2\e^{-r_{j-1}}$. Combining the bounds yields
\begin{align}\label{eq:late-rejection}
 \sqrt{p_j}
 &\le2\e^{-r_j}+2\e^{-r_{j-1}}\notag\\
 &\le4\e^{-r_{j-1}},
 \qquad
 p_j\le16\e^{-2r_{j-1}}=16\e^{-r_j}.
\end{align}
Thus the probability of a late rejection decreases exponentially with $r_j$, even though the cost of reaching that stage increases with $r_j$.

We now turn this probability bound into a bound on the total query cost. For one complete attempt, define a nonnegative random variable $Y$ as follows. If the first rejection occurs at stage $j$, let $Y=\sum_{i=0}^jr_i$. If the attempt accepts every stage, let $Y=0$. Thus $Y$ measures the total verifier parameters used by a rejected attempt. The parameter cost of rejecting at stage $j$ satisfies $\sum_{i=0}^jr_i<2r_j$. We use this bound to estimate $\mathbb E[\e^{Y/4}]$, which will control the probability that many rejected attempts have an unusually large total cost.

The rejection events at different stages are disjoint. A stage-zero rejection has probability at most one and gives $Y=r_0$. For $j\ge1$, its probability is $p_j\le16\e^{-r_j}$ by~\eqref{eq:late-rejection}. We therefore obtain
\begin{align}\label{eq:mgf}
 \mathbb E[\e^{Y/4}]
 &\le1+\e^{r_0/2}
      +\sum_{j=1}^J p_j\e^{r_j/2}\notag\\
 &\le1+\e^{r_0/2}
      +16\sum_{j\ge1}\e^{-r_j/2}
 =:K_*<\infty.
\end{align}
The first term accounts for accepting attempts, for which $Y=0$. The remaining series converges because $r_j=2^jr_0$. Since $r_0$ is an absolute constant, $K_*$ is also an absolute constant independent of $J$, $M$, and the problem parameters.

To use~\eqref{eq:mgf} for several attempts, imagine generating $M$ independent complete attempts in advance, even if the actual algorithm would stop after an earlier acceptance. Let $Y_1,\ldots,Y_M$ be their rejected-attempt costs. Conditional on the fixed accurate energy estimate, the candidates and verification outcomes of different attempts are independent. Hence
\[
 \mathbb E\!\left[\exp\!\left(\frac14\sum_{i=1}^M Y_i\right)\right]
 =\prod_{i=1}^M\mathbb E[\e^{Y_i/4}]
 \le K_*^M.
\]
Define the verification budget
\begin{equation}\label{eq:budget}
 B_*=4\left(M\log K_*+\log\frac4\varepsilon\right)
 =O\!\left(\log\frac1\varepsilon\right).
\end{equation}
Applying Markov's inequality to the nonnegative random variable $\exp(\tfrac14\sum_iY_i)$ gives
\begin{align}\label{eq:budget-tail}
 \Pr\!\left[\sum_{i=1}^M Y_i>B_*\right]
 &\le\e^{-B_*/4}
   \mathbb E\!\left[\exp\!\left(\frac14\sum_{i=1}^M Y_i\right)\right]\notag\\
 &\le\e^{-B_*/4}K_*^M
 =\frac\varepsilon4.
\end{align}
Thus, except with probability at most $\varepsilon/4$, the total verifier parameter cost of rejected attempts is at most $B_*$. This is stronger than an expected-cost bound because it gives a specific budget that fails with a controlled probability.

We can now impose a deterministic cutoff. During the algorithm, maintain a counter equal to the sum of the parameters $r_j$ of all verification stages executed so far. Before running a new stage, check whether its parameter would make the counter exceed $B_*+2r_J$. If it would, stop and output a fixed normalized system state. We also return this fallback state if all $M$ attempts reject. To bound the chance of using the fallback, suppose that one of the first $M$ complete attempts accepts and that $\sum_{i=1}^M Y_i\le B_*$. Every attempt before the first acceptance is rejected, so these earlier attempts cost at most $B_*$ in total. The first accepting attempt uses less than $2r_J$. Therefore the counter never exceeds $B_*+2r_J$ before this acceptance, and the algorithm returns the accepting output rather than the fallback. It follows from~\eqref{eq:attempt-number} and~\eqref{eq:budget-tail} that
\[
 \Pr[\text{fallback}]
 \le\Pr[\text{all $M$ attempts reject}]
     +\Pr\!\left[\sum_{i=1}^M Y_i>B_*\right]
 \le\frac\varepsilon2.
\]
The fallback state need not be a ground state. Its total contribution to the excited-state weight is nevertheless at most the fallback probability, which is at most $\varepsilon/2$.

Finally, we count sparse-matrix queries on every possible execution. The algorithm prepares at most $M$ candidates, each at cost $C_{\rm cand}=O(\sqrt{s}/(\gamma\Delta))$ by~\eqref{eq:candidate}. By the cutoff, the sum of the verifier parameters is at most $B_*+2r_J$, and a verifier with parameter $r$ costs $O(\sqrt{s}r/\Delta)$ sparse-matrix queries by~\eqref{eq:verifier-cost}. Consequently,
\begin{align}\label{eq:restart-cost}
 C_{\rm restart}
 &=O\!\left(MC_{\rm cand}
        +\frac{\sqrt{s}}\Delta(B_*+2r_J)\right)\notag\\
 &=O\!\left(\frac{\sqrt{s}}{\gamma\Delta}
                  \log\frac1\varepsilon\right),
\end{align}
where the second line uses $M,B_*,r_J=O(\log(1/\varepsilon))$ and $\gamma\le1$. This is a worst-case bound because the cutoff prevents any branch from exceeding the stated verifier budget. Conditional on~\eqref{eq:gsp-energy-estimate}, accepted outputs contribute excited-state weight at most $\varepsilon/4$, and fallback occurs with probability at most $\varepsilon/2$. In the next subsection we account for the possible failure of the initial energy estimate and establish the unconditional GSP guarantee~\eqref{eq:gsp-error}.

\subsection{Proof of Theorem~\ref{thm:upper}}\label{sec:upper-completion}

We have already proved the GSEE upper bound in~\eqref{eq:gsee-upper}. For GSP, the candidate-preparation and restart arguments were carried out under the assumption that an estimate $\widetilde E_0$ satisfies~\eqref{eq:gsp-energy-estimate}. We now obtain this estimate using the GSEE algorithm and account for the possibility that it is inaccurate. We will also prove the trace-distance guarantee in Theorem~\ref{thm:upper}. Throughout the proof, let $\Pi=\Pi_0(H)$ and $\Pi_\perp=I-\Pi$.

\begin{proof}
First, run the GSEE algorithm of Section~\ref{sec:gsee-upper} with energy error $\delta=\Delta/32$ and failure probability $\varepsilon/4$. Let $\widetilde E_0$ be its output, and define the event
\[
 \mathcal E=\left\{|\widetilde E_0-E_0(H)|\le\frac\Delta{32}\right\}.
\]
By~\eqref{eq:gsee-upper}, this step satisfies $\Pr(\mathcal E^c)\le\varepsilon/4$ and uses
\[
 O\!\left(\frac{\sqrt{s}}{\gamma(\Delta/32)}
               \log\frac4\varepsilon\right)
 =O\!\left(\frac{\sqrt{s}}{\gamma\Delta}
               \log\frac1\varepsilon\right)
\]
sparse-matrix queries. We then use the reported value $\widetilde E_0$ to prepare candidates and run the bounded restart procedure of Section~\ref{sec:restart}. The algorithm does not know whether $\mathcal E$ occurred, so it follows the same procedure for every reported estimate.

We first analyze the output conditional on an accurate estimate. Fix any value $\widetilde E_0$ satisfying $\mathcal E$. The candidate guarantee~\eqref{eq:candidate} and the verifier guarantees of Lemma~\ref{lem:sparse-verifier} then apply, so the restart analysis of Section~\ref{sec:restart} holds for this fixed value. There are two ways in which the restart procedure can terminate. It either returns the system register of an attempt whose verifiers all accepted, or it returns the fixed fallback state because the attempt limit or verification budget was reached.

Let $\rho_{\rm acc}$ and $\rho_{\rm fb}$ be the corresponding unnormalized output density operators, averaged over the candidate-preparation and verification outcomes for this fixed estimate. Their traces are the probabilities of the two outcomes, and $\rho_{\rm acc}+\rho_{\rm fb}$ is a normalized density operator. We use unnormalized operators because their traces already include the probabilities of the branches they describe.

For an attempt that accepts every verifier,~\eqref{eq:attempt-error} bounds its unnormalized excited-state weight by $4\e^{-2r_J}$. The algorithm can return an accepting output from at most one of its first $M$ attempts. Summing over the possible accepting attempts and using the choice of $r_J$ in~\eqref{eq:last-stage}, we obtain
\[
 \Tr(\Pi_\perp\rho_{\rm acc})
 \le4M\e^{-2r_J}\le\frac\varepsilon4.
\]
On the other hand,~\eqref{eq:attempt-number} bounds the probability that all $M$ attempts reject by $\varepsilon/4$, and~\eqref{eq:budget-tail} bounds the probability that the rejected attempts exceed the verification budget by another $\varepsilon/4$. The cutoff argument in Section~\ref{sec:restart} therefore gives $\Tr\rho_{\rm fb}\le\varepsilon/2$. Since $0\preceq\Pi_\perp\preceq I$, the excited-state weight of the fallback branches is at most their total probability, namely
\[
 \Tr(\Pi_\perp\rho_{\rm fb})
 \le\Tr\rho_{\rm fb}\le\frac\varepsilon2.
\]
Combining the two bounds shows that, for every fixed estimate satisfying $\mathcal E$, the complete restart output has excited-state weight at most $3\varepsilon/4$.

We now remove the conditioning on $\mathcal E$. Let $\rho$ be the final state averaged over all outcomes, including the initial energy estimate, and let $\rho_{\rm good}$ and $\rho_{\rm bad}$ be its normalized conditional states on $\mathcal E$ and $\mathcal E^c$, respectively. If $\Pr(\mathcal E^c)=0$, the second state can be chosen arbitrarily. By the preceding calculation, $\Tr(\Pi_\perp\rho_{\rm good})\le3\varepsilon/4$. We do not need any guarantee on the state produced when the energy estimate is inaccurate, because every density operator satisfies $\Tr(\Pi_\perp\rho_{\rm bad})\le1$. Therefore,
\begin{align}
 \Tr(\Pi_\perp\rho)
 &=\Pr(\mathcal E)\Tr(\Pi_\perp\rho_{\rm good})
   +\Pr(\mathcal E^c)\Tr(\Pi_\perp\rho_{\rm bad})\notag\\
 &\le\frac{3\varepsilon}{4}+\Pr(\mathcal E^c)
 \le\frac{3\varepsilon}{4}+\frac\varepsilon4
 =\varepsilon.
\end{align}
Since $\Pi+\Pi_\perp=I$ and $\Tr\rho=1$, this implies $\Tr(\Pi\rho)=1-\Tr(\Pi_\perp\rho)\ge1-\varepsilon$, which is the unconditional GSP guarantee in~\eqref{eq:gsp-error}.

It remains to count the sparse-matrix queries. We bounded the cost of the initial energy estimate above. By~\eqref{eq:restart-cost}, the restart procedure has the same asymptotic worst-case cost, because the number of candidates is bounded and the verifier counter imposes a deterministic cutoff. These bounds hold even if the energy estimate is inaccurate. Adding the two costs gives
\[
 Q_{\rm sparse}^{\GSP}
 =O\!\left(\frac{\sqrt{s}}{\gamma\Delta}
                  \log\frac1\varepsilon\right),
\]
proving~\eqref{eq:gsp-upper-main}.

We finally prove the trace-distance statement in Theorem~\ref{thm:upper}. Run the algorithm just established with weight-error parameter $\varepsilon^2$ in place of $\varepsilon$, and denote its output again by $\rho$. It satisfies $p:=\Tr(\Pi\rho)\ge1-\varepsilon^2>0$. We will construct a state $\sigma$ supported on the ground space and show that $\rho$ is within trace distance $\varepsilon$ of $\sigma$.

Choose a normalized purification $\ket\Psi$ of $\rho$ on the system and a reference register $R$. The squared norm of its projection onto the ground space is
\[
 \norm{(\Pi\otimes I_R)\ket\Psi}^2
 =\bra\Psi(\Pi\otimes I_R)\ket\Psi
 =\Tr(\Pi\rho)=p.
\]
We can therefore normalize this projection and define
\[
 \ket\Phi=\frac{(\Pi\otimes I_R)\ket\Psi}{\sqrt p},
 \qquad \sigma=\Tr_R(\ket\Phi\bra\Phi).
\]
Because $(\Pi\otimes I_R)\ket\Phi=\ket\Phi$, the reduced state satisfies $\Pi\sigma\Pi=\sigma$, so it is supported entirely on the ground space. Moreover, the definition of $\ket\Phi$ gives
\[
 \braket\Psi\Phi
 =\frac{\bra\Psi(\Pi\otimes I_R)\ket\Psi}{\sqrt p}
 =\sqrt p.
\]
By~\eqref{eq:pure-trace-distance}, the trace distance between the two pure states is $\sqrt{1-p}$. Taking the partial trace over $R$ cannot increase trace distance, and hence
\begin{equation}
 \frac12\norm{\rho-\sigma}_1
 \le\frac12\norm{\ket\Psi\bra\Psi-\ket\Phi\bra\Phi}_1
 =\sqrt{1-p}\le\varepsilon.
\end{equation}
This proves the trace-distance guarantee. Replacing $\varepsilon$ by $\varepsilon^2$ changes the logarithmic factor in the query complexity from $\log(1/\varepsilon)$ to $\log(1/\varepsilon^2)=2\log(1/\varepsilon)$, so the asymptotic sparse-query bound remains unchanged. This completes the proof of Theorem~\ref{thm:upper}.
\end{proof}

Combining Theorem~\ref{thm:upper} with the lower bounds of Theorems~\ref{thm:gsee} and~\ref{thm:gsp}, we obtain, in their common parameter ranges and for sufficiently large dimension, the matching query complexities
\begin{equation}
 Q_{\rm sparse}^{\GSEE}=\Theta\!\left(\frac{\sqrt{s}}{\gamma\delta}
                         \log\frac1\varepsilon\right),
 \qquad
 Q_{\rm sparse}^{\GSP}=\Theta\!\left(\frac{\sqrt{s}}{\gamma\Delta}
                        \log\frac1\varepsilon\right).
\end{equation}
These statements apply in the reversible in-place sparse-location model defined in Section~\ref{sec:sparse-matrix-access}. Indeed, the upper bounds use that model, while the lower bounds remain valid even when the stronger in-place location oracles are supplied. The upper bounds allow degenerate ground spaces, and the GSP guarantee holds for the unconditional output state, including all fallback branches. The same GSP query bound also suffices to achieve trace-distance error at most $\varepsilon$ from a state supported on the ground space.

\section{Discussion}\label{sec:discussion}

We have established matching lower and upper bounds for guided ground-state energy estimation and ground-state preparation in the reversible in-place sparse-location model. 
These bounds determine the optimal joint
dependence of the sparse-query complexity on the Hamiltonian sparsity $s$, guiding-state overlap $\gamma$, energy-estimation error $\delta$ or spectral gap $\Delta$, and failure probability $\varepsilon$, up to constant factors. In particular, the factor $\sqrt{s}$ is both necessary and sufficient, and the lower bounds hold even for stoquastic Hamiltonians. 
Thus, our results settle the optimal joint sparse-query complexity of both problems in this model.

One remaining question is whether the same lower bounds hold when the ground state is promised to be unique. Our lower-bound construction allows degenerate ground spaces, whereas the matching upper bounds do not require uniqueness. In particular, establishing the full $\log(1/\varepsilon)$ lower-bound dependence under a unique-ground-state promise would strengthen our results.

Another direction is to establish the same upper bounds in the standard sparse-access model. Our lower bounds already hold in this model, but the upper bounds use reversible in-place row and column location oracles and their inverses. It remains to determine whether the same query complexities can be achieved
without this additional access.

Finally, our lower-bound Hamiltonians have exact finite binary entries, but their encoding lengths are not necessarily bounded by $O(\log D)$. They are also not required to be geometrically local or to decompose into terms acting on a fixed number of qubits. Establishing comparable lower bounds under these additional restrictions would extend our results to more constrained classes of Hamiltonians.

\section*{Use of AI tools}

The author developed the main algorithmic ideas and provided detailed proof sketches. In particular, the lower-bound constructions for GSEE and GSP were inspired by the earlier work of the author on quantum linear systems in Ref.~\cite{Patel2026}. ChatGPT (Sol 5.6) assisted in developing these sketches into detailed mathematical proofs and in writing and revising the manuscript, while ChatGPT (Astra) was used for technical audits. The manuscript underwent multiple rounds of proofreading and technical checks with AI assistance, including the identification and correction of several technical errors. Throughout this process, the author independently reviewed and checked the proofs multiple times. Responsibility for the final manuscript and its claims rests with the author.

\appendix

\section{Proofs for the Boolean Query Lower Bound}\label{app:boolean}

We first prove the two Boolean query lower bounds used in Section~\ref{sec:boolean}. The first establishes the adversary bound for a single instance of \textsc{Position-Parity}, while the second proves that the outer promise in~\eqref{eq:outer} requires a number of queries proportional to $k\sqrt L$ when the allowed error is exponentially small in $k$.

\begin{proof}[Proof of Lemma~\ref{lem:parity-adversary}]
Let $L$ be even, and partition the domain $\mathcal S_L$ into $\mathcal E=\{e_q:q\text{ even}\}$ and $\mathcal O=\{e_q:q\text{ odd}\}$. Each set has $L/2$ elements. Define the real symmetric matrix $\Gamma$, indexed by $\mathcal S_L$, by $\Gamma_{xy}=1$ if one of $x,y$ belongs to $\mathcal E$ and the other to $\mathcal O$, and $\Gamma_{xy}=0$ otherwise. Since $g_L$ takes different values on the two sets, $\Gamma$ satisfies the constraint in~\eqref{eq:adversary}.

Ordering the inputs in $\mathcal E$ before those in $\mathcal O$, we have $\Gamma=\bigl(\begin{smallmatrix}0&J\\J&0\end{smallmatrix}\bigr)$, where $J$ is the $(L/2)\times(L/2)$ all-ones matrix. To verify the norm, set $n=L/2$ and let $e=n^{-1/2}(1,\ldots,1)^T$. Then $J=nee^T$, so $Je=ne$ and $Jv=0$ whenever $e^Tv=0$. Consequently $\Gamma(e\oplus e)=n(e\oplus e)$ and $\Gamma(e\oplus(-e))=-n(e\oplus(-e))$, while $\Gamma$ vanishes on vectors whose two components are orthogonal to $e$. Its only nonzero eigenvalues are $n$ and $-n$, and so, $\norm\Gamma=L/2$.

Fix a bit position $q\in\mathbb Z_L$. Two distinct bit strings $e_p,e_{p'}\in\mathcal S_L$ differ at position $q$ exactly when $p=q$ or $p'=q$. Consequently, $\Gamma\circ D_q$ has nonzero entries only between $e_q$ and the $L/2$ bit strings whose marked positions have parity opposite to $q$. It is therefore the adjacency matrix of a star graph with $L/2$ leaves, together with isolated vertices. Let $\ket c$ be its central vertex and $\ket\ell$ the normalized uniform vector over its $L/2$ leaves. The star graph maps $\ket c$ to $\sqrt{L/2}\ket\ell$ and $\ket\ell$ to $\sqrt{L/2}\ket c$, and vanishes on the remaining orthogonal subspace. Its two nonzero eigenvalues are therefore $\pm\sqrt{L/2}$, giving $\norm{\Gamma\circ D_q}=\sqrt{L/2}$ for every $q$. Substituting into~\eqref{eq:adversary} yields
\begin{equation}
  \Adv^\pm(g_L)\ge
  \frac{L/2}{\sqrt{L/2}}=\sqrt{L/2},
\end{equation}
as claimed.
\end{proof}

\bigskip

We next prove Lemma~\ref{lem:outer}.

\begin{proof}[Proof of Lemma~\ref{lem:outer}]
Let $t=\lceil k/2\rceil$ and $r=k-t+1$, so that $r\ge k/2$. Recall that the promise in~\eqref{eq:outer} restricts the Hamming weight $|z|$ to either $0$ or one of the integers $t,t+1,\ldots,k$.

We first construct coefficients $\lambda_t,\ldots,\lambda_k$ such that, for every polynomial $p$ of degree less than $r$, its value $p(0)$ can be expressed as a linear combination of its values $p(t),\ldots,p(k)$. This is possible because a polynomial of degree less than $r$ is uniquely determined by its values at $r$ distinct points. We will use these coefficients to construct a function on the promised Boolean inputs that is orthogonal to all multilinear polynomials of degree less than $r$.

For every integer $j\in\{t,\ldots,k\}$, define
\begin{equation}\label{eq:app-interpolation}
  \lambda_j=\prod_{\substack{h=t\\h\ne j}}^k
  \frac{-h}{j-h},
  \qquad
  Z=1+\sum_{j=t}^k|\lambda_j|.
\end{equation}
Here $Z$ is a normalization factor that we will use when defining this function. We now verify that these coefficients satisfy
\begin{equation}\label{eq:app-lagrange}
  p(0)=\sum_{j=t}^k\lambda_j p(j)
\end{equation}
for every polynomial $p$ of degree less than $r$.
Define $L_j(x)=\prod_{h=t,h\ne j}^k(x-h)/(j-h)$. This polynomial has degree $r-1$, with $L_j(j)=1$ and $L_j(h)=0$ at all other nodes. For any polynomial $p$ of degree less than $r$, the difference $p(x)-\sum_{j=t}^k p(j)L_j(x)$ has degree less than $r$ and vanishes at all $r$ nodes. It is therefore zero. Evaluating at $x=0$, and observing that $L_j(0)=\lambda_j$, gives~\eqref{eq:app-lagrange}.

To simplify the expression for $\lambda_j$ in~\eqref{eq:app-interpolation}, observe that the product in its numerator is $(-1)^{k-t}k!/((t-1)!j)$, while the product in its denominator is $(j-t)!(-1)^{k-j}(k-j)!$. Dividing these products gives
\begin{equation}
  \lambda_j=(-1)^{j-t}\frac{t}{j}
  \binom{k}{t}\binom{k-t}{j-t}.
\end{equation}
Since $t/j\le1$ and $\binom{k}{t}\le2^k$, summing their absolute values gives
\begin{equation}\label{eq:app-Z-bound}
  Z\le1+\binom{k}{t}2^{k-t}
   \le1+2^{2k-t}\le2^{2k}.
\end{equation}
The last inequality holds for all $k\ge1$.

Now, define a real function $\phi$ on $\{0,1\}^k$ by
\begin{equation}\label{eq:app-phi}
  \phi(z)=
  \begin{cases}
    Z^{-1},&z=0^k,\\
    -\lambda_{|z|}/\bigl(Z\binom{k}{|z|}\bigr),
      &|z|\ge t,\\
    0,&\text{otherwise}.
  \end{cases}
\end{equation}
In sums over all Boolean strings below, we extend $F_k$ by setting $F_k(z)=0$ outside its promise. 
This does not affect these sums because $\phi$ vanishes there.
Now, by construction, we have that $\sum_z|\phi(z)|=1$, and because $F_k(0^k)=1$ and $F_k(z)=0$ for $|z|\ge t$, we have $\sum_z\phi(z)F_k(z)=1/Z$.

We now prove that $\phi$ is orthogonal to every multilinear polynomial of degree less than $r$. Every such polynomial is a linear combination of monomials, so it is enough to prove that
\begin{equation}
    \sum_z \phi(z)p(z)=0
\end{equation}
for a monomial $p(z)=\prod_{i\in T}z_i$, where $T\subseteq[k]$ and $d:=|T|<r$.

For each integer $j\in\{0,\ldots,k\}$, let $\bar p(j)$ denote the average value of $p(z)$ over all Boolean strings of Hamming weight $j$. Thus,
\begin{equation}
    \bar p(j)=\frac{1}{\binom{k}{j}}
    \sum_{\substack{z\in\{0,1\}^k\\|z|=j}}p(z).
\end{equation}
To compute this average, observe that $p(z)=1$ precisely when $z_i=1$ for every $i\in T$. Since $T$ contains $d$ positions, a string of Hamming weight $j\ge d$ satisfies this condition if its remaining $j-d$ ones are chosen from the other $k-d$ positions. There are $\binom{k-d}{j-d}$ such strings, out of $\binom{k}{j}$ strings of Hamming weight $j$. Therefore,
\begin{equation}\label{eq:layer-average}
    \bar p(j)
    =\frac{\binom{k-d}{j-d}}{\binom{k}{j}}
    =\frac{j(j-1)\cdots(j-d+1)}
    {k(k-1)\cdots(k-d+1)}.
\end{equation}
For $j<d$, no string of Hamming weight $j$ can have ones at all $d$ positions in $T$, so $\bar p(j)=0$. The expression on the right-hand side of~\eqref{eq:layer-average} also equals zero in this case. When $d=0$, both sides equal one, with the empty product understood to be one. Hence~\eqref{eq:layer-average} holds for every $j\in\{0,\ldots,k\}$.

The important observation is that $\bar p(j)$ is a polynomial in $j$ of degree $d<r$. We can therefore apply the interpolation identity~\eqref{eq:app-lagrange} to obtain
\begin{equation}
    \bar p(0)=\sum_{j=t}^k\lambda_j\bar p(j).
\end{equation}
On the other hand, by the definition of $\phi$, we have
\begin{align}
    \sum_z\phi(z)p(z)
    &=\frac1Z\left(
        p(0^k)-\sum_{j=t}^k\lambda_j\bar p(j)
      \right)\\
    &=\frac1Z\left(
        \bar p(0)-\sum_{j=t}^k\lambda_j\bar p(j)
      \right)\\
    &=0,
\end{align}
where the second equality uses $p(0^k)=\bar p(0)$, and the last equality follows from the interpolation identity.
We have thus proved the claim for every monomial of degree less than $r$. Since every multilinear polynomial of degree less than $r$ is a linear combination of these monomials, the claim follows by linearity.

Now suppose that an algorithm $\mathcal A$ makes at most $Q$ Boolean queries and computes $F_k\circ g_L^{(k)}$ with error probability at most $\eta\le2^{-2k-1}$ on every promised input.

Our first goal is to use this algorithm to construct a function $P:\{0,1\}^k\to[0,1]$ that approximates $F_k$ on its promise. We will then show that the values of $P$ contain information about the parity of some subset of at least $k/2$ of the bits $z_1,\ldots,z_k$. This will allow us to construct an algorithm that predicts the XOR of the answers to at least $k/2$ \textsc{Position-Parity} instances, to which we can apply Lemma~\ref{lem:xor}.

Recall that each Boolean instance is a singleton string $x^{(i)}\in\mathcal S_L$, and its answer is $z_i=g_L(x^{(i)})$. For a fixed $z\in\{0,1\}^k$, independently choose each $x^{(i)}$ uniformly from the singleton strings satisfying $g_L(x^{(i)})=z_i$. In other words, the marked position is chosen uniformly among the even positions when $z_i=0$, and uniformly among the odd positions when $z_i=1$.

Define $P(z)$ to be the probability that $\mathcal A$ outputs $1$, averaged over these choices of singleton inputs and over the internal randomness and measurements of $\mathcal A$.

The function $P$ is defined for every $z\in\{0,1\}^k$, including those outside the promise of $F_k$. To ensure this, we return an arbitrary bit if the algorithm attempts to make more than $Q$ queries. Otherwise, we allow it to complete its computation and return its output. This does not change the algorithm on promised inputs, where its query count is already at most $Q$, but ensures that the modified algorithm is defined on every tuple of singleton inputs.

Whenever $z$ satisfies the promise of $F_k$, the assumed error guarantee implies
\begin{equation}
    |P(z)-F_k(z)|\le\eta.
\end{equation}
Indeed, if $z=0^k$, then $F_k(z)=1$, so the algorithm outputs $1$ with probability at least $1-\eta$. If $|z|\ge t$, then $F_k(z)=0$, so it outputs $1$ with probability at most $\eta$.

We next use the function $\phi$ constructed above in~\eqref{eq:app-phi}. Recall that $\phi$ is supported only on strings satisfying the promise of $F_k$, has $\ell_1$ norm one, and satisfies $\phi(0^k)=1/Z$. Since $F_k(z)=1$ only at $z=0^k$ within its promise, we have
\begin{equation}
    \sum_z\phi(z)F_k(z)=\phi(0^k)=\frac1Z.
\end{equation}
Furthermore, because $|P(z)-F_k(z)|\le\eta$ on the support of $\phi$,
\begin{align}
    \left|\sum_z\phi(z)(P(z)-F_k(z))\right|
    &\le\sum_z|\phi(z)|\,|P(z)-F_k(z)|\\
    &\le\eta\sum_z|\phi(z)|\\
    &=\eta.
\end{align}
Combining these two observations gives
\begin{equation}\label{eq:app-correlation}
    \sum_z\phi(z)P(z)
    \ge\frac1Z-\eta
    \ge\frac1{2Z}.
\end{equation}
For the last inequality, recall from~\eqref{eq:app-Z-bound} that $Z\le2^{2k}$. Since $\eta\le2^{-2k-1}$, we have $\eta\le1/(2Z)$.

We now use~\eqref{eq:app-correlation} together with the orthogonality property of $\phi$ to show that $P$ is correlated with the parity of at least $k/2$ of its input bits.

To make this precise, consider a subset $S\subseteq[k]$. The XOR of the bits indexed by $S$ is $\bigoplus_{i\in S}z_i$. Define
\begin{equation}
    \chi_S(z)=(-1)^{\sum_{i\in S}z_i},
\end{equation}
so that $\chi_S(z)=1$ when this XOR is zero and $\chi_S(z)=-1$ when it is one.

We want to determine whether the output of the algorithm is correlated with $\chi_S(z)$ when $z$ is chosen uniformly from $\{0,1\}^k$. Recall that $P(z)$ is the probability that the algorithm outputs $1$. We therefore consider
\begin{equation}
    \widehat P(S)
    =2^{-k}\sum_{z\in\{0,1\}^k}P(z)\chi_S(z).
\end{equation}
This quantity is called the Fourier coefficient of $P$ corresponding to $S$.

For a nonempty set $S$, a nonzero value of $\widehat P(S)$ means that the output probability of the algorithm is correlated with the parity of the bits indexed by $S$. We will show that there exists a subset $S$ containing at least $k/2$ coordinates for which $|\widehat P(S)|$ is sufficiently large. We will then use this correlation to construct an algorithm that predicts the corresponding XOR with success probability greater than $1/2$.

We next show that the functions $\chi_S$, for $S\subseteq[k]$, form an orthonormal basis for the real-valued functions on $\{0,1\}^k$. For any two subsets $S,T\subseteq[k]$, we have
\begin{equation}
    2^{-k}\sum_{z\in\{0,1\}^k}\chi_S(z)\chi_T(z)
    =
    \begin{cases}
        1,&S=T,\\
        0,&S\ne T.
    \end{cases}
\end{equation}
Indeed, when $S=T$, the product equals one for every $z$. When $S\ne T$, there is some coordinate belonging to exactly one of the two sets. Pairing each string $z$ with the string obtained by flipping this coordinate gives two terms with opposite signs, so their contributions cancel.

There are $2^k$ subsets of $[k]$, and hence $2^k$ such orthonormal functions. Since the space of real-valued functions on $\{0,1\}^k$ has dimension $2^k$, they form an orthonormal basis. Consequently, we can express $P$ as
\begin{equation}
    P(z)=\sum_{S\subseteq[k]}\widehat P(S)\chi_S(z).
\end{equation}

We now substitute this expansion into~\eqref{eq:app-correlation} and use the orthogonality property of $\phi$ to show that at least one coefficient $\widehat P(S)$ with $|S|\ge r$ must be nonzero.
Interchanging the two sums gives
\begin{equation}
    \sum_z\phi(z)P(z)
    =\sum_{S\subseteq[k]}\widehat P(S)
      \sum_z\phi(z)\chi_S(z).
\end{equation}
For convenience, define
\begin{equation}
    a_S=\sum_z\phi(z)\chi_S(z).
\end{equation}

Recall that, for Boolean variables $z_i\in\{0,1\}$, we have $(-1)^{z_i}=1-2z_i$. Therefore,
\begin{equation}
    \chi_S(z)=\prod_{i\in S}(1-2z_i),
\end{equation}
which is a multilinear polynomial of degree $|S|$.

We proved above that $\phi$ is orthogonal to every multilinear polynomial of degree less than $r$. Applying that result to $\chi_S$ gives $a_S=0$ whenever $|S|<r$.
For every other subset $S$, we have the bound
\begin{equation}
    |a_S|
    \le\sum_z|\phi(z)|\,|\chi_S(z)|
    =\sum_z|\phi(z)|=1,
\end{equation}
where we used $|\chi_S(z)|=1$.

Thus, all Fourier coefficients corresponding to fewer than $r$ coordinates disappear from the correlation, and the remaining coefficients satisfy
\begin{align}
    \frac1{2Z}
    &\le\left|\sum_z\phi(z)P(z)\right|\\
    &=\left|\sum_{\substack{S\subseteq[k]\\|S|\ge r}}
             a_S\widehat P(S)\right|\\
    &\le\sum_{\substack{S\subseteq[k]\\|S|\ge r}}
             |\widehat P(S)|.
\end{align}

There are at most $2^k$ subsets of $[k]$. Hence, if every coefficient in the final sum had absolute value less than $1/(2^{k+1}Z)$, that sum would be less than $1/(2Z)$, contradicting the inequality above.
Therefore, there exists a subset $S\subseteq[k]$ with $\ell:=|S|\ge r\ge k/2$ such that
\begin{equation}\label{eq:app-fourier-bound}
    |\widehat P(S)|
    \ge\frac1{2^{k+1}Z}
    \ge2^{-3k-1}
    =:\alpha,
\end{equation}
where the second inequality again uses $Z\le2^{2k}$.

We have thus identified a set $S$ containing at least half of the $k$ coordinates for which the corresponding Fourier coefficient of $P$ has absolute value at least $\alpha$. We next show how this coefficient can be used to predict the XOR of the \textsc{Position-Parity} answers indexed by $S$.
Fix the subset $S$ satisfying~\eqref{eq:app-fourier-bound}, and consider arbitrary singleton inputs $x^{(i)}$, for $i\in S$. We want to compute
\begin{equation}
    G=\bigoplus_{i\in S}g_L(x^{(i)}).
\end{equation}
Our goal is to construct an algorithm for $G$ using the assumed algorithm $\mathcal A$ as a subroutine.

The difficulty is that $\mathcal A$ was used to define $P(z)$ by averaging over uniformly random singleton inputs of each prescribed parity, whereas the inputs $x^{(i)}$ to our XOR algorithm are arbitrary. We therefore randomize their marked positions without losing the information about their parities.

For every $i\in S$, independently choose a uniform random bit $b_i\in\{0,1\}$. Then choose a uniformly random permutation $\pi_i$ of $\mathbb Z_L$ with the following property: if $b_i=0$, the permutation preserves the parity of every position, and if $b_i=1$, the permutation interchanges the even and odd positions.

Apply $\pi_i$ to the marked position of $x^{(i)}$, obtaining another singleton input $y^{(i)}$. Because $\pi_i$ either preserves or reverses parity according to $b_i$, we have
\begin{equation}
    g_L(y^{(i)})=g_L(x^{(i)})\oplus b_i.
\end{equation}
Moreover, the permutation is chosen uniformly from all permutations with the specified parity behavior. Consequently, conditioned on the value of $g_L(y^{(i)})$, the marked position of $y^{(i)}$ is uniformly distributed over all positions of that parity.

For every $i\notin S$, independently choose $y^{(i)}$ to be a uniformly random singleton input, without using any of the original inputs. Write
\begin{equation}
    Y=(y^{(1)},\ldots,y^{(k)}),
    \qquad z_i'=g_L(y^{(i)}).
\end{equation}

We now check the distribution of $Y$. For $i\in S$, the bit $z_i'=g_L(x^{(i)})\oplus b_i$ is uniform because $b_i$ is uniform, regardless of the original marked position. For $i\notin S$, the bit $z_i'$ is also uniform because $y^{(i)}$ is a uniformly random singleton string and $L$ is even. Since all choices are independent, $z'$ is uniform on $\{0,1\}^k$.

Furthermore, conditioned on any fixed value of $z'$, the strings $y^{(i)}$ are independent and uniformly distributed among singleton inputs with parity $z_i'$. This is exactly the distribution used in the definition of $P(z')$.

We can also simulate Boolean queries to $Y$ using queries to the original inputs indexed by $S$. For $i\in S$, the relation between the two strings is
\begin{equation}
    y_q^{(i)}=x_{\pi_i^{-1}(q)}^{(i)}.
\end{equation}
Since $\pi_i$ is known, a query to $y_q^{(i)}$ can be simulated by computing $\pi_i^{-1}(q)$ reversibly and querying the corresponding bit of $x^{(i)}$. A second query may be used to uncompute temporary registers. For $i\notin S$, the string $y^{(i)}$ was sampled by the algorithm and is therefore known, so its queries require no access to the original inputs. These operations can be performed coherently when the query indices are in superposition. Thus, each query to $Y$ requires at most two queries to the original tuple.

Run $\mathcal A$ on the randomized input $Y$, using at most $2Q$ queries to the original inputs, and let $A\in\{0,1\}$ be its output.

We now use $A$ and the random bits $b_i$ to produce an estimate $B$ of $G$. Define $B\in\{0,1\}$ by
\begin{equation}\label{eq:app-xor-output}
    (-1)^B
    =\operatorname{sgn}(\widehat P(S))
       (2A-1)(-1)^{\sum_{i\in S}b_i}.
\end{equation}
The sign $\operatorname{sgn}(\widehat P(S))$ is well-defined because~\eqref{eq:app-fourier-bound} guarantees that $\widehat P(S)\ne0$.

We next calculate the probability that $B=G$. From
$z_i'=g_L(x^{(i)})\oplus b_i$ for $i\in S$, we have
\begin{equation}
    \chi_S(z')
    =(-1)^{\sum_{i\in S}z_i'}
    =(-1)^{G+\sum_{i\in S}b_i}.
\end{equation}
Combining this with~\eqref{eq:app-xor-output} gives
\begin{equation}
    (-1)^{B+G}
    =\operatorname{sgn}(\widehat P(S))
       (2A-1)\chi_S(z').
\end{equation}

Take the expectation over the random permutations, the bits $b_i$, the independently sampled inputs outside $S$, and the randomness and measurements of $\mathcal A$. Since the conditional distribution of $Y$ given $z'$ is exactly the distribution defining $P(z')$, we have
\begin{equation}
    \mathbb E[2A-1\mid z']=2P(z')-1.
\end{equation}
Therefore,
\begin{align}
    \mathbb E[(-1)^{B+G}]
    &=\operatorname{sgn}(\widehat P(S))
      \mathbb E[(2A-1)\chi_S(z')]\\
    &=\operatorname{sgn}(\widehat P(S))
      \mathbb E[(2P(z')-1)\chi_S(z')].
\end{align}

Recall that $z'$ is uniform on $\{0,1\}^k$. Since $S$ is nonempty, the character $\chi_S$ takes the values $1$ and $-1$ equally often, so $\mathbb E[\chi_S(z')]=0$. By the definition of the Fourier coefficient,
\begin{equation}
    \mathbb E[P(z')\chi_S(z')]
    =2^{-k}\sum_{z'}P(z')\chi_S(z')
    =\widehat P(S).
\end{equation}
Substituting these identities yields
\begin{align}
    \mathbb E[(-1)^{B+G}]
    &=2\operatorname{sgn}(\widehat P(S))
       \widehat P(S)\\
    &=2|\widehat P(S)|.
\end{align}

Finally, $(-1)^{B+G}$ equals $1$ when $B=G$ and $-1$ when $B\ne G$. Hence
\begin{equation}
    \mathbb E[(-1)^{B+G}]
    =\Pr[B=G]-\Pr[B\ne G]
    =2\Pr[B=G]-1.
\end{equation}
Combining this with the preceding equality gives
\begin{equation}
    \Pr[B=G]
    =\frac12+|\widehat P(S)|
    \ge\frac12+\alpha.
\end{equation}

This success-probability bound holds for every fixed tuple of original singleton inputs, because the randomization argument did not use any assumption about their marked positions. Also, the set $S$ and the sign of $\widehat P(S)$ depend only on the algorithm $\mathcal A$, not on the input tuple. They can therefore be fixed in advance.

We have consequently constructed an algorithm that computes the XOR of $\ell\ge k/2$ independent instances of $g_L$, with success probability at least $1/2+\alpha$, using at most $2Q$ Boolean queries.

To apply the XOR lemma (Lemma~\ref{lem:xor}), we express the success probability in terms of its parameter $\rho$. Recall that the XOR lemma allows error probability at most
\begin{equation}
    \frac{1-\rho^{\ell/2}}2,
\end{equation}
which corresponds to success probability at least
\begin{equation}
    \frac12+\frac{\rho^{\ell/2}}2.
\end{equation}
Our constructed algorithm has success probability at least $1/2+\alpha$, so we choose $\rho$ such that
\begin{equation}
    \rho^{\ell/2}=2\alpha.
\end{equation}
Using $\alpha=2^{-3k-1}$ from~\eqref{eq:app-fourier-bound}, this gives
\begin{equation}
    \rho=(2\alpha)^{2/\ell}
    =2^{-6k/\ell}.
\end{equation}
Since $\ell\ge k/2$, we have $6k/\ell\le12$ and therefore
\begin{equation}
    \rho\ge2^{-12}.
\end{equation}
In particular, $\rho$ is bounded below by an absolute positive constant independent of $k$ and $L$.

The constructed XOR algorithm satisfies the error condition of Lemma~\ref{lem:xor} and uses at most $2Q$ queries. Applying that lemma gives
\begin{equation}
    2Q\ge\frac{\ell\rho}{8}\Adv^\pm(g_L).
\end{equation}
Using $\ell\ge k/2$, $\rho\ge2^{-12}$, and the Position-Parity adversary bound $\Adv^\pm(g_L)\ge\sqrt{L/2}$ from Lemma~\ref{lem:parity-adversary}, we obtain
\begin{align}
    2Q
    &\ge\frac{k}{16\cdot2^{12}}\Adv^\pm(g_L)\\
    &\ge\frac{k}{16\cdot2^{12}}\sqrt{\frac L2}
    =\Omega(k\sqrt L).
\end{align}
Dividing by two and absorbing the resulting constant into the $\Omega$ notation gives
\begin{equation}
    Q=\Omega(k\sqrt L).
\end{equation}
This proves Lemma~\ref{lem:outer} with an absolute constant $c_*>0$.
\end{proof}

\section{Proof of Lemma~\ref{lem:spectral}}\label{app:spectrum}

Recall from~\eqref{eq:expander} that $P$ is real symmetric, $P\ket u=\ket u$, $\norm P\le1$, and every eigenvalue of $P$ on $\ket u^\perp$ is at most $1-c$. Therefore $K^0=-P$ has unique ground-state energy $-1$, ground state $\ket u$, and spectral gap at least $c$.

Diagonalize $P$ in an orthonormal eigenbasis. Its eigenvalues lie in $[-1,1]$, so the eigenvalues of $B_-=(1+h)I-P$ lie in $[h,2+h]$. In the same basis, the eigenvalues of $B_-^{-1}$ are their reciprocals, in $[(2+h)^{-1},h^{-1}]$. This proves that $(2+h)^{-1}I\preceq B_-^{-1}\preceq h^{-1}I$. In particular, we have $\bra aB_-^{-1}\ket a\ge1/(2+h)>0$, so
\begin{equation}
  0<\beta_a^-
  =\frac1{\bra aB_-^{-1}\ket a}
  \le2+h\le3.
\end{equation}

As shown immediately before Lemma~\ref{lem:spectral},
$K_a^-$ has an eigenvalue $-1-h$ with eigenvector
$B_-^{-1}\ket a$. It remains to show that this is the
unique ground-state energy and that the spectral gap is
at least $h$. For this, we show that $K_a^-$ has at most
one eigenvalue strictly below $-1$.
Since $-1-h<-1$, this will establish that $-1-h$ is the unique ground-state energy and that the spectral gap is at least $h$. Write the eigenvalues of $K_a^-$ in nondecreasing order as $\mu_1\le\mu_2\le\cdots$. For every $\ket v\perp\ket a$, the definition $K_a^-=K^0-\beta_a^-\ket a\bra a$ gives
\begin{equation}\label{eq:ortho-complement-a-inequaltiy}
\bra vK_a^-\ket v=\bra vK^0\ket v\ge-\norm v^2,
\end{equation}
where the inequality follows from $K^0=-P$ and $\norm P\le1$. Suppose, to the contrary, that two orthonormal eigenvectors $v_1,v_2$ have eigenvalues $\mu_1,\mu_2<-1$. Choose complex numbers $\alpha,\beta$, not both zero, solving the single homogeneous equation $\alpha\braket a{v_1}+\beta\braket a{v_2}=0$. Then $v=\alpha v_1+\beta v_2$ is nonzero and orthogonal to $\ket a$, but
\begin{equation}\label{eq:rank-one-contradiction}
 \bra vK_a^-\ket v=|\alpha|^2\mu_1+|\beta|^2\mu_2
 <-|\alpha|^2-|\beta|^2=-\norm v^2.
\end{equation}
This contradicts the inequality in~\eqref{eq:ortho-complement-a-inequaltiy} established above for every $\ket v\perp \ket a$. Hence at most one eigenvalue lies below $-1$, or equivalently $\mu_2(K_a^-)\ge-1$. The eigenvalue $-1-h$ lies strictly below $-1$, so it is the unique ground-state energy and
\begin{equation}
  \operatorname{gap}(K_a^-)
  =\mu_2(K_a^-)-(-1-h)
  \ge h.
\end{equation}

It remains to bound the guiding overlap. Write $\ket a=N^{-1/2}\ket u+\ket{a_\perp}$, where $\ket{a_\perp}\perp\ket u$ and $\norm{a_\perp}^2=1-1/N\le1$. Since $B_-\ket u=h\ket u$, while every eigenvalue of $B_-$ on $\ket u^\perp$ is at least $c+h$, we have
\begin{equation}
  B_-^{-1}\ket a
  =\frac1{\sqrt N h}\ket u+B_-^{-1}\ket{a_\perp},
  \qquad
  \norm{B_-^{-1}\ket{a_\perp}}
  \le\frac{\norm{a_\perp}}{c+h}
  \le\frac1c.
\end{equation}
The two terms are orthogonal, since $B_-$ preserves $\ket u^\perp$. Normalizing $B_-^{-1}\ket a$ to obtain $\ket{w_a^-}$ therefore gives
\begin{equation}
  |\braket u{w_a^-}|^2
  =\frac{1/(Nh^2)}{1/(Nh^2)+\norm{B_-^{-1}\ket{a_\perp}}^2}
  \ge\frac{1/(Nh^2)}{1/(Nh^2)+1/c^2}
  =\frac1{1+Nh^2/c^2}.
\end{equation}
This proves all claims.

\section{Fixed Support and Simulation of Sparse Oracles}\label{app:oracles}

\begin{proof}[Proof of Lemma~\ref{lem:oracles}]
Recall that $H_x$ acts on $\mathcal A\otimes\mathcal B$, with basis states $\ket{a,b}$ indexed by $(a,b)\in[N]\times\mathbb Z_m$. We first determine which matrix entries are nonzero.

Consider two basis states $\ket{a,b}$ and $\ket{a',b'}$. If $a\ne a'$, only the term $-P\otimes V$ in~\eqref{eq:lift} contributes to their matrix entry. Since every entry of $V$ equals $1/m$, we have
\begin{equation}
  \bra{a,b}H_x\ket{a',b'}
  =-\frac{P_{aa'}}m.
\end{equation}
Thus, for $a\ne a'$, every entry in the block indexed by $(a,a')$ is nonzero if $P_{aa'}>0$, and every entry is zero if $P_{aa'}=0$. These locations are independent of $x$.

Next, consider a diagonal block, corresponding to $a=a'$. Its input-independent part is
\begin{equation}\label{eq:app-diagonal-block}
  MI_m-(M+P_{aa})V.
\end{equation}
We show that every entry in this block is nonzero, regardless of the Boolean input.

First, consider a diagonal entry with $b=b'$. Recall that $W_a(x)$ is a linear combination of shifts $S^j$ and $S^{-j}$ with odd $j$. Since $m$ is even, no odd $j\in\mathbb Z_m$ is congruent to zero modulo $m$. Hence these shifts have zero diagonal entries, and $W_a(x)$ does not contribute to the diagonal. Every diagonal entry of the full block is therefore
\begin{align}
  M-\frac{M+P_{aa}}m
  &\ge M-\frac{M+1}{m}\notag\\
  &\ge M-\frac{M+1}{4}\notag\\
  &=\frac{3M-1}{4}>0,
\end{align}
where we used $0\le P_{aa}\le1$, $m\ge4$, and $M=17/2$.

Now consider an off-diagonal entry with $b\ne b'$. The input-independent part contributes
\begin{equation}\label{eq:app-fixed-entry}
  -\frac{M+P_{aa}}m<0.
\end{equation}
Moreover, every entry of $W_a(x)$ is nonpositive, because~\eqref{eq:w} expresses it as a sum of permutation matrices with nonpositive coefficients. Adding $W_a(x)$ therefore cannot make this entry zero. Thus every entry in a diagonal block is nonzero, independently of $x$.

We can now count the nonzero entries in each row. For a fixed row $(a,b)$, the diagonal block contributes exactly $m$ nonzero entries. Each vertex $a'\ne a$ satisfying $P_{aa'}>0$ contributes another $m$ nonzero entries. Hence the number of nonzero entries in this row is
\begin{equation}
  m\left(1+|\{a'\ne a:P_{aa'}>0\}|\right)
  \in[m,(d+1)m],
\end{equation}
where the upper bound follows because each row of $P$ has at most $d$ nonzero entries. Since $H_x$ is symmetric, the same bound holds for every column. In particular, its nonzero-entry locations depend only on the fixed matrix $P$ and the register dimension $m$, not on $x$.

We next show how to simulate a sparse-value query to $H_x$ using Boolean queries to $x$.
For $a\ne a'$, the entry is $-P_{aa'}/m$, so its value is known independently of $x$. It therefore requires no Boolean queries.
For $a=a'$, the only input-dependent contribution comes from $W_a(x)$. Let
\begin{equation}
  t=b-b'\pmod m,
  \qquad
  t'=b'-b\pmod m,
\end{equation}
where $t,t'\in\mathbb Z_m$. Since $m$ is even and $t'\equiv-t\pmod m$, the integers $t$ and $t'$ have the same parity.
Recall that $S^j\ket{b'}=\ket{b'+j\bmod m}$. Therefore,
\begin{equation}
  \bra bS^j\ket{b'}=1
  \quad\Longleftrightarrow\quad j=t,
\end{equation}
and similarly,
\begin{equation}
  \bra bS^{-j}\ket{b'}=1
  \quad\Longleftrightarrow\quad j=t'.
\end{equation}
Substituting these identities into~\eqref{eq:w}, and recalling that only odd values of $j$ appear in that sum, gives
\begin{equation}\label{eq:app-entry-query}
  \bra bW_a(x)\ket{b'}
  =\begin{cases}
    0,&t\text{ even},\\[2pt]
    -\dfrac{\beta_a^-}{2}
    \bigl(x_{a,t}+x_{a,t'}\bigr),
      &t\text{ odd}.
  \end{cases}
\end{equation}
This expression depends on at most two Boolean bits. It also covers the case $t=t'=m/2$ when $m/2$ is odd. In that case, both shifts contribute the same bit, and the expression becomes $-\beta_a^-x_{a,t}$, as required by $S^{m/2}=S^{-m/2}$.

We now describe the reversible implementation of the value oracle. Let $i=(a,b)$ and $i'=(a',b')$ denote the row and column indices of the requested matrix entry. When $a=a'$ and $t$ is odd, define the two Boolean query addresses
\begin{equation}
  q_1=(a-1)m+t,\qquad
  q_2=(a-1)m+t'.
\end{equation}
Thus $x_{q_1}=x_{a,t}$ and $x_{q_2}=x_{a,t'}$, using the indexing convention introduced in Section~\ref{sec:lift}. When the entry does not depend on $x$, we may instead choose fixed valid addresses and ignore their returned bits.

Let $f(i,i',b_1,b_2)$ be the known function that returns the exact binary encoding of the requested matrix entry when the two Boolean bits have values $b_1,b_2$. The formulas above determine this function completely from the row and column indices, the fixed matrix $P$, and the known coefficients $\beta_a^-$.

We implement the value oracle by first querying the two Boolean bits, then computing the encoded matrix entry, and finally uncomputing the Boolean query registers. Suppressing the registers used to compute the query addresses, the circuit acts as
\begin{equation}\label{eq:value-query-compute-uncompute}
  \begin{aligned}
  \ket{i,i',z,0,0}
  &\longmapsto
    \ket{i,i',z,x_{q_1},x_{q_2}}\\
  &\longmapsto
    \ket{i,i',
      z\oplus f(i,i',x_{q_1},x_{q_2}),
      x_{q_1},x_{q_2}}\\
  &\longmapsto
    \ket{i,i',
      z\oplus[(H_x)_{ii'}],0,0}.
  \end{aligned}
\end{equation}
The first arrow uses two Boolean queries. The second uses reversible arithmetic on known coefficients and requires no additional Boolean queries. The third uses two more Boolean queries to return the query registers to zero. All temporary registers, including those used to compute $q_1$ and $q_2$, are also uncomputed.

The query addresses are determined reversibly from the row and column indices, so the same circuit works when these indices are in superposition. A controlled value query is obtained by controlling only the update to the value register in the middle step, while still computing and uncomputing the Boolean bits. This also uses at most four Boolean queries. Since the value oracle acts by XOR, it is its own inverse, and inverse queries have the same cost.

The entries of $H_x$ are rational and can be represented exactly using finite binary encodings of their numerators and denominators. For the final Hamiltonian family, the common rescaling of Appendix~\ref{app:dyadic} makes every entry dyadic, so the same computation instead returns its exact dyadic encoding. In both cases, the encoded entry is determined by the same two Boolean bits, and the Boolean query cost remains at most four.

Finally, we construct the sparse-location oracles. Since the nonzero-entry locations of $H_x$ are independent of $x$, we can choose the same ordered list of nonzero column positions for every Boolean input. We pad this list with fixed distinct zero-entry positions, if necessary, to obtain $s$ positions as required by~\eqref{eq:pos}.

For the reversible in-place row oracle, extend this list to a permutation of all column labels, as in~\eqref{eq:row-inplace}. Since the entire list is independent of $x$, so are this permutation and its inverse. The column location oracles are constructed analogously using the known column support. Thus every out-of-place position query and every reversible in-place row or column location query, including controlled queries and inverses, can be implemented without querying $x$.

This proves all claims of Lemma~\ref{lem:oracles}.
\end{proof}
\section{Common Dyadic Rescaling}\label{app:dyadic}

We justify the common rescaling used in Section~\ref{sec:gsee-assembly} to ensure that every matrix entry in the final Hamiltonian family has an exact finite binary representation.

\begin{lemma}[Common dyadic rescaling]\label{lem:dyadic}
Let $\mathcal C$ be a finite set of rational numbers. There exists a rational number $\sigma\in(1/2,1]$ such that $\sigma c$ is dyadic for every $c\in\mathcal C$. Multiplication of a finite rational Hamiltonian family by this common positive factor preserves all eigenspaces, nonzero-entry locations, and signs of entries, while multiplying every eigenvalue difference by $\sigma$. In particular, it preserves stoquasticity.
\end{lemma}

\begin{proof}
Choose a positive integer $R$ that is a common denominator of every element of $\mathcal C$, and let $b=\lceil\log_2 R\rceil$. Then
\begin{equation}
  2^{b-1}<R\le2^b.
\end{equation}
Define $\sigma=R/2^b$, so that $1/2<\sigma\le1$. Since $Rc$ is an integer for every $c\in\mathcal C$, we have
\begin{equation}
  \sigma c=\frac{Rc}{2^b},
\end{equation}
which is dyadic.

Now consider any Hamiltonian $H$ in the family. If $H\ket f=E\ket f$, then
\begin{equation}
  (\sigma H)\ket f=(\sigma E)\ket f.
\end{equation}
Since $\sigma>0$, the eigenvectors and the ordering of the eigenvalues remain unchanged, while every difference between eigenvalues is multiplied by $\sigma$.

Furthermore, $(\sigma H)_{ij}=0$ if and only if $H_{ij}=0$, so the nonzero-entry locations are unchanged. Multiplication by a positive scalar also preserves the signs of all real matrix entries. Hence stoquasticity is preserved.
\end{proof}

\section{Proof of the Cayley transducer lemma}\label{app:cayley-transducer}

\begin{proof}[Proof of Lemma~\ref{lem:cayley-upper}]
We construct the transducer explicitly, verify its action on the public and private spaces, and then bound the number of sparse queries needed to compose several copies. The starting point is the sparse representation of Ref.~\cite{BRSL2026}; the compression identities below show how this representation produces the particular Cayley transform~\eqref{eq:phase-map}.

We first spell out the sparse representation so that its two required identities can be checked directly. Let $n=D$ be the dimension of $H$. In the notation of~\eqref{eq:transduction}, set $\mathcal H_{\rm pub}=\mathbb C^n$ and $\mathcal H_{\rm priv}=\mathbb C^n\otimes\mathbb C^n$. Their bases are $\{\ket i\}$ and $\{\ket{i,j}\}$, respectively. Let $\pi_i$ be the row permutation from~\eqref{eq:row-inplace}. The following two operators act on $\mathcal H_{\rm priv}$:
\begin{equation}\label{eq:entrywise-representation}
  L=\sqrt{s}\sum_{i,j=1}^n H_{ij}\ket{i,j}\bra{j,i},
  \qquad
  T\ket{i,\ell}=\ket{i,\pi_i(\ell)}.
\end{equation}
The operator $L$ is Hermitian because $H_{ji}=\overline{H_{ij}}$, and $T$ is exactly the row-location unitary. Define $J:\mathcal H_{\rm pub}\to\mathcal H_{\rm priv}$ and the Hermitian operator $K_0$ on $\mathcal H_{\rm priv}$ by
\begin{equation}\label{eq:uniform-embedding}
  J\ket i=\ket i\otimes\frac1{\sqrt{s}}\sum_{\ell=1}^s\ket\ell,
  \qquad K_0=T^\dagger LT.
\end{equation}
In particular, $TJ\ket i=s^{-1/2}\sum_{\ell=1}^s\ket{i,\pi_i(\ell)}$. The normalization gives $\braket{Ji}{Jj}=\delta_{ij}$, so $J^\dagger J=I_n$; here $\ket{Ji}$ abbreviates $J\ket i$. Let $\mathcal N_i=\{\pi_i(1),\ldots,\pi_i(s)\}$. The matrix element between public basis vectors is
\begin{equation}\label{eq:compression-entry}
 \begin{aligned}
 \bra iJ^\dagger T^\dagger LTJ\ket j
 &=\frac1s\sum_{a\in\mathcal N_i}\sum_{b\in\mathcal N_j}
          \bra{i,a}L\ket{j,b}\\
 &=\frac{H_{ij}}{\sqrt{s}}\,
          \mathbf1_{\{j\in\mathcal N_i\}}\mathbf1_{\{i\in\mathcal N_j\}}
 =\frac{H_{ij}}{\sqrt{s}}.
 \end{aligned}
\end{equation}
Here $\mathbf1_{\{\mathcal P\}}$ equals one when the condition $\mathcal P$ holds and zero otherwise. For the last equality, if $H_{ij}\ne0$, Hermiticity gives $H_{ji}\ne0$ and both indicators are one. If $H_{ij}=0$, both sides vanish regardless of the padding locations. Thus
\begin{equation}\label{eq:cayley-representation}
  J^\dagger K_0J=\frac H{\sqrt{s}},
  \qquad
  J^\dagger K_0^2J
   =\sum_{i=1}^n\left(\sum_{j=1}^n|H_{ij}|^2\right)\ket i\bra i
   \preceq I.
\end{equation}
For the second identity, $L^2\ket{i,j}=s|H_{ij}|^2\ket{i,j}$, and the uniform average in~\eqref{eq:uniform-embedding} cancels the factor $s$. The row list includes every nonzero entry, so padding adds only zeros. Finally, $\sum_j|H_{ij}|^2=\bra iH^2\ket i\le\norm H^2\le1$, which proves the inequality.

We also verify the constant query cost needed below. For any known real scalar $\alpha$, the Cayley transform of $\alpha L$ acts separately on each one-dimensional space $\operatorname{span}\{\ket{i,i}\}$ and each two-dimensional space $\operatorname{span}\{\ket{i,j},\ket{j,i}\}$ with $i<j$. On a two-dimensional space, write $z=\alpha\sqrt{s}H_{ij}$ and
\begin{equation}\label{eq:entrywise-cayley}
  C_z=\begin{pmatrix}0&z\\\overline z&0\end{pmatrix},
  \qquad
  (C_z+\iu I)(C_z-\iu I)^{-1}
  =\frac{(|z|^2-1)I+2\iu C_z}{1+|z|^2}.
\end{equation}
This follows by multiplying numerator and denominator by $C_z+\iu I$ and using $C_z^2=|z|^2I$. On a diagonal space the corresponding factor is $(\alpha\sqrt{s}H_{ii}+\iu)/(\alpha\sqrt{s}H_{ii}-\iu)$.

For $i<j$, encode the ordered pairs $(i,j)$ and $(j,i)$ as the same pair $(i,j)$ together with an orientation bit $o=0$ or $o=1$. This is a reversible relabeling of basis states. Query $H_{ij}$ using the stored ordered pair with $i<j$, apply~\eqref{eq:entrywise-cayley} to $o$, and clear the value register by a second query. For $i=j$, use the one-dimensional phase above. Finally undo the relabeling. All operations are controlled by the stored pair, so the construction applies to arbitrary superpositions of indices. The canonical indices are unchanged by the two-dimensional operation, so the value register can be cleared. This uses two value queries and gates determined by the returned encoding. Conjugating by $T$ and $T^\dagger$ gives the Cayley transform of $\alpha K_0$ using two additional location queries. The inverse and controlled versions have the same constant query cost. As throughout the paper, these are query bounds; we do not count the cost of compiling the entry-dependent gates. This is the entrywise sparse representation underlying Ref.~\cite{BRSL2026}.

We rescale $K_0$ by a factor of four and define
\begin{equation}\label{eq:cayley-defs}
  K=\frac{K_0}{4},
  \qquad B=J^\dagger KJ=\frac{H}{4\sqrt{s}},
  \qquad C_K=(K+\iu I)(K-\iu I)^{-1},
  \qquad P=JJ^\dagger.
\end{equation}
Here $K,C_K,P$ act on $\mathcal H_{\rm priv}$, whereas $B$ acts on $\mathcal H_{\rm pub}$. For each real eigenvalue $\lambda$ of $K$, $\lambda-\iu\ne0$ and $|(\lambda+\iu)/(\lambda-\iu)|=1$. Thus $K-\iu I$ is invertible and $C_K$ is unitary. Also,~\eqref{eq:cayley-representation} implies $J^\dagger K^2J\preceq I/16$. The rescaling is useful because this inequality will give a constant catalyst bound.

We next record an identity for $C_K$. Directly from its definition,
\begin{equation}
  I-C_K
  =\bigl[(K-\iu I)-(K+\iu I)\bigr](K-\iu I)^{-1}
  =-2\iu(K-\iu I)^{-1},
\end{equation}
and hence
\begin{equation}\label{eq:cayley-resolvent}
  (I-C_K)^{-1}=\frac{I+\iu K}{2}.
\end{equation}
This identity will let us express the action of the larger-space unitary $C_K$ using only its compression $B=J^\dagger KJ$.

Define the unitary candidate and the catalyst map by
\begin{equation}\label{eq:swap-transducer}
  R=
  \begin{pmatrix}
    0&J^\dagger\\
    J&I-P
  \end{pmatrix},
  \qquad
  \Gamma=(I+\iu K)J(I+\iu B)^{-1}.
\end{equation}
Here $R$ acts on $\mathcal H_{\rm pub}\oplus\mathcal H_{\rm priv}$, and $\Gamma$ maps the first summand to the second. Every private vector decomposes uniquely as $J\eta+w$ with $J^\dagger w=0$, since $P=JJ^\dagger$ projects onto $J\mathcal H_{\rm pub}$. The block matrix gives the explicit action
\begin{equation}\label{eq:swap-action}
 R\bigl(\xi\oplus(J\eta+w)\bigr)
       =\eta\oplus(J\xi+w).
\end{equation}
Thus it exchanges the two copies of the public space and leaves $w\in\ker J^\dagger$ fixed. Since $P=JJ^\dagger$ is an orthogonal projector, $J^\dagger(I-P)=0$ and $(I-P)J=0$. Hence $R^\dagger=R$ and
\begin{equation}\label{eq:swap-unitarity}
  R^2=
  \begin{pmatrix}
    J^\dagger J&J^\dagger(I-P)\\
    (I-P)J&JJ^\dagger+(I-P)^2
  \end{pmatrix}
  =\begin{pmatrix}I&0\\0&I\end{pmatrix}.
\end{equation}
Thus $R$ is unitary. The inverse in the definition of $\Gamma$ exists because $B$ is Hermitian.

We show that $\Gamma$ satisfies two identities. Using $J^\dagger J=I$ and $J^\dagger KJ=B$, we first obtain
\begin{equation}\label{eq:gamma-compression}
  J^\dagger\Gamma
  =J^\dagger(I+\iu K)J(I+\iu B)^{-1}
  =(I+\iu B)(I+\iu B)^{-1}=I.
\end{equation}
Using~\eqref{eq:cayley-resolvent} then gives
\begin{equation}\label{eq:gamma-resolvent}
  (I-C_K)\Gamma
  =2(I+\iu K)^{-1}(I+\iu K)J(I+\iu B)^{-1}
  =2J(I+\iu B)^{-1}.
\end{equation}
We will use the first identity to recover the public input, and the second to show that the private vector is returned unchanged.

Consider the action of $R(I\oplus C_K)$ on $\ket\xi\oplus\Gamma\ket\xi$. By the block form of $R$, its public component is $J^\dagger C_K\Gamma\ket\xi$. Since $C_K\Gamma=\Gamma-(I-C_K)\Gamma$, equations~\eqref{eq:gamma-compression} and~\eqref{eq:gamma-resolvent} give
\begin{equation}\label{eq:cayley-public-action}
  J^\dagger C_K\Gamma\ket\xi
  =\left[I-2(I+\iu B)^{-1}\right]\ket\xi.
\end{equation}
Its private component is $J\ket\xi+(I-P)C_K\Gamma\ket\xi$. Equation~\eqref{eq:gamma-resolvent} implies $(I-P)(I-C_K)\Gamma=0$, since $(I-P)J=0$, and therefore $(I-P)C_K\Gamma=(I-P)\Gamma$. Combining this with $P\Gamma=JJ^\dagger\Gamma=J$ yields
\begin{equation}\label{eq:cayley-private-action}
  J\ket\xi+(I-P)C_K\Gamma\ket\xi
  =P\Gamma\ket\xi+(I-P)\Gamma\ket\xi
  =\Gamma\ket\xi.
\end{equation}
Thus, the private component is exactly unchanged. The public component in~\eqref{eq:cayley-public-action} has the opposite sign from the desired transform, so we correct that sign by defining $S=(-I\oplus I)R(I\oplus C_K)$. We then have
\begin{equation}\label{eq:cayley-identity}
  S(\ket\xi\oplus\Gamma\ket\xi)
  =\left[2(I+\iu B)^{-1}-I\right]\ket\xi
   \oplus\Gamma\ket\xi.
\end{equation}

It remains to identify the public unitary. Since $B$ commutes with $I+\iu B$, we have
\begin{equation}
  2(I+\iu B)^{-1}-I
  =(I-\iu B)(I+\iu B)^{-1}.
\end{equation}
For every real $b$, the scalar identity $(1-\iu b)/(1+\iu b)=\exp(-2\iu\arctan b)$ holds, so functional calculus gives
\begin{equation}
  (I-\iu B)(I+\iu B)^{-1}
  =\exp(-2\iu\arctan B)=U_H.
\end{equation}
Equation~\eqref{eq:cayley-identity} is therefore the exact transduction relation~\eqref{eq:transduction} for $U_H$.

To bound its cost, use the definition of $\Gamma$ and the Hermiticity of $K$ and $B$ to calculate
\begin{align}
  \norm{\Gamma\ket\xi}^2
  &=\bra\xi(I-\iu B)^{-1}
    J^\dagger(I-\iu K)(I+\iu K)J
    (I+\iu B)^{-1}\ket\xi\notag\\
  &=\bra\xi(I-\iu B)^{-1}
    (I+J^\dagger K^2J)(I+\iu B)^{-1}\ket\xi\notag\\
  &\le\frac{17}{16}\norm{\ket\xi}^2.
  \label{eq:catalyst-bound}
\end{align}
To see the last inequality directly, put $y=(I+\iu B)^{-1}\xi$. The preceding line is $\bra y(I+J^\dagger K^2J)\ket y\le(17/16)\norm y^2$. Each eigenvalue of $(I+\iu B)^{-1}$ has modulus $1/\sqrt{1+b^2}\le1$ for a real eigenvalue $b$ of $B$. Hence $\norm y\le\norm\xi$. Thus, the catalyst has squared norm at most $17/16$ on every normalized input.

The implementing unitary $S$ consists of $R$, the known sign correction, and one application of $C_K$. The embedding $J$ in~\eqref{eq:uniform-embedding}, and hence the swap $R$ in~\eqref{eq:swap-transducer}, use no matrix queries. The construction following~\eqref{eq:entrywise-cayley}, with $\alpha=1/4$, implements $C_K$ using $O(1)$ sparse queries. Inverting~\eqref{eq:transduction} gives $S^\dagger(\ket y\oplus\Gamma U_H^\dagger\ket y)=U_H^\dagger\ket y\oplus\Gamma U_H^\dagger\ket y$. Thus the inverse transducer has catalyst map $\Gamma U_H^\dagger$, whose norm obeys the same bound because $U_H^\dagger$ is unitary. For a controlled public action, take the identity transducer on the inactive control branch and the transducer just constructed on the active branch; the required controlled sparse oracles are included in the model of Section~\ref{sec:sparse-matrix-access}.

Finally, consider a circuit with $q\ge1$ matrix-dependent queries $U_1,\ldots,U_q$, each equal to $U_H$, $U_H^\dagger$, or a controlled version. Write its unitary as
\begin{equation}\label{eq:interleaved-circuit}
  \mathcal U=G_qU_qG_{q-1}U_{q-1}\cdots G_1U_1G_0,
\end{equation}
where the $G_j$ make no sparse-matrix queries. They may include calls to the guiding-state oracle, whose cost is not part of the sparse-query count. Any measurements and classical control can first be stored coherently in additional registers. At each query position, the choice of target register and oracle direction can be implemented coherently using controlled swaps and controlled queries to the oracle and its inverse. Branches requiring fewer queries are padded with inactive controlled queries. Thus, all branches can be represented by a common query schedule, increasing the query count by at most a factor of two. Here, $q$ denotes the length of this schedule.

Apply $G_0$ directly and group each $G_j$ with $U_j$. If $S_j$ transduces $U_j$ with catalyst map $\Gamma_j$, then $(G_j\oplus I)S_j$ transduces $G_jU_j$ with the same catalyst. For a normalized initial vector $\xi$, set $\xi_1=G_0\xi$ and $\xi_{j+1}=G_jU_j\xi_j$ for $1\le j<q$. These are the states before the successive components, with all work registers included, and $\norm{\xi_j}=1$ because every factor is unitary. The uniform catalyst bound~\eqref{eq:catalyst-bound} remains valid on these enlarged spaces. Proposition~9.10 of Ref.~\cite{Belovs2024} gives a transducer for the $q$ components with squared catalyst norm
\begin{equation}\label{eq:composed-catalyst-cost}
  W\le\sum_{j=2}^{q}\norm{\ket{\xi_j}}^2
       +\sum_{j=1}^q\norm{\Gamma_j\ket{\xi_j}}^2
    \le(q-1)+\frac{17q}{16}=O(q).
\end{equation}
The catalyst lies in a direct sum with one copy of the public space for each $j=2,\ldots,q$ and one private space for each component $j=1,\ldots,q$. Its components are the intermediate vectors $\xi_j$ and the component catalysts $\Gamma_j\xi_j$. Distinct summands are orthogonal, so their squared norms add as displayed; no tensor product of these vectors is required. One invocation of the composed implementing unitary coherently selects a component operation. It does not execute all $q$ matrix queries in sequence. Concretely, the selection has the form $\sum_{j=1}^q\ket j\bra j\otimes S'_j$, where $S'_j=(G_j\oplus I)S_j$. The component spaces are embedded in a common space, with the unitaries extended by the identity, and shorter oracle sequences are padded by inactive calls. Before each of the constant number of oracle positions, known controlled gates route the selected component's query registers into a common oracle register. One controlled call to the required oracle type then handles every component label in superposition. There are only constantly many oracle types and calls per component. The selected $G_j$ may require many gates, but it makes no sparse queries. Thus selection costs $O(1)$ sparse queries.

By Theorem~3.2 of Ref.~\cite{Belovs2024}, a transducer with catalyst bound $W$ can be converted into a circuit with Euclidean output error at most $\tau>0$ using $O(1+W/\tau^2)$ invocations of its implementing unitary. With any fixed $\tau$ and~\eqref{eq:composed-catalyst-cost}, the total is $O(q)$ sparse queries. The bound holds uniformly over inputs and untouched reference registers, because~\eqref{eq:catalyst-bound} is an operator bound. Discarding auxiliary registers or performing the deferred measurements cannot increase trace distance. This proves Lemma~\ref{lem:cayley-upper}.
\end{proof}

\bibliography{ref}
\bibliographystyle{alpha}

\end{document}